  \documentclass{aastex}
  \usepackage{graphicx,epsfig,emulateapj5}
\newcommand{\etal}{{et al.~}}

\newcommand{\bq}{\begin{equation}}
\newcommand{\eq}{\end{equation}}

\def\gtsim{\lower.5ex\hbox{$\buSildrel > \over\sim$}}
\def\ltsim{\lower.5ex\hbox{$\buildrel < \over\sim$}}
\def\arcsec{^{\prime\prime}}

\def\sun{\mbox{$_\odot$}}
\def\apjl{ApJL}
\def\apj{ApJ}
\def\apjs{ApJS}

\def\araa{ARAA}

\def\aap{A\&A}

\begin{document}
\submitted{Accepted by The Astrophysical Journal}

\title
{The Central Region of Barred Galaxies: Molecular Environment, Starbursts, and 
Secular  Evolution}
\author {Shardha Jogee\altaffilmark{1}, Nick Scoville\altaffilmark{2}, 
\& Jeffrey D. P. Kenney\altaffilmark{3}} 
\authoremail{sj@astro.as.utexas.edu, nzs@astro.caltech.edu, kenney@astro.yale.edu}
\altaffiltext{1}{
Department of Astronomy, University of Texas at Austin, 
1 University Station C1400, Austin, TX 78712-0259}
\altaffiltext{2}{Division of Physics, Mathematics, and Astronomy, MS 105-24, 
California Institute of Technology, Pasadena, CA 91125}
\altaffiltext{3}{Yale University Astronomy Department, New Haven, CT 06520-8101}
\begin{abstract}
Stellar bars drive gas  into the circumnuclear (CN) region of galaxies. To investigate the fate of the CN gas and star formation (SF), we study a sample of barred non-starbursts  and  starbursts with high-resolution CO, optical, H$\alpha$, RC,  Br$\gamma$, and  $HST$ data, and find the following. (1) The inner kpc of bars differs markedly from the outer disk. It hosts molecular gas  surface densities $\Sigma_{\rm gas-m}$ of 500-3500  M$_{\tiny \sun}$ pc$^{-2}$,  gas mass fractions of  10 to 30\%, and epicyclic frequencies of several 100--1000 km~s$^{-1}$~kpc$^{-1}$. Consequently, in the CN region, gravitational instabilities can only grow at high gas densities and on short timescales, explaining in part why powerful starbursts reside there. (2) Across the sample,  we find bar pattern speeds with upper limits of  43 to 115 km~s$^{-1}$~pc$^{-1}$  and  outer inner Lindblad resonance (OILR) radii of $>$ 500 pc. (3) Barred starbursts and  non-starbursts have CN SFRs of  3--11 and 0.1--2 M$_{\tiny \sun}$ yr$^{-1}$, despite similar CN gas mass. $\Sigma_{\rm gas-m}$ in the starbursts is larger  (1000--3500 $M_{\tiny \sun}$ pc$^{-2}$) and close to the  Toomre critical density over  a large region. (4) Molecular gas makes up 10\%--30\% of the  CN dynamical mass, and fuels large CN SFRs in the starbursts, building young, massive, high $V/\sigma$ components. Implications for secular  evolution along the Hubble sequence are discussed.
\end {abstract}

\keywords{galaxies: starburst--- galaxies: ISM ---galaxies: structure --- galaxies:kinematics and dynamics---galaxies: evolution}

\section{Introduction}
It is widely recognized that  non-axisymmetries,  such as 
large-scale stellar bars,  
facilitate  the radial transfer of angular momentum in 
disk galaxies, thus  driving their dynamical and  
secular evolution  (e.g., Pfenniger \& Norman 1990; Friedli \& Benz 1995;
Athanassoula 1992; Kormendy 1993;  Kormendy \&  Kennicutt 2004; review by Jogee 2004 
and references therein). 
Numerous studies (e.g., Knapen et al. 2000; Eskridge et al. 2000) 
have  shown  that the  majority ($>$ 70 \%) 
of nearby disk galaxies  host large-scale stellar bars. 
Using $HST$ Advanced Camera for Surveys (ACS) data, 
Jogee \etal (2004a,b)  find that strong bars remain frequent  
from the present day out to lookback times of 8 Gyr ($z \sim$~1), and infer that
bars are long-lived features with a lifetime well above 2 Gyr.  A non-declining
optical bar fraction out to $z \sim$~1 is also confirmed by 
Elmegreen \etal  (2004). 
These findings suggest that bars can influence a galaxy over a significant
part of its lifetime. Compelling evidence that   bars efficiently 
drive  gas from the outer disk into the inner kpc 
comes from the larger central molecular  gas concentrations 
observed in barred galaxies compared to unbarred galaxies
(Sakamoto \etal 1999).

However, few  high resolution studies based on  large samples 
exist of the fate of gas once it reaches the inner kpc of a bar.
In this study, we  use  high 
resolution  ($2 \arcsec$; $\sim$~200 pc) CO  ($J$=1--0) 
observations, optical, NIR, H$\alpha$, radio continuum (RC),  
Br$\gamma$,  and archival $HST$ 
data in order to characterize  the molecular environment, the onset 
of starbursts, and the   dynamical evolution  in 
the  circumnuclear  region of barred galaxies.
Our sample consists of ten  carefully selected  nearby 
barred non-starbursts  and starbursts, including some of the  most 
luminous  starbursts within 40 Mpc.
Given the time-consuming nature of   interferometric $2 \arcsec$
CO observations, this is one of the largest  sample studied 
at this resolution.
CO observations of  nearby galaxies have in the past often 
been limited to a a few individual systems,  and it is only in  
recent years  that we have seen a systematic mapping of  
sizable  samples  (10--12) of   nearby galaxies  (e.g.,  Baker 2000;  Jogee 1999).    
Complementary surveys such as the  
$7''$ resolution  BIMA CO(1--0) Survey of  Nearby Galaxies  (SONG; 
Regan  et al. 2001; Helfer et al. 2001) or the NMA--OVRO  $4''$ 
resolution CO(1--0)  survey of galaxies  (Sakamoto et al.  1999)
are better suited  for studying extended structures or the  
average properties within the inner kpc.

The main sections of the paper are as follows.
$\S$ 2 outlines  the sample selection and $\S$ 3  the 
properties of stellar bars in the sample galaxies.
The imaging  and  interferometric CO  observations are
covered in $\S$ 4.  
$\S$ 6 highlights the extreme  molecular environment which has built up 
in the inner kpc of barred galaxies and discusses its implication.
In $\S$ 7, we estimate the SFR in the circumnuclear region using  
different  tracers  such as  Br$\gamma$, RC, and FIR luminosities
In $\S$ 8, we compare the  circumnuclear molecular gas  
in the barred starbursts and non-starbursts in order to 
investigate  why they have such different SFR/$M_{\tiny \rm H2}$ in 
the inner few kpc.
In $\S$ 9, we push these investigations further by comparing 
the observed circumnuclear gas surface density  to the critical 
density for the  onset of gravitational  instabilities. 
In $\S$ 10, we investigate where the molecular gas has piled 
up with respect to  the  dynamical  resonances of the bar. 
In $\S$ 11, we  discuss the results  within the context of 
bar-driven  secular evolutionary  scenarios.
For readers interested in specific galaxies, $\S$ 12 describes 
the molecular gas distribution and kinematics of 
individual galaxies. 
$\S$ 13 summarizes our main results.

\section{Sample Selection}

In this study, we wish to investigate the properties of molecular
gas in the circumnuclear  region (inner 1--2 kpc)  of barred galaxies, 
the parameters which control  when they form stars, 
and  the reasons  for the large range in SFR/M$_{\tiny H2}$
 seen in the inner few kpc.  
In order to effectively address these questions, we will  carry out 
high resolution ($2\arcsec$) interferometric CO  observations  ($\S$ 2) 
of the sample  galaxies. 
However, when selecting the sample prior to such observations,  
we necessarily have to use existing  low resolution  
($45\arcsec$) single dish CO luminosities 
(e.g., Young et al. 1995)  as a first measure of the  mass of  cold 
molecular hydrogen  
(Dickman, Snell, \& Schloerb 1986; Scoville \& Sanders 1987). 
We obtain a  measure of the massive   SF rate   within 
the same $45\arcsec$ aperture from the 
RC    luminosity  ($L_{\tiny \rm RC}$) 
at  1.5 GHz  (e.g., Condon et al. 1990). 
Empirical evidence such as the FIR-RC  correlation  
(Condon et al. 1982) 
the H$\alpha$- RC  correlation (Kennicutt 1983), as well 
as simple physical models (e.g., Condon 1992) suggest that the 
RC   luminosity  at  1.5 GHz is a good tracer of the massive SFR 
on sufficiently large scales.  
The FIR-radio correlation also  appears to hold locally 
within individual galaxies at least on scales  $\ge$ 1 kpc 
(Condon  1992). 

We   selected ten sample galaxies  which 
satisfy the following criteria: 
(i) 
\rm 
We avoided systems with very low  CO luminosities, which  require 
inordinate investments of time for interferometric CO observations. 
Instead, we picked galaxies which are fairly gas-rich but still  
span an order of magnitude in circumnuclear molecular gas content   
(a few $\times$ 10$^{8}$--10$^{9}$ M$_{\tiny \sun}$) and  a wide range 
in SFR per unit mass of gas. These galaxies were chosen from  
different parts of the 
$L_{\tiny \rm RC}$/$L_{\tiny \rm CO}$ versus $L_{\tiny \rm CO}$ 
plane (Fig. 1);
(ii)
\rm
They have moderate  distances  ($D$=10 to 34 Mpc) 
so that the $2\arcsec$ interferometric CO observations ($\S$ 2) 
can accurately  resolve the gas distribution and kinematics 
on scales of 100-300 pc;
(iii) 
\rm 
They have inclinations below 
70 $\deg$ so that  we can minimize extinction problems problems and  
get a good handle on the gas kinematics;
(iv)
\rm 
They have  complementary high resolution observations  such as  
Brackett-$\gamma$ fluxes (Puxley, Hawarden, \& Mountain  1988,  1990), 
and/or 5 GHz RC observations 
(Saikia et al. 1994) 
for tracing the SF activity at  resolutions comparable to 
that of the $2 \arcsec$ interferometric CO observations; 
(v)
\rm  
We excluded  major mergers with strong morphological distortions 
in the inner kpc region  as the associated highly non-axisymmetric gas 
distributions and large non-circular motions often preclude an 
accurate dynamical analysis. This is discussed further in $\S$ 3.

Table~1  gives the general properties of the sample.
Table~2  shows $L_{\tiny \rm CO}$, $L_{\tiny \rm RC}$, and 
the luminosities  of different  SF tracers within various apertures.  
Fig.~1 shows  the sample galaxies plotted in the 
$L_{\tiny \rm RC}$/$L_{\tiny \rm CO}$ versus $L_{\tiny \rm CO}$
plane. The term  `starburst' 
is used in a multitude of ways in the literature: in some 
cases, a large SF luminosity alone is used to identify a 
star-forming region as a starburst, while in 
other cases the duration of the burst is also taken into account.  
In this work, the term circumnuclear 
starburst refers to  a luminous and short-lived episode of 
 SF  in the inner  kpc region. 
We adopt a working definition where the circumnuclear starbursts 
have a ratio ($L_{\tiny \rm RC}$/$L_{\tiny \rm CO}$) 
above 0.5,  which corresponds to a
gas consumption timescale ($M_{\tiny \rm H2}$/SFR)  
below 1 Gyr, under standard assumptions.
The Milky Way has a 
(SFR/$M_{\tiny \rm H2}$) ratio   above  10$^{-9}$  yr$^{-1}$, and would 
be a non-starburst. 
The circumnuclear starbursts in our sample  (NGC 470, NGC 2782, 
NGC 3504, NGC 4102, and NGC 4536) are   marked on Fig. 1 
and are among the most luminous  starbursts within  40 Mpc. 
While  they are  selected on the basis 
of RC  luminosities, the  circumnuclear starbursts  also 
have a large  FIR luminosity,  
comparable to that of the prototypical starburst  M82, 
which is marked 
for reference on Fig. 1. 
Four of the  starbursts (NGC 470, NGC 2782, NGC 3504, and NGC 4102)  
in fact belong to  Devereux's (1989) sample of the 
twenty brightest nearby (15 $< D <$ 40 Mpc) circumnuclear starbursts 
selected on the basis of a high far-infrared luminosity 
($L_{\tiny \rm FIR} > 2.9 \times 10^{9 }L_{\tiny \sun}$) and  central 
10 $\micron$ luminosity ($L_{10 \micron}$ $>$  $6 \times 10^{8}  L_{\tiny \sun}$).

Note  that in Figure 1, \it
the SFR per unit mass of  molecular gas 
in the central $45\arcsec$ spans more than an order of magnitude 
for a given total  mass of  molecular gas .
\rm
Can this range  be due to the different distance  of 
the sample galaxies causing  the $45\arcsec$ beam  to encompass 
regions of different sizes ? We believe this is unlikely 
because  most of the SF activity is concentrated well 
inside the $45\arcsec$ beam, and 80\% of the galaxies have
comparable distances in the range 10--20 Mpc.
In fact, a  similar range in  the SFR per unit mass of  
molecular gas is obtained if the central 10 micron luminosity 
(Giuricin et al. 1994;  Devereux 1989) is used instead of the radio
continuum  to trace the SFR. 
It is also valid to ask whether  variations in the 
CO-to-H$_{2}$ conversion factor  $\chi$  
between the galaxies  might reduce the range in 
(SFR/$M_{\tiny \rm H2}$) on Fig. 1.
We refer the reader to $\S$ 5 for a discussion on  $\chi$,
but note here that  the  somewhat lower conversion 
factor in starbursts suggested by some studies (Aalto et al. 1994, 1995), 
would in fact  \it increase \rm  rather than reduce  
the range in (SFR/$M_{\tiny \rm H2}$)  in Fig. 1.
Therefore, it appears that the range in  SFR/$M_{\tiny \rm H2}$
within the sample is a truly inherent property.

\section{Stellar Bars and Tidal Interactions in the Sample}

It is widely held that large-scale stellar bars, whether tidally or spontaneously  
induced, efficiently drive gas from the outer disk into the inner kpc via 
gravitational torques, thereby building large gas concentrations  ($\S$ 1). 
In fact, bars are found to  be the main driver of gas inflows into 
the circumnuclear region  even in  interacting systems,
 namely,  in a large fraction  of  minor mergers, 
most intermediate 1:3 mass ratio mergers and  the early phases of most  
major mergers(Jogee  2004 and references therein).
The properties of the sample galaxies, which  
have moderate to large amounts of circumnuclear  
molecular gas  in the range of a few $\times$ (10$^{8}$--10$^{9}$) 
M$_{\tiny \sun}$, are consistent with this picture.
While we have excluded systems with  strong morphological distortions ($\S$ 2), 
all of the sample galaxies, except for NGC 6951, show some evidence for tidal 
interaction in the recent past, or are in environments 
where they could potentially have  interacted with or accreted
other galaxies.  All of them host large-scale  stellar bars 
and  oval distortions, which is some cases appear to be tidally triggered.

Table 3  illustrates the evidence for recent tidal interactions.
NGC 470 is interacting with the large S0 galaxy NGC 474. 
NGC 2782  appears to have undergone 
a close, recent interaction with another smaller disk galaxy, 
and  its  properties have been modeled with an 
intermediate mass ratio (1:4) interaction by Smith (1994). 
NGC 3359  has a nearby companion. NGC 3504, NGC 4314, and 
NGC 4102 dwell in groups, while NGC 4569 and NGC 4536 are part of 
the Virgo cluster. 
As described below, NGC 4569 shows several indications of a
recent tidal disturbance. 

There is evidence for stellar bars and  oval distortions in all 
the  sample galaxies, according to our optical $R$-band and 
NIR images, as well as  the  Third 
Reference Catalog of Bright Galaxies (RC3; de Vaucouleurs et al. 1991; 
Table 3).Figs. 2a-c illustrate  the prominent bars in  
NGC 3504,  NGC 4314, and NGC 6951.
In other cases such as NGC 4102,  NGC 4536, or NGC 2782, there is 
less  conspicuous evidence for a  large-scale bar.
Both the $K$ (Fig. 2d) and $R$-band images of NGC 4102 reveal 
a weak oval distortion, with a surface brightness profile 
reminiscent of a lens (Jogee \& Kenney 1996). This may be the remnant 
of a weakened or destroyed bar .
In   NGC 4536,  despite the large  inclination (66 $\deg$), 
prominent spiral arms in the outer disk, appear  to
branch out from an oval structure, suggesting  the presence of 
a large-scale stellar bar. 
Isophotal analysis of NIR images  also  indicate the presence 
of nested nuclear stellar bars in  NGC 2782 (Jogee, Kenney, 
\& Smith 1999;  $\S$ 12) and NGC 4314 (Friedli et al. 1996; 
$\S$ 12).

In the non-starburst NGC 3359 and NGC 4569, 
the  stellar bar may have been recently tidally  triggered. 
After comparing simulation results 
(Friedli, Benz, \& Kennicutt  1994) with  the observed 
properties of NGC 3359, 
such as the presence of HII regions along the bar and the  
abundance gradient profile, Martin \& Roy (1995) claim 
that the  bar  in NGC 3359,  is younger  than a Gyr.
A nearby  companion (Ball 1986) may have tidally triggered this bar.
In NGC 4569, the K-band isophotes  ($\S$ 8.1) in the inner 
$45\arcsec$ (3.6 kpc) are significantly more elongated than those of 
the outer disk, and suggest the presence of a stellar bar-like feature 
that lies at a 
P.A. of $\sim$ 15 $\deg$ and extends outside the boxy bulge of diameter 
$30\arcsec$  (2.9 kpc). The bar  is more elongated to the 
north than to the south and the bulge isophotes also show east-west asymmetries. 
Furthermore, optical images (e.g., Sandage \& Bedke 1994) show a warp 
in the outer stellar disk.
A merger event or tidal interaction is likely to be responsible for 
the asymmetries in the stellar and CO ($\S$ 8.1) morphologies.
Ram pressure stripping which  was previously invoked to account for the 
HI properties (Kenney \& Young 1989; Cayatte et al. 1994) primarily affects  
the diffuse atomic gas only.

\section{Observations}

\subsection{CO (J=1-$>$0) observations}

Table 4  describes  the CO observations and channel  maps.
For most of the  sample galaxies 
(NGC 470, NGC 2782, NGC 3310, NGC 4102, NGC 4536, NGC 4569, and 
NGC 3359),  2$''$  observations  of the CO (J=1-$>$0) line whose
rest frequency is 115.27 GHz were made between 1995 and 2002  
with the  OVRO millimeter-wave interferometer (Padin et al. 1991).  
The other four were observed previously by Kenney \etal  (1992; 
NGC 3351 and NGC 6951; 1993; NGC 3504)) and  Benedict \etal 
(1996; NGC 4314).   As of 1996, the array 
consists of six 10.4-m antennae with a  primary HPBW 
of 65 $''$ at 115  GHz. 
The seven galaxies were observed in the low, equatorial, and high 
resolution configurations with projected baselines ranging 
from 15 to 242 m. Data were obtained simultaneously 
with an analog continuum correlator of bandwidth 1 GHz and 
a digital spectrometer set-up  which produced four independent modules 
that each have 32 channels with a velocity resolution of 
 10.4 km s$^{-1}$. 
For our observations, the modules were partially overlapping and covered a 
total bandwidth of 1200 km s$^{-1}$ with 116 frequency channels. 
We corrected for temporal phase variations by interspersing 
integrations on the galaxy with observations of  a phase 
calibrator (typically a quasar) every 20-25 minutes.
Passbands were calibrated on the bright quasars
3C273, 3C84, and 3C345. The absolute flux scale,  determined 
from observations of  Uranus, Neptune, and 3C273, 
is accurate to 20\%.  
The passband and flux calibration of the data were carried out using the 
Owens Valley millimeter array software  (Scoville et al. 1993). 
We used the NRAO AIPS software to Fourier transform the calibrated
uv data and deconvolve the channel maps with the `CLEAN' algorithm
as implemented in the AIPS tasks `MX' and `IMAGR'. 
Channel maps with both  uniform and natural weighting 
were made for the galaxies. 
The naturally weighted maps capture more of the low signal-to-noise 
emission, but have lower spatial resolution.

\subsection{Optical observations}

We observed  the ten sample galaxies using the  
Wisconsin Indiana Yale NOAO (WIYN) 3.5-m  
telescope at Kitt Peak National Observatory (KPNO), 
and the 0.9-m  telescope at KPNO.
On the WIYN telescope we used  a 2048 x 2048 S2KB CCD with a plate scale 
of 0.2$''$/pixel, and a  field of view of $6.8'$x $6.8'$.
On the 0.9-m, the 2048 x 2048 T2KA CCD had a plate scale of $0.68''$/pix, 
and field of view of $23.2'$ x $23.2'$.
The galaxies were imaged with broad-band Harris $ B V R I$   filters, 
 narrow-band filters centered on the  appropriately redshifted 
H$\alpha$+[N II]  6563 6583 $ \rm \AA$ emission lines, and 
narrow off-line filters centered on a wavelength $\sim$ 
80  $\rm \AA$ redward of the H$\alpha$+[NII] line. 
The latter images are used to subtract stellar continuum from the 
H$\alpha$+[NII] image, and the continuum bandpass in this 
region is sufficiently close to the H$\alpha$+[NII] emission line 
that spectral color correction terms are generally negligible 
(Taylor et al. 1994).
Standard fields of Landolt stars (e.g., 
PG0918+029, PG1047+003, PG1323-086, WOLF 629, SA110-502)  
were observed throughout the night over a range of airmasses 
for extinction correction. 
For some galaxies, where we had photometric conditions, 
exposures of the galaxy and of spectrophotometric standards at similar 
airmasses  were taken for flux calibrating the  H$\alpha$+[NII] observations. 
The IRAF Package was used for data reduction. After 
bias-subtraction  and flat-fielding, 
the  images in each filter were registered 
using unsaturated field stars, combined, and cleaned free of cosmic rays. 
After sky subtraction, the off-line narrow-band image was 
subtracted from the on-line image to obtain continuum-free  H$\alpha$+[NII] 
line images. Where available, observations of  spectrophotometric standards
were used to  flux calibrate the line. 

\section{Circumnuclear Molecular Gas Content}

The mass of molecular hydrogen ($M_{\rm H2}$) is estimated 
 from  the total flux  using  the relation 
(e.g., Scoville et al. 1987; Kenney \& Young 1989)

\begin{equation}
\frac {M_{\rm H2}} {(M_{\tiny \sun})} = 1.1 \times 10^4 
\ (\frac {\chi}{2.8 \times 10^{20} })\ (D^2) \ (\int{S_{\tiny \rm CO}}dV)
\end{equation}

where  $D$ is the  distance  in Mpc, $ \int{S_{\tiny \rm CO}}dV$  is the 
integrated CO line flux  in Jy km s$^{-1}$, $\chi$ is  the  CO-to-H$_{\rm 2}$ 
conversion factor, defined as  the ratio of the 
beam-averaged column density of hydrogen to the integrated 
CO brightness temperature ($N(H_{2}$)/$I_{\tiny \rm CO}$).

The CO-to-H$_{\rm 2}$ conversion 
factor $\chi$ depends on many factors  including 
the metallicity of the gas, 
the dust column density, the ambient radiation field, 
the physical conditions  in the gas, and the optical thickness of the line.  
All the sample targets are massive spirals and  are 
likely to have solar or super-solar metallicities 
(Vila-Costas \& Edmunds 1992).
In this regime, the CO line is optically thick and  
there is only a weak dependence of  $\chi$ on 
metallicity  (e.g., Elmegreen 1989) so  that our
use of a single $\chi$  value across the sample is 
justified.

Values of $\chi$   for  molecular clouds in the 
Milky Way range from 
 1.8  to   4.5  $ \times  10^{20}$  (K km s$^{-1}$)$^{-1}$, 
based on different methods such as 
comparisons  of the integrated $^{13}$CO intensities  
with the visual extinction  (e.g., Dickman 1975), 
observations of $\gamma$ ray fluxes (e.g., Bloemen et al.  1986; 
Strong et al. 1988),  and  virial equilibrium considerations 
(e.g., Dickman et al. 1986; Scoville et al. 1987;   Solomon et al. 1987).  
It is arguable whether a a Milky Way conversion factor applies to 
the actively star-forming circumnuclear regions of galaxies,  
where the temperatures and densities are much higher. 
For clouds  in virial  equilibrium,  $\chi$ depends on  
 $\rho^{0.5}$/$T_{\rm B}$  (e.g., Scoville \& Sanders 1987),  where  $T_{\rm B}$ is the 
CO brightness temperature averaged over the cloud and 
$\rho$ is the density. One might argue, therefore, that 
the effects of elevated temperatures and densities will 
partially  offset each other. On the other hand,  the 
gas may not be in virial equilibrium and multiple-line studies 
combined with radiative transfer modeling  (e.g., Wall \& Jaffe 1990; 
Helfer \& Blitz 1993; Aalto et al. 1994, 1995) have suggested 
that $\chi$ is lower by a factor of $\sim$ 3 in the centers
of some starburst galaxies, as compared to the Milky Way. 
However, given the absence of multiple-line studies or 
determinations of $\chi$ in our  sample galaxies, 
we adopted in this paper  the  value of $ 2.8 \times 10^{20}$  
(K km s$^{-1}$)$^{-1}$  that was derived for the inner Galaxy  
(Bloemen et al. 1986),  while bearing in mind the
possible uncertainties.

Table 5  shows the CO flux and the mass of molecular hydrogen detected
by the interferometric observations. In most galaxies, we capture a 
large  fraction (60 to 90 \%) of the  CO  flux 
detected  by single dish  observations.  
We find  massive gas 
concentrations -- $3 \times  10^{8}$ to  $ 2 \times  
10^{9}$ $M_{\tiny \sun}$ 
of molecular hydrogen --in  the inner 2  kpc radius of the
sample galaxies. 
In particular,   many galaxies with different RC   
luminosities within the central $45\arcsec$  (e.g., NGC  6951 and NGC  
4569 as compared to NGC  2782, NGC  3504, and NGC  4102)  can 
have quite comparable circumnuclear molecular hydrogen  
content ($\sim$ $ 1-2 \times  10^{9}$ $M_{\tiny \sun}$).

Fig. 3 shows the CO total intensity (moment 0) maps, 
the size of the synthesized beam and the position of 
the large-scale stellar bar for the sample galaxies.
There is a wide variety of  CO morphologies ranging 
from  relatively axisymmetric annuli or disks 
(NGC 4102, NGC 3504, NGC 4536,  NGC 470, and 
NGC 4314) to  elongated double- peaked  and spiral  morphologies 
(starburst NGC 2782 and non-starbursts NGC 3351 and  
NGC 6951).
We refer the reader to  $\S$ 8 for  a  comparison of the molecular 
gas properties in the starbursts and non-starbursts, 
and to  $\S$ 12 for a description of the individual galaxies.

\section{The extreme  molecular environment in the inner kpc of barred 
         spirals and its implications}
+

Table 6 illustrates how the molecular environment that has developed  
in the circumnuclear region  of the barred spirals  differs markedly from 
that present in the outer disk of galaxies. The implications for the inner
kpc are outlined below.

\medskip
\bf
(i) Gas densities and mass fraction:  
\rm
We compute the  molecular gas surface densities
($\Sigma_{\rm gas-m}$) by deprojecting the moment 0 map and  
computing the  azimuthally averaged surface density of  molecular hydrogen  
($\Sigma_{\rm \tiny H2}$) as a function of radius  in the 
galactic plane, assuming a standard CO-to-H$_{\rm 2}$ conversion 
factor. We add in  the contribution of He  using 
$\Sigma_{\rm gas-m}$ =  $A_{\rm Z}$ $\times$ $\Sigma_{\rm \tiny H2}$, 
where $A_{\rm Z}$~=~1.36  for an assumed solar 
metallicity (Z = 0.02, X=0.72, Y=0.26).  
The peak  gas surface density $\Sigma_{\rm gas-m}$ in the
inner kpc of these galaxies  ranges from 500-3500 
$M_{\tiny \sun}$ pc$^{-2}$.

To estimate  what fraction of the dynamical mass is in molecular 
form, we first calculate the  total  dynamical mass ($M_{\rm dyn}$) interior 
to radius R from  the circular speed V$_{c}$  :
\begin{equation}
M_{\rm dyn} (R) = \frac{ V_{\rm c}^{2}(R) \  R \ \beta} {G} 
\end{equation} 
\noindent
where  G is the gravitational constant and 
$\beta$ depends on the shape of the mass distribution. 
$\beta$=1 for a spherically symmetric mass distribution and is 
slightly lower for flattened configurations.
For an exponential disk, we can overestimate the true 
dynamical mass by at most 1.3  if we assume $\beta$=1 
(Binney \& Tremaine 1987).
We  estimate  $V_{\rm c}$ and 
the CO rotation curve in the central 1-2 kpc radius 
from the CO  position-velocity (p-v) plots. We can more effectively 
identify different  kinematic components  (e.g., circular
motions, azimuthal streaming motions, radial inflow motions, and
vertical outflow motions) in gas of different intensities from 
p-v plots along the kinematic major and minor axes than  from 
moment 1 maps, which reflect only  the intensity-weighted average 
values. In particular, we estimate  $V_{\rm c}$ by taking the  CO 
velocity at which the intensity levels peak at each radius
along the kinematic major axis. The values of $V_{\rm c}$
estimated in this manner show 
good agreement with the rotation curve derived from 
H$\alpha$ emission spectra (Jogee 1999) in cases where both
data are available (e.g.,  NGC 4102, NGC 4536, NGC 3351, NGC 4569, 
NGC 6951). In regions where the gas shows evidence of 
non-circular motions  in the p-v plots, estimates of  
$V_{\rm c}$ will  be somewhat uncertain. We therefore derive 
$V_{\rm c}$ only for galaxies where the kinematics are relatively 
ordered and  dominated by circular motions ($\S$ 8).
 
In summary, we find that the inner kpc of the 10 barred spirals in our 
sample hosts  peak molecular gas surface densities of 
500-3500  M$_{\tiny \sun}$ pc$^{-2}$  and molecular gas mass 
fractions of  10 to 30 \%. 
\it 
These  densities are at least an order of magnitude higher 
than the  typical average  atomic and  molecular gas 
surface  densities  (1-100  $M_{\tiny \sun}$ pc$^{-2}$) 
in the outer disk of normal spirals 
\rm  (Deharveng et al. 1994; 
Kennicutt 1998),  or  the  gas surface density 
(200  $M_{\tiny \sun}$ pc$^{-2}$) of  a typical giant molecular 
cloud in the outer disk of the Milky Way (Scoville \& Sanders 1987). 
Such large molecular gas surface densities and   mass fraction in the 
inner kpc can 
\it 
enhance the self gravity and clumpiness  of the gas 
and produce stronger  gravitational coupling between the gas and the 
stars
\rm  (e.g., Shlosman et al. 1989; Jog \& Solomon 1984).
The resulting  two-fluid disk will be more unstable 
to gravitational instabilities than  a purely  stellar one-fluid disk 
(Jog \& Solomon 1984; Elmegreen 1995) .

Furthermore, as the gas becomes more clumpy 
dynamical friction might become an increasingly important transport 
mechanism at low  radii. 
For  instance, in the case of a  gas clump which has a  mass M 
and a speed v at 
radius r,  the timescale t$_{df}$ on which dynamical 
friction operates is  $\propto$ (r$^2$ v/M ln$\Lambda$), 
where ln$\Lambda$ is the Coulomb logarithm (Binney \& Tremaine 1987). 
For M= 10$^8$  $M_{\tiny \sun}$, v $\sim$ 300 km s$^{-1}$, 
r=200 pc, t$_{df}$  is as short as $\sim  3 \times 10^{6}$ years.

\medskip
\bf
(ii) Onset and growth of gravitational instabilities: 
\rm
For a thin  differentially rotating gas disk, support against 
gravitational  instabilities is provided by pressure forces on 
small scales, and by Coriolis forces from rotation on large scales.
The Coriolis forces depend on the epicyclic frequency of 
oscillations $\kappa$ which is a function of the circular 
speed V$_{\rm c}$: 

\begin{equation} 
\kappa = \sqrt{ \frac{2V_{\rm c}}{R} (\frac{V_{\rm c}}{R}+ 
\frac{dV_{\rm c}}{dR})}.  
\end{equation}  

We  estimate $\kappa$  from the CO  and  H$\alpha$ rotation curves,
bearing in mind that uncertainties exist  in regions where there are  
non-circular motions. However, even uncertainties of a factor of a 
few do not change the basic expectations of very high  epicyclic 
frequencies  $\kappa$  of a few 100 to several 1000  km s$^{-1}$ 
kpc$^{-1}$) in the inner 500 pc.
These very high  epicyclic frequencies coupled 
with the moderately large gas velocity dispersions \rm 
(10-40  km s$^{-1}$)  \it
cause the critical density  ($\Sigma_{\rm crit}$) 
for the onset of gravitational instabilities  to shoot 
up to several 100-1000 M$_{\tiny \sun}$ pc$^{-2}$.
\rm 
Although hard to trigger, once triggered, 
these instabilities  have a short 
growth timescale (t$_{\tiny \rm GI}$) of  a few Myrs 
for the most  unstable wavelength:  

\begin{equation}
t_{\tiny \rm GI} = 
\frac { \sigma} { \pi \rm G \ \Sigma_{\rm gas}}
\end{equation}

In other words, 
\it   
the molecular environment and dynamical parameters of the
inner kpc naturally force gravitational  instabilities 
to set in and to grow 
in a  ``burst'' mode  defined by  high density and short 
timescales.
\rm 
This  may explain, at least in part, why the most intense
starbursts  tend to be found in the inner kpc of galaxies. 

It is also important to note that the 
growth timescale (t$_{\tiny \rm GI}$) of a few Myr in the
inner kpc is comparable to the lifetime of an OB star. 
\rm
As a result, the fraction  of molecular gas  mass 
converted into stars 
before  cloud complexes  are disrupted by massive stars 
can be higher in the circumnuclear region than in the outer disk.
The high pressure, high turbulence ISM may  also favor more 
massive clusters as suggested by some authors
(e.g., Elmegreen et al. 1993).  It is relevant that 
80 \%  of the sample galaxies show super star clusters.

\medskip
\bf 
(iii) ULIRGs as scaled-up local starbursts?
\rm
In the circumnuclear region  of the ULIRG Arp 220 
(Scoville, Yun, \& Bryant 1997), the conditions 
observed seem to be even more extreme than
those in the inner kpc of the local barred starbursts: a  gas  
surface density $\sim$~10$^{4}$ $M_{\tiny \sun}$ pc$^{-2}$),  
a velocity dispersion $\sigma$~$\sim$~90 km s$^{-1}$, and 
an epicyclic frequency $\kappa$~$\sim$~several 1000 km s$^{-1}$  kpc$^{-1}$ 
(Table 6). 
Observations of these quantities at comparable resolution do not
exist for other ULIRGs, but if Arp 220 is a proto-typical ULIRG, then
Table 6 suggests  that 
\it 
the circumnuclear molecular environment of ULIRGs is a scaled-up 
version of that developed by the local weakly-interacting starbursts 
in our sample.
\rm
This scaled-up  molecular environment may be the result of the 
the stronger external triggers acting on ULIRGs, which are generally 
strongly interacting systems, often involving mergers of comparably 
massive spirals. 
Such interactions can pile gas to very high densities in the inner 
kpc and  produce larger turbulent  linewidths, thereby 
boosting the critical density $\Sigma_{\rm crit}$ at which 
gravitational instabilities set in and ensuring 
the conditions for a super-starburst.

\section {The Circumnuclear Star Formation Rate}

When using the $45\arcsec$ single dish CO map as a measure of the 
circumnuclear gas content (e.g., in $\S$ 2), it was appropriate
to use the  RC  luminosity within a similar 
$45\arcsec$ aperture as a measure of the circumnuclear SFR.
The  $45\arcsec$ aperture corresponds to an aperture with a 2 kpc 
radius at the mean sample distance of 20 Mpc. 
However, now that we have obtained high resolution ($2\arcsec$)
interferometric CO maps, we need to resolve the SF activity and 
estimate the SFR within the radius ($R_{\tiny \rm CO}$), 
which encompasses most of the 
detected CO emission.  We denote this as SFR$_{\rm CO}$. 
We can then verify whether the large range 
in SFR/$M_{\tiny \rm H2}$ seen within the sample when using  
$45\arcsec$ averaged data persists even at higher resolution.
$R_{\tiny \rm CO}$  is estimated by determining where the 
azimuthally averaged gas surface density falls below a few \% 
of its peak value, and ranges from 500 to 1680 pc (Table 8). 

We estimate  SFR$_{\rm CO}$ within  $R_{\tiny \rm CO}$ 
by considering an ensemble of SFRs derived from different tracers 
which each may offer 
a different advantage in terms of resolution, apertures, and 
extinction.
For instance, a well-established physical framework  exists linking 
H$\alpha$ recombination lines to the UV continuum photons from 
massive young stars (e.g., Kennicutt 1998),  but H$\alpha$  fluxes 
can be strongly affected by extinction in the  dusty  circumnuclear region 
where gas densities reach  several  100 to 1000  M$_{\tiny \sun}$ pc$^{-2}$.
Conversely, tracers such as the  far-infrared  
or RC   luminosity  are less affected by extinction, but
have  lower  spatial resolution such that they may not always 
resolve the  SF within $R_{\tiny \rm CO}$. 

Following Condon (1992) and Kennicutt (1998), we estimate in sections 
(i) to (iii) the SFRs from the global FIR luminosity,  the  Br$\gamma$ 
luminosity, the non-thermal and  thermal components of the RC  emission 
which we denote, respectively,  as SFR$_{\tiny \rm FIR}$,  
SFR$_{\tiny Br\gamma}$, SFR$_{\tiny RC-N}$, and 
SFR$_{\tiny RC-T}$.   Table 7 summarizes the values
and apertures for each tracer.

  \medskip
  \noindent
  \bf
  \it
  (i)
  \rm
  \it
  SFR based on the  the global FIR luminosity (SFR$_{\tiny \rm FIR}$): 
   \rm
  The global  FIR  luminosity traces the massive SFR 
  over the entire galaxy, under the assumption that most 
  of FIR emission between  40 and 120 $\micron$ emanates 
  from warm dust heated by massive stars. 
  SFR$_{\tiny \rm FIR}$ was computed from
  
\begin{equation}
(\frac {\rm SFR} {M_{\tiny \sun} \rm yr^{-1}})
= 4.3 \times  \rm SFR (M \ge 5 M_{\tiny \sun})
= \gamma  \times  (\frac {L_{\rm FIR}} {10^{10} L_{\tiny \sun}})
\end{equation}

where   SFR ($ M \ge 5 M_{\tiny \sun}$) is the rate of formation 
of stars with mass $ M \ge 5 M_{\tiny \sun}$.  An  
extended Miller-Scalo IMF (Kennicutt 1983) 
with an upper mass cut-off  $M_{\tiny \rm U}$ of 100  $M_{\tiny \sun}$ 
is  assumed. We adopt $\gamma$= 3.9  (Condon 1992). 
As  shown in  Table 7, the global SFR ranges from 6 to 14 $M_{\tiny \sun}$ 
yr$^{-1}$ for the starbursts and  from 0.2 to 4  $M_{\tiny \sun}$ 
yr$^{-1}$ for the non-starbursts. 
These values apply to the entire galaxy and therefore 
serve as an upper limit to the SFR  in  the circumnuclear 
region.

\medskip
\noindent
\bf
\it
(ii)
\rm
\it
SFR based on  the non-thermal   and thermal component of the RC   emission 
(SFR$_{\tiny RC-N}$,  SFR$_{\tiny \rm RC-T}$):  
\rm
The thermal free-free emission from the photoionized  HII regions 
and the non-thermal RC  emission 
produced by relativistic electrons which are accelerated by 
supernova remnants are both related to the massive SFR. 
The RC   flux density at 1.5 and 4.9   GHz is dominated 
by non-thermal emission. 
We estimate the non-thermal fraction to be 0.88 at 1.5 GHz 
and 0.77 at 4.9  GHz following the approximation by 
Condon \& Yin (1990) : 

\begin{equation}
(\frac {S} {S_{\tiny \rm T}}) 
= 1 \ + \ 10 \times  (\frac {\nu} {\rm GHz})^{0.1 - \alpha} 
\end{equation}

where the non-thermal index  $\alpha$ is $\sim$ 0.8.
From  the non-thermal component of the RC  luminosity 
density ($L_{\rm N\nu}$), we estimate the star  formation  rate  
SFR$_{\tiny \rm RC-N}$ within the radius $R_{\tiny \rm RC}$ using 

\begin{equation}
(\frac {SFR_{\rm RC-N} } {M_{\tiny \sun} \rm yr^{-1}}) 
= 4.3 \times 0.019 \times (\frac {\nu} {\rm GHz})^{\alpha} \
(\frac {L_{\rm N\nu }} {10^{20} \ \rm W \rm Hz^{-1}})
\end{equation}

This equation assumes the Galactic 
relationship between the supernova rate and the RC luminosity density 
(Condon 1992). As shown in Table 7, SFR$_{\tiny \rm RC-N}$ applies to 
the region of radius R$_{\tiny \rm RC}$ which covers  the inner 
1-2 kpc, except for NGC 3359 and NGC 4314 where larger apertures apply. 
SFR$_{\tiny \rm RC-N}$  ranges from 3 to 14  $M_{\tiny \sun}$ 
yr$^{-1}$ for the starbursts and  from 0.1 to 2  $M_{\tiny \sun}$ 
yr$^{-1}$ for the non-starbursts. 
When estimating SFRs in NGC 6951, we excluded 
the luminosity contribution from the inner 100 pc radius 
to avoid contamination from the Seyfert nucleus 
that is known to exist from optical line ratios
(Munoz-Tunon \etal 1989)

We estimate  the SF  rate  (SFR$_{\tiny \rm RC-T}$) 
from the thermal component of the RC  emission  ($L_{\rm T\nu}$) 
using

\begin{eqnarray}                               \label{eqitot_a}
(\frac {SFR_{\rm RC-T} } {M_{\tiny \sun} \rm yr^{-1}})  & =  & 
4.3 \times 0.176 \ \times (\frac {L_{\rm T\nu }} {10^{20} \ \rm W \rm Hz^{-1}}) \ \times 
\nonumber  \\	                 
&   &  (\frac{T_{\rm e}} {10^{4} \ \rm K})^{-0.45} \ (\frac {\nu} {\rm GHz})^{0.1} 
\end{eqnarray}.

SFR$_{\tiny \rm RC-T}$ is slightly lower than  SFR$_{\tiny \rm RC-N}$, but 
the  difference is only at the 5 \% level.
It is encouraging that the same trend is seen in estimates based 
on radio continuum  (SFR$_{\tiny \rm RC-N}$) and FIR  
(SFR$_{\tiny \rm FIR}$) light: the starbursts have larger SFRs
(3 to 14 $M_{\tiny \sun}$ yr$^{-1}$)  
than the non-starbursts (0.1 to 2 $M_{\tiny \sun}$ yr$^{-1}$). 
Figure 4 shows SFR$_{\tiny \rm RC-N}$ plotted against 
SFR$_{\tiny \rm FIR}$. 
SFR$_{\tiny \rm RC-N}$ is comparable to SFR$_{\tiny \rm FIR}$  
in  all galaxies except NGC 2782 and NGC 3504, where 
SFR$_{\tiny \rm RC-N}$ is 10 to 30 \% higher. 
This could be caused by a variety of reasons. 
The dust temperature  and emissivities could be different from 
those  assumed such that 
SFR$_{\tiny \rm FIR}$ underestimates the true global SFR.  
Another possibility is that  part  of the non-thermal RC emission is 
coming from starburst-driven outflows, where the empirical Galactic 
relationship between the supernova rate and the RC luminosity 
density might  not hold. 
In fact, in NGC 2782,  the high ($1''$) resolution RC map  
(Saikia et al. 1994)  shows that a large fraction of the RC emission 
emanates from double outflow shells which extend 850 pc 
away from the central starburst (Jogee et al. 1998). Furthermore, 
while  there is no evidence from optical line ratios for an active 
nucleus in these two galaxies,  one cannot rule out the possibility
that  part of the non-thermal RC emission may be due to a hidden 
active nucleus.

\medskip
\noindent
\bf
\it
(iii) 
SFR based on  Br$\gamma$  fluxes (SFR$_{\tiny Br\gamma}$): 
\rm
The  Br$\gamma$ recombination line  at 2.16 $\micron$ 
traces Lyman continuum photons from massive stars.
Published  Br$\gamma$ fluxes (e.g., Puxley et al. 1990) 
are only available for four starbursts NGC 2782,  NGC 3504, 
NGC 4102,  and NGC 4536 and for the non-starburst NGC 3351. 
As shown in  Table 7, 
 the radius $R_{\tiny Br\gamma}$  of the region over which 
Br$\gamma$  observations were made varies between 500 to  1600 
pc depending on the galaxy.  
SFR$_{\tiny Br\gamma}$ is 1 to 6 $M_{\tiny \sun}$ yr$^{-1}$ 
for the starbursts and  0.5 $M_{\tiny \sun}$ yr$^{-1}$ for 
the non-starburst NGC 3351,  as  computed from : 

\begin{equation}
(\frac {\rm SFR}{M_{\tiny \sun} \rm yr^{-1}}) 
= 11.97 \times 10^{-54} \ (\frac {N_{\rm Ly}} {\rm s^{-1}})
\end{equation}

SFR$_{\tiny \rm RC-N}$ 
is larger by a factor of 2 to 5 compared to SFR$_{\tiny Br\gamma}$ 
in  NGC 2782, NGC 3504, NGC 4102, and NGC 4536, 
where both   SF rates are measured  over comparable radii. 
It is likely that SFR$_{\tiny Br\gamma}$  underestimates the
true SFR as the  Br$\gamma$ data  and ionization rate  (N$_{\rm Ly}$) 
quoted by Puxley et al. (1990) have  not been corrected 
for extinction at 2.2 $\micron$.
Dust within an internally dusty HII region can absorb Lyman continuum 
photons before they ionize any H atom, and dust along the line of sight 
can absorb part of the  Br$\gamma$ recombination  photons. 

\medskip 
Table 7 summarizes the SFR estimated (SFR$_{\tiny \rm FIR}$, 
SFR$_{\tiny Br\gamma}$, SFR$_{\tiny RC-N}$, and SFR$_{\tiny RC-T}$) 
from these tracers  and the aperture over which it applies.
From these values, we make a best estimate for SFR$_{\rm CO}$,
the extinction-corrected SFR within the radius ($R_{\tiny \rm CO}$) 
over which  CO emission is detected in the moment maps.
In the case of NGC 470, NGC 2782, NGC 4102, NGC 4536, and NGC 4569, 
we consider SFR$_{\tiny \rm RC-N}$ to be 
a reasonable estimate  for SFR$_{\rm CO}$ since 
the RC  maps show that 
most of the RC emission is concentrated within $R_{\tiny \rm CO}$. 
For the remaining galaxies NGC 3504, NGC 3351, and NGC 4314, 
($R_{\tiny \rm RC}$/$R_{\tiny \rm CO}$)  is 0.6, 1.7, and 6.0 
respectively. 
In NGC 3504, 
SFR$_{\tiny \rm RC-N}$  is a reasonable approximation 
for the  SF rate  within $R_{\tiny \rm CO}$ (1300 pc) 
because the H$\alpha$ map  shows that more than 90 \% of H$\alpha$ 
flux within  $R_{\tiny \rm CO}$ originates from the inner 
800 pc radius.
In the case of NGC 3351, $R_{\tiny \rm CO}$ is $\sim$ 600 pc, which
is close to $R_{\tiny \rm  Br\gamma}$ $\sim$ 500 pc. 
We therefore   adopt  
SFR$_{\rm CO}$ as  0.5 $M_{\tiny \sun}$ yr$^{-1}$ based on the Br$\gamma$ 
estimates. 
In the case of NGC 4314, we consider  SFR$_{\tiny \rm RC-N}$ $\sim$  
0.1  $M_{\tiny \sun}$ yr$^{-1}$  within $R_{\tiny \rm RC}$ $\sim$ 3600 pc 
as an upper limit to the  SF rate  within 
$R_{\tiny \rm CO}$ $\sim$ 600 pc.

Table 8  shows  the resulting $R_{\tiny \rm CO}$,  SFR$_{\rm CO}$ and 
the mass of molecular hydrogen  ($M_{\tiny \rm H2}$) within this 
radius. The resulting 
\it 
circumnuclear SFR 
ranges from 3 to 11 $M_{\tiny \sun}$ yr$^{-1}$  
in  the starbursts and from 0.1 to 2 $M_{\tiny \sun}$ yr$^{-1}$ in  
the non-starbursts.  
\rm
\rm 
It is evident that  while non-starbursts such as  NGC 4569  and  
NGC 6951 and  starbursts such as  NGC 4102, NGC 4536, and NGC 2782  
all host  several times  $10^{9}$  $M_{\tiny \sun}$ of molecular  hydrogen, 
their  SFRs are substantially different.  
\rm 
A similar comparison applies between  
the starburst NGC 470 and  the non-starbursts NGC 3351 and NGC 4314 
which host  several times   $10^{8}$  $M_{\tiny \sun}$ of molecular  
hydrogen.

\section{Comparison of Molecular Gas in the Starbursts and Non-Starbursts}

Table 8 shows that 
\it 
barred galaxies   with 
comparable amount of molecular gas in the
inner 1-2 kpc radius  can show an order of magnitude range in 
the average SFR per unit mass of gas  (SFR/$M_{\tiny \rm H2}$) 
within this region. 
\rm 
Among galaxies which host  several times  $10^{9}$  
$M_{\tiny \sun}$ of molecular  hydrogen  in the inner 2 kpc, 
starbursts such as  NGC 4102, NGC 4536, and NGC 2782 
have an  average (SFR/$M_{\tiny \rm H2}$)  of order  
a few times $10^{-9}$  yr$^{\rm -1}$  while 
non-starbursts NGC 4569  and  NGC 6951 have  
 (SFR/$M_{\tiny \rm H2}$) $\sim$ a few  $\times 10^{-10}$ yr$^{\rm -1}$.
We now turn to a comparison of the  circumnuclear molecular gas  
in the barred starbursts and non-starbursts  with the overall goal 
of understanding why they have such 
different SFR per unit mass of gas  (SFR/$M_{\tiny \rm H2}$) in 
the inner few kpc.

While we focus here on a comparison of the starbursts and 
non-starbursts (Figs. 3,5,6,7), we refer any  interested 
reader to $\S$ 11  for  a description  of each \it 
individual \rm  galaxy.
Figures 3a--b show a wide variety of molecular gas   morphologies  
\rm 
including   relatively axisymmetric annuli or disks
(starbursts NGC 4102, NGC 3504, NGC 4536,  NGC 470, and 
non-starburst NGC 4314),  
elongated double- peaked  and spiral  morphologies 
(starburst NGC 2782 and non-starbursts NGC 3351 and  
NGC 6951), and 
highly extended distributions (non-starburst NGC 4569)
\rm 
In all of the sample galaxies (Figs. 3a--b)  except for
the non-starburst NGC 4569,  the molecular gas 
is concentrated  in the inner kpc radius.
Figure 5  compares  the spatial distribution of  SF activity (shown
as greyscale) and  molecular gas (shown as CO contours). 
The x and y axes represent respectively, 
$M_{\tiny \rm H2}$  and (SFR/$M_{\tiny \rm H2}$), as defined in Table 8.
In the case of NGC 4102, NGC 2782 and NGC 6951, we show the SF activity 
with existing high resolution  ($1.0\arcsec$--$1.5\arcsec$)   1.49 and 4.89 
GHz RC maps. Strictly  speaking, in the case of NGC 6951, the RC map 
shows not just the star-forming ring of   $4''$ (360 pc) radius, but
also  the Seyfert nucleus in the  inner  $1''$ (90 pc).
For the remaining galaxies in Fig. 5, we use high resolution 
($1.0\arcsec$ --  $1.5\arcsec$) H$\alpha$ maps as there 
are no RC maps of resolution comparable to the CO. 
Our conclusions  are given below.

\subsection{The Type I Non-Starburst NGC 4569}

The non-starburst NGC 4569  differs from  the other sample galaxies 
in that it has a  highly extended molecular gas distribution with 
complex kinematics and large linewidths  of $\sim$ 150 km s$^{-1}$.
These properties  of NGC 4569  are illustrated in Figs. 6a--d.
A large fraction of the gas  lies outside the inner kpc radius 
and extends out to a  radius of 20 $''$  (1.7 kpc),  
at a similar P.A. ($\sim$ 15 $\deg$) as the large-scale 
stellar bar (Figs. 6a--b).
The extended gas in NGC4569  has  highly  non-circular 
kinematics, as shown by the p-v plots along the 
kinematic major  and   minor axes (Figs. 6c--d). 
Along the kinematic major axis, 
velocities are generally positive (i.e., above 
the systemic value of -235  km~s$^{-1}$) on the northeastern side, 
and  negative on the southwest side. 
However,  at a radius of $5''$ (400 pc), near the feature
marked `N1' in  Fig. 6b, the velocities 
change from +100  km~s$^{-1}$ to a forbidden velocity
of -75 km~s$^{-1}$.  This indicates the presence 
of non-circular motions caused by in-plane 
azimuthal streaming motions or/and vertical motions out of
the plane. 

As outlined in $\S$ 2.2,  NGC 4569 shows signs of optical and NIR  
asymmetries suggestive of a recent tidal interaction or minor merger 
and it also hosts a large-scale stellar bar. 
Thus, we suggest that the extended gas in NGC 4569 which is 
generally elongated along the bar and shows strong streaming 
motions,  is responding to gravitational torques from both the  
interaction  and the (induced or spontaneous) stellar bar.
\it
Taken together, the optical, NIR, and  CO properties of 
NGC 4569 suggest it is in the early phases of 
bar-driven/tidally-driven gas inflow.
\rm 
NGC 4569  is reminiscent of NGC 7479 which  hosts  large amounts of gas 
with non-circular kinematics  along a large-scale 
stellar bar  and show evidence for  a minor merger 
(Laine \& Heller 1999).  However, the gas kinematics in NGC 4569 
look more disturbed in the inner 400 pc than in NGC 7479.

There is intense SF  in the central $2''$ (170 pc) radius,
but the  gas  which is extended  along the bar and has 
disturbed kinematics shows no appreciable SF.
The properties of  
NGC 4569  stand in sharp contrast to the starburst NGC 4102,  
where most of the circumnuclear molecular gas 
has piled up  in the inner $4''$ (340 pc) radius, shows 
predominantly circular kinematics,  and is undergoing 
intense SF.   The low SF efficiency in NGC 4569 
is likely related to the fact that large velocity gradients 
and shear  in gas streaming along the bar can prevent clouds from 
being self-gravitating. Additionally, 
gas moving in  a bar potential can experience large tidal forces 
(Elmegreen 1979; Kenney \& Lord 1991) and 
tidal heating (Das \& Jog 1995) which slow down gravitational collapse. 
Several other galaxies show such  low  SF per unit mass of gas along
their  bars  e.g.,  NGC 7723 (Chevalier \& Furenlid 1978), NGC 1300,  
NGC  5383 (Tubbs 1982),  NGC 7479 (Laine et al. 1999) 
or in M83 where  the ratio of UV to CO luminosities 
is unusually low despite abundant CO  (Handa, Sofue, \& Nakai 1991). 

In summary, we therefore suggest that 
\rm
the low SFR per unit gas mass of the non-starbursts NGC 4569 
over the inner few kpc is due to the fact that 
\it it is in the early stages
of bar-driven/tidally-driven gas inflow, where most of its 
molecular gas has an extended distribution, highly non-circular 
kinematics and an associated large local shear which is 
not conducive to SF.
\rm 
The barred galaxy NGC 7479  studied by  Laine et al. (1999)
may also be in this  early evolutionary phase.
In  NGC 7479, only low levels of  SF are seen along 
the bar despite the presence of 
several  $\times 10^{9}$  $M_{\tiny \sun}$ of molecular
gas in  a dust lane along the bar (Laine et al. 1999). 
This gas shows large non-circular motions,  a large velocity gradient,  
and a projected velocity gradient of at least 100 
km~s$^{-1}$ (Laine et al. 1999). 
We shall henceforth use the term 
\bf `Type I non-starburst' \rm 
to  denote this kind of system,  where a large fraction of the
molecular gas is still inflowing toward the inner kpc, along 
the leading edges of the  bar, exhibiting 
large non-circular motions and not forming stars efficiently.

\subsection{ The Starbursts and Type II  Non-Starbursts}

The remaining starbursts (NGC 4102, NGC 3504, NGC 4536, NGC 470) 
and non-starbursts  (NGC 6951, NGC 3351, NGC 4314)  in our sample 
do not show such extreme kinematics or extended gas distributions
as NGC 4569. In all of them,  the molecular gas detected 
is concentrated  within the inner kpc radius and
the velocity field in the inner 500 pc radius is generally 
dominated by circular motions, with occasional weaker 
non-circular components (see $\S$ 11 for details). 
Most of the  circumnuclear molecular gas
seems to have reached the inner kpc radius of the bar
and  only a small fraction of it seems to  be still inflowing 
along the stellar bar. 
The latter gas component shows up in the form
of  faint gas streams 
which extend out and intersect  the dust lanes on  
the leading edges of the  large-scale  bar and  
show  non-circular motions. Such gas streams are seen in 
clearly in non-starburst NGC 6951 (Fig. 15a--b) and also at fainter
levels in  NGC 3351 (Fig. 15c--d), NGC 4314 (Fig. 15e--f),  and 
starburst NGC 4102 (Fig. 14a--b).  
We conclude that in contrast to NGC 4569, the remaining  
barred starbursts and non-starbursts  seem to be 
\it 
in significantly later stages of bar-driven  inflow  where most 
of the molecular has  already settled into the inner kpc
of the barred potential  and no longer shows extreme 
non circular motions.  
\rm 
We henceforth refer to  these kind of  non-starbursts  
as 
\it  
\bf `Type II non-starbursts'. \rm

What is the difference between  the starbursts and  
Type II non-starbursts? 
To answer this question, we use the two-dimensional distribution 
of SF  and gas  in Fig. 5, as well as the  azimuthally averaged  
gas and SFR surface densities in Fig. 7a--b.
Fig. 7a  shows the radial variation of 
the deprojected azimuthally-averaged  molecular 
gas surface density ($\Sigma_{\rm gas-m}$)  derived in $\S$ 5.
In order to display  meaningful values, quantities are plotted starting 
at a radius $\ge$ half the size of the synthesized CO beam which is 
typically $2''$ or 150 pc. 
Fig. 7b plots  the azimuthally-averaged  molecular 
gas surface density  SFR per unit area 
($\Sigma_{\tiny SFR}$)  in the galactic plane   
as a function of radius.  
\it 
We find that the  starbursts show  larger 
gas surface densities  (1000-3500 $M_{\tiny \sun}$ 
pc$^{-2}$) in the inner 500 pc radius compared to the 
Type II non-starbursts for a given CO-to-H$_{\rm 2}$ 
conversion factor. 
\rm 
In Fig. 7a, the Type II non-starbursts have peak $\Sigma_{\rm gas-m}$ 
reaching only 500-950 $M_{\tiny \sun}$  pc$^{-2}$. 
We also note  from Fig 6a and 6b that in the central 
600 pc  radius,  $\Sigma_{\rm gas-m}$  and $\Sigma_{\tiny SFR}$  
follow each other more closely in the case of the starbursts.

Is the presence of apparently higher gas surface densities in  
the starbursts a real effect? 
We can rule out artifacts   caused by resolution effects 
since we have similar  linear resolution (in pc) for  starbursts and 
non-starbursts with comparable circumnuclear molecular gas content 
(e.g., NGC 4102, and NGC 3504 versus NGC 6951,  or NGC 2782 and NGC 
4536 versus  NGC 4569). 
Another concern is that the  higher  
$\Sigma_{\rm gas-m}$  in the  starbursts may be due 
to  an increase in CO emissivity  caused by the  higher temperatures 
associated with the intense  SF. 
However,  potential variations in CO emissivity cannot 
account  for the observed differences in 
$\Sigma_{\rm gas-m}$ since   CO peaks do  not 
coincide with the  H$\alpha$ and RC peaks 
in several galaxies  e.g., NGC 6951, NGC 2782, and 
NGC 3351 (Fig. 5).
Thus,  the  starbursts appear to have an intrinsically larger 
molecular gas  surface density in the central 500 pc than  
the non-starbursts.
This difference of a factor of 3--4  in  $\Sigma_{\rm gas-m}$  can 
in turn lead to \it large \rm differences  in the SF
\it  
if gas in the starbursts is above a  critical density for SF while 
a large part of the  gas in the non-starbursts is sub-critical. 
\rm 
We address this possibility quantitatively in $\S$ 9 by comparing  
$\Sigma_{\rm gas-m}$ to different theoretical critical densities 
which are believed to be relevant for the onset of SF.

\section{ Theoretical Models for SF applied to Circumnuclear
 Starbursts and Type II Non-starbursts}

We compare the observed circumnuclear 
gas surface density  $\Sigma_{\rm gas-m}$  to the critical density 
relevant for the onset of SF  with the overall goal of understanding
some of the differences between the  starbursts and 
Type-II non-starbursts.
In particular, we consider theoretical scenarios which assume that 
the onset of gravitational or axisymmetric 
instabilities (Safronov 1960; Toomre 1964; Goldreich \&
Lynden-Bell 1965a;  Elmegreen 1979)  is relevant for the growth of 
clouds and the onset of SF. 
Such models have achieved a fair degree of empirical 
success in the outer disks of  Sa-Sc spirals, gas-rich E and  S0s s 
(Kennicutt 1989; Eder 1990; Kennicutt 1998b).  Extensions of such 
models to include magneto-Jeans instabilities have also been 
developed (Kim \etal 2002).

For a thin  differentially rotating gas disk, support against gravitational  
instabilities is provided 
by pressure forces on small scales, and by Coriolis forces from rotation 
on large scales.  One can show that axisymmetric instabilities set 
in when the gas surface density  $\Sigma_{\rm gas}$  
exceeds a critical density such that the 
Safronov (1960)/Toomre (1964) $Q$  parameter falls below 1, where  

\begin{equation}
\rm Q = \frac{\Sigma_{\rm crit}}{\Sigma_{\rm gas }} 
= \frac{ \alpha \kappa \ \sigma } { \pi \ \rm G \ \Sigma_{\rm gas}} \le 1
\end{equation}

Here,  $\omega$ is the angular frequency, 
$\Sigma_{\rm gas}$ is the gas 
surface density, $\sigma$ is the gas velocity dispersion, 
$\kappa$  is the epicyclic frequency, and G is 
the gravitational constant. 
For an infinitely  thin disk, $\alpha$ is 1, but for 
disks of finite thickness, $\alpha$ is somewhat 
larger (Larson 1985).  Empirically, Kennicutt (1989)  
found that  $\alpha$ is $\sim$ 0.7 in the outer disks of  
Sc spirals.  A two-fluid disk made of both gas and  
stars  is always more unstable and will have a lower  $\alpha$ 
than a purely gaseous or  stellar one-fluid disk (Jog \& Solomon1984; 
Elmegreen 1995).

We discuss the gravitational instability models in the context
of the type II  non-starbursts NGC 4314, NGC 3351, and NGC 6951, 
and the starbursts NGC 4102, NGC 4536 and NGC 3504. 
We cannot apply these simply axisymmetric models to 
other sample galaxies as they  have highly non-axisymmetric 
gas distributions  (e.g., the starburst NGC 2782  and Type I non-starburst 
NGC 4569)   and non-circular kinematics  (e.g., NGC 4569).
We ignore the contribution of  atomic HI  
because we  have limited high resolution HI data for our
sample galaxies and furthermore,  molecular hydrogen is believed 
to dominate the ISM  in the inner kpc of spirals 
(e.g., Scoville \& Sanders 1987; Tacconi  \& Young 1986).

A more vicious problem is   to get a good estimate for 
the velocity dispersion  $\sigma$ in the central ($2\arcsec$), 
especially when the $2\arcsec$  CO beam encompasses a steeply rising 
rotation curve.  In the latter case, beam smearing  
artificially increases the observed $\sigma$ and to 
a lesser degree  impacts  $\kappa$ and $\Sigma_{\rm gas}$   in ways 
that depend on the intrinsic intensity and velocity field (Jogee 1999).
Given that corrections for beam-smearing are model-dependent 
and  not necessarily unique, we decide  not to apply any such 
corrections,  and instead perform the instability analyzes  
(Fig. 8) only for radii exceeding  $2 \arcsec$ 
where the effect of beam-smearing  is much less severe.  
In non-starbursts and starbursts alike, we therefore present 
results over radii of $\sim$~$2\arcsec$--$8\arcsec$  (Fig. 8).
The lower limit of $2 \arcsec$ corresponds to 180 pc for  
NGC 6951, NGC 4102 and NGC 4536, and to 100 pc for NGC 4314 and NGC 3351.
Our results are summarized below.


\medskip
\noindent
\bf 
(i) Type II non-starburst NGC 4314: 
\rm 
In NGC 4314,  HII regions are concentrated within  an annulus between  
a radius of $5\arcsec$ and  $7\arcsec$ (240 and 400 pc) 
where $\Sigma_{\rm gas}$  is $\sim$ 800 $M_{\tiny \sun}$ pc$^{-2}$ 
(Fig. 8). 
Interior to this annulus,  there are no HII regions 
although $\Sigma_{\rm gas}$   drops by only a factor of 2-3 
(Fig 8).
This lack of SF activity is not an artifact caused by extinction 
in the $H\alpha$ image or by resolution effects since neither 
the $J$-$K$ image (Friedli et al. 1996)  nor the high resolution 
$HST$ images (Benedict et al.  1996)  show elevated levels 
of dust or young stars at $r<$ 250 pc. 
Thus, SF appears truly  suppressed interior to the annulus of 
HII regions.
The instability analysis (Fig. 8) shows that 
\it  $Q$ reaches its lowest value, between 1 and 2, in the
star-forming annulus \rm  between r=$5\arcsec$ and $7\arcsec$  (240 
and 400 pc)  and it rises sharply at lower and higher radii.
At $r \ge 7 \arcsec$  where we cannot measure $\sigma$,  we  
adopted  a lower limit of 10  km s$^{-1}$  based on the fact that a 
relatively constant value of 10 to 6 km s$^{-1}$ has been 
measured in the outer disks of spirals (e.g., Dickey et al. 1990). 
Thus, values estimated for $Q$  at  $r \ge  7 \arcsec$ 
outside the SF annulus are \it lower \rm limits and the true 
rise  in $Q$  is likely to be even sharper than shown in Fig. 8.
Interior to  the star-forming annulus, between r = $2\arcsec$ and  
$5\arcsec$ (100 to 240 pc),  $Q$ increases by more than a factor of 
3, indicating  highly sub-critical gas densities. 
It is unlikely that this rise in $Q$  is  primarily  due to beam-smearing  
because the observed $\sigma$ increases by at most 1.5  interior to 
the SF annulus. 
(The mean value of  $\sigma$ = 25 km~s$^{-1}$  inside the SF ring and 
15 km~s$^{-1}$  in the ring).

\medskip
\noindent
\bf 
(ii) Type II non-starburst NGC 3351: 
\rm 
In  NGC 3351,  the SF  activity  is concentrated within  an annulus  
between a radius of $6\arcsec$ and  $9\arcsec$  (300 and 450 pc) 
where $\Sigma_{\rm gas}$  is $\sim$  550  $M_{\tiny \sun}$ 
pc$^{-2}$  (Fig. 8).  
At lower radii, between  $2\arcsec$ and  $6\arcsec$  (100 and 450 pc),
there is a large amount of molecular gas and  $\Sigma_{\rm gas}$  
shows no significant change. Yet  no  HII regions are seen  (Fig. 8). 
The absence of observed  SF activity at $ r < 6\arcsec$  is not 
caused by extinction  since  
near-infrared  $K$-band  observations of NGC 3351 show 
several hot spots (likely  due to  young $K$ supergiants) 
in the ring of HII regions, but a relatively smooth distribution 
suggestive  of older stars further in (Elmegreen et al. 1997). 
We find that in NGC 3351 \it  $Q$ reaches its minimum 
value of $\sim$ 1.5 in the annulus of HII regions  \rm 
between a radius of $6\arcsec$ and  $9\arcsec$  
(300 and 450 pc; Fig 8). 
Further in,  between a radius of 
$2\arcsec$ and  $6\arcsec$  (100 and 450 pc),  
$Q$ rises  sharply to $\sim$  6  
suggesting  sub-critical gas densities.
We again note that this increase in $Q$  by a factor 
of $>$ 3  is not primarily due to beam-smearing  since 
the observed $\sigma$ does not show a similar increase. 
$\sigma$ has a mean value of 20 km~s$^{-1}$ in the SF ring 
and  25 km~s$^{-1}$  inside it.

\medskip
\noindent
\bf 
(iii) Type II non-starburst NGC 6951: 
\rm 
In  NGC 6951,   a similar  trend in the $Q$ parameter exists.
The massive CO peaks at a radius of $6''$ (550 pc) 
host a local molecular gas surface density of 2000  $M_{\tiny \sun}$ 
pc$^{-2}$ and an  azimuthally averaged  $\Sigma_{\rm gas}$ of 1000 
$M_{\tiny \sun}$ pc$^{-2}$ (Fig. 8).
Yet, the CO peaks are not associated with SF either in 
H$\alpha$ and RC  maps.
Instead, SF  is concentrated in a ring of radius $\sim$  $3''$ (300 pc). 
Kohno et  al. (1999) find that $Q$ is close to 1 in the star-forming 
annulus.

\medskip
\noindent
\bf  
(iv) The starbursts  NGC 4102, NGC 4536: 
\rm
The circumnuclear SF in the starbursts NGC 4102 and NGC 4536 
is distributed over a wide annulus between 100 and 700 pc. 
Although the ground-based $H\alpha$  or RC images (e.g., Fig. 5) 
may give the misleading impression that Sf extends all the way
into the center, $HST$ images of both galaxies show a  smooth 
distribution  of old stars within  the central 100 pc 
($1\arcsec$)  radius. 
Adopting the same approach for the starbursts 
NGC 4102 and NGC 4536 as for the non-starbursts,
we  perform the instability analysis 
from  $r \sim$~$2\arcsec$--$8\arcsec$ (Fig. 8).
As before, we exclude the central  $r \sim$~$2\arcsec$ radius where 
the effect of beam smearing is severe. 
We also exclude larger radii  ($r > 8\arcsec$) where 
there are faint gas streams which connect to the large-scale 
dust lanes on the leading edges of the bar because the Toomre analysis
is not valid for such non-axisymmetric gas distributions. 

The instability analysis (Fig. 8)  shows that 
\it \rm 
in the starbursts NGC 4102 and NGC 4536, 
the Toomre $Q$ parameter remains  $\sim$  1-2  
from  $r \sim$~$2\arcsec$--$8\arcsec$ (200--700 pc)  
in the region of SF
\it \rm 
despite a factor of 
2--3 variation in  $\sigma$  and a factor of 10 
variation in both the gas surface density ($\Sigma_{\rm gas}$) and  
the epicyclic frequency ($\kappa$).
\rm 
Indeed, between a radius of  200 and 700 pc, 
$\sigma$  ranges from 40 to 
15 km s$^{-1}$,  $\Sigma_{\rm gas}$  from 2200 to 200 
$M_{\tiny \sun}$ pc$^{-2}$,   and $\kappa$ from 1500 to 200  
km s$^{-1}$  kpc$^{-1}$.
A value of  $Q \sim 1$  is also found in the starburst  NGC 3504 
by Kenney et al  (1993).
Furthermore, inspection of Fig. 8 shows that
the $Q$  $\sim$  1-2  annulus  spans $\sim$ 450 pc  (e.g., 250--700 pc) 
in the starbursts  NGC 4102 and NGC 4536, but only  
$\sim$ 150 pc (e.g., 300--450 pc)   in the non-starbursts 
NGC 4314 and NGC 3351. It thus appears that 
\it \rm
the annulus  over which Q  remains $\sim$  1-2  is 3 times 
wider in the starbursts  
than in the non-starbursts  suggesting 
that the former host a larger amount of gas which is close to 
the critical density.

\medskip
\noindent 
\bf 
(v) Overall trends:
\rm 
The analyzes in $\S$ (i)--(iv) suggest that 
\it  
the Toomre $Q$ parameter   remains  
$\sim$  1-2   in the region of SF despite the  large 
dynamic range in molecular gas properties, namely 
a factor of a few  variation in  $\sigma$ 
and  an order of magnitude  variation 
in the gas surface density ($\Sigma_{\rm gas}$) and 
the epicyclic frequency ($\kappa$).
\rm 
It is  true that the analyzes involve  many  uncertain 
quantities such as  the CO-to-H$_{\rm 2}$  conversion factor 
and should, therefore,  be taken  with a grain of salt.
However, with such a large variation in quantities such as 
$\Sigma_{\rm gas}$ and $\kappa$,  $Q$ could easily
have ended up having almost any value, and the fact that it 
remains close to 1 can hardly be fortuitous.
Instead, the evidence presented  strongly suggests that 
\it \rm 
the onset of gravitational instabilities as characterized by $Q$ 
plays an important role in controlling the onset of SF in 
the inner kpc of spirals.
\rm 
At first sight, this may seem surprising since 
large-scale gravitational instabilities 
have been invoked  for the collapse of  atomic gas 
into molecular clouds  within the outer disk   of galaxies, but 
molecular gas in the inner kpc  is  already  in 
the form  of  clouds.  However, large-scale gravitational 
instabilities may be relevant for SF even in the inner kpc 
because they help to aggregate  molecular clouds  
into large  complexes where the clouds  can thereafter  grow  
through local processes involving accretion and 
collisions (e.g., Kwan 1979; Scoville et al. 1986; 
Gammie, Ostriker, \& Jog 1991; Elmegreen 1990).

The results presented in  $\S$ (i)--(iv) also suggest 
at least a partial explanation for the differences in 
SFR per unit mass of gas between the  non-starbursts  
and starbursts. 
In both Type II non-starbursts  and starbursts, 
the $Q$ parameter  reaches its minimum value of 
$\sim$  1--2 in the  annuli of SF. 
However,
\it  \rm 
the annulus  over which Q  remains $\sim$  1-2  is 
three times wider in the starbursts than in the non-starbursts, 
suggesting that 
\it 
 starbursts host a larger amount of gas which 
is close to the critical density. 
\rm 
Inside and outside the annuli of SF, the Type II  non-starbursts 
host large gas concentrations  where gas  densities are well 
below  $\Sigma_{\rm crit}$  ($Q >>$ 2),  and SF appears  inhibited.

\section{ Bar Pattern Speeds and Dynamical  Resonances of the Bar}

A barred potential is made up of different families of periodic stellar
orbits  characterized by a (conserved) Jacobi energy, $E_{\rm J}$, 
a combination of energy and angular momentum (e.g., Binney \& Tremaine 1987). The most
important families are those aligned with the bar major axis (so-called $x_1$
orbits) or with its minor axis ($x_2$ orbits) 
(Contopoulos \& Papayannopoulos 1980). 
The $x_1$   family extends between the center and the
bar's corotation radius. 
If the central mass concentration of a bar is large enough, it 
can also develop one or more inner Lindblad resonances (ILRs). 
The exact locations and even the number of ILRs can be inferred reliably 
from non-linear orbit analysis (e.g., Heller \& Shlosman 1996) based on the
knowledge of the galactic potential. 
The abrupt change in orientation by $\pi/2$  at each  resonance 
is restricted to 
(collisionless) stellar orbits. The gas-populated orbits can change their
orientation only gradually due to shocks induced by the finite gas pressure.
The gas response to bar torquing leads to the formation of
large-scale offset shocks and a subsequent gas inflow. The latter
stalls in the inner few 100 pc  because  shocks  associated with the  bar 
weaken, and gravitational torques in the vicinity of  ILRs  may even reverse  
(e.g., Combes \& Gerin 1985; Shlosman et al. 1989; Athanassoula 1992).

Here we use the CO, optical, and NIR  data of the sample galaxies 
to constrain two fundamental properties of their stellar bar: its  pattern speed 
($\Omega_{\rm p}$)  and the  location of its dynamical resonances.  
The pattern speed  influences the interaction of the bar with the dark matter halo 
(e.g., Debattista \& Sellwood 2000; Athanassoula 2002; Berentzen, Shlosman, 
\& Jogee 2005) and  the location of  its resonances. We use two empirical  methods: 
\it
(I) Kinematic method  based on the epicyclic  approximation: 
\rm
For a weak bar,   the potential can be expressed as the linear sum of 
an axisymmetric term and  a smaller non-axisymmetric term, and 
epicycle theory can be used to solve for the equations of motion of 
a star. In such a case, simple relationships exist between the location
of its  resonances and dynamical frequencies, such as $\Omega_{\rm p}$, 
$\Omega$, and $\kappa$.
We derive  $\Omega$ and  $\kappa$  from the CO  and  H$\alpha$ rotation 
curves as described in $\S$ 5, bearing in mind that uncertainties exist 
in regions where there are  non-circular motions.  
We fit ellipses to the NIR and optical images in order to estimate 
the location of the bar end (e.g., Jogee \etal 1999; 2004a). Next, we 
set upper limits on the pattern speed $\Omega_{\rm p}$ by assuming
that  a bar ends  inside or at the corotation resonance  (CR), 
where $\Omega$= $\Omega_{\rm p}$.  This widely used assumption 
is justified by the  fact that between the ultra-harmonic resonance and 
CR, chaotic orbits dominate, while  beyond the CR, the orbits become 
aligned perpendicular to the bar.  We derive upper limits on $\Omega_{\rm p}$  in the
range of  43 to 115   km s$^{-1}$  kpc$^{-1}$  across the sample 
(Table 9; Figs.  9a--b). 

At the ILRs, there is a match between the natural frequency of radial oscillations 
($\kappa$), and the forcing frequency ($\Omega$ - $\Omega_{\rm p}$), 
 \begin{equation}
\Omega  -  \Omega_{\rm p} = \pm \kappa/2
\end{equation}
where  $\Omega$ is the angular  frequency and $\Omega_{\rm p}$ is the bar pattern speed.
For a weak bar, if the  peak of ($\Omega$ - $\kappa$/2)  exceeds $\Omega_{\rm p}$ 
(Figs.  9a--b), then one or more ILRs must exist. This condition is satisfied 
in the starbursts NGC 4102, NGC 4536, and NGC 3504, and in the  non-starbursts 
NGC 4314, NGC 6951, and NGC 3351.
We use the  upper limit on $\Omega_{\rm p}$ in Fig. 9a--b to 
derive lower and upper limits,  respectively, for the radius 
of the OILR ($R_{\tiny \rm OILR}$) and IILR 
($R_{\tiny \rm IILR}$). These are shown in Table 9.

\it
(II) Method based on  the morphology of circumnuclear dust  lanes: 
\rm 
In the presence of  x$_{2}$ orbits associated with ILRs, 
the two dust lanes which are associated with the 
leading edges  of a  bar 
do not cross the center of the galaxy, but are offset along 
the bar minor axis, in the direction  of x$_{2}$ orbits 
(e.g.,  Athanassoula  1992; 
Byrd et al 1994; Piner et al. 1995). 
The separation of the offset dust lanes provides a lower
limit to  $2 \times R_{\tiny \rm OILR}$. 
Figs. 10a-c illustrate the case for 
the type II  non-starburst NGC 6951.
The large-scale stellar bar  of  NGC 6951 has a semi-major axis 
of $28''$ 
(5.2 kpc; Fig. 10a) and  two relatively straight dust lanes  
are visible on its leading edge  (Fig. 10b).
As they approach the circumnuclear region, 
these dust lanes do not cross the center of the galaxy, but 
instead  are offset  along the bar minor axis. 
These dust lanes connect to the  spiral-shaped CO arms 
where the two  CO peaks lie almost along the minor axis of the 
stellar bar (Fig. 10c) 
The CO and dust morphology are consistent with  the accumulation 
of gas  near the OILR. 
Similar pairs of offset dust lanes are seen in  the 
starbursts NGC 3504 and NGC 4102, and in the type II  non-starbursts
NGC 3351 and NGC 4314.

\medskip

A comparison of the lower limits on $R_{\tiny \rm OILR}$  
 (whether determined by method I or II)  with the azimuthally-averaged 
molecular gas surface density $\Sigma_{\rm gas-m}$  (Fig. 7a)  shows that 
\it 
the starbursts and  Type II non-starbursts  have 
their peak gas surface density  and
most of their circumnuclear gas inside or  near the  OILR 
of the stellar bar.
\rm
Similar results are reported in studies of individual systems 
(e.g., Kenney et al. 1992;  Knapen 1995a,b; 
Jogee 1999, 2001; Jogee et  al. 2002a,b). 
However, the CO data do not enable us to determine whether the gas 
has  flown past the IILR since  the CO resolution (100-200 pc) is 
comparable to the upper limits on $R_{\tiny \rm IILR}$. However, 
we note that  the $HST$ images of many sample galaxies 
(e.g., NGC 4102, NGC 3504, NGC 4536, NGC 4314, NGC 3351, 
NGC 6951)  show that SF does not extend inside the central
100 pc radius or so; this may   be due to 
the IILR blocking further gas inflow. It is interesting
that NGC 2782 which hosts a nuclear stellar bar 
($\S$ 11; Jogee et al. 1999) is the only starburst where 
$HST$  images  show the SF peaking at the center.

\section{ Bar-Driven Secular Evolutionary Scenarios}

In this penultimate section, we  discuss 
potential evolutionary  connections between the 
barred starbursts and non-starbursts  in our sample and how they
fit into scenarios for bar-driven dynamical evolution.

The evidence presented in $\S$ 10  (based on the dynamical frequencies 
and dust morphologies) suggest that  the  
starbursts and type II non-starbursts  have piled up 
most of  their circumnuclear  gas  inside  or  near the OILR 
of the large-scale stellar bar.  
In contrast, the Type I non-starburst such as 
NGC 4569 has an 
extended gas distribution extending out to $r \sim $2 kpc 
($\S$ 7), with a large fraction of the gas 
 still inflowing   along the bar  and exhibiting 
large non-circular motions.

We suggest that a barred galaxy would show up as a  
Type I non-starburst in the early stages of bar-driven inflow 
where  large amounts of gas are still along the bar, 
experiencing shocks, torques, and large non-circular motions.
A schematic illustration is shown in Fig. 11. 
Examples of such systems might 
be NGC 4569 ($\S$ 7; this work), NGC 7479 ($\S$ 7; Laine et al. 
1999), NGC 7723 (Chevalier \& Furenlid 1978), NGC 1300,  and 
NGC  5383 (Tubbs 1982).
During this phase, SF is suppressed or highly inefficient 
due to   large velocity gradients, shear, and tidal 
forces (Elmegreen 1979) in the gas along the bar.
The low  SFR/$M_{\tiny \rm H2}$ along a strong  bar, as typified by
the Type I non-starburst, is fundamental in enabling a bar to efficiently
increases the central mass  concentration in the inner kpc of 
galaxies. 
It is only when the  gas inflow rate along the  bar
exceeds the SFR over the same region, that an efficient mass 
buildup happens.
Available estimates of gas inflow rates along large-scale stellar
bars  range from 1 to 4  $M_{\tiny \sun}$ yr$^{-1}$  
(Quillen et al. 1995;  Laine, Heller, 
\&  Shlosman 1998; Regan, Teuben, \&  Vogel 1997).
If we assume  a conservative net inflow rate of 1  
$M_{\tiny \sun}$ yr$^{-1}$  along the bar into  the inner  
kpc radius, then over 1 Gyr a bar can increase the dynamical 
mass  in the inner kpc by  1 $ \times  10^{9}$ $M_{\tiny \sun}$
or  a few \%.
For instance,  in our sample galaxies  the 
dynamical mass ($\S$ 5)  enclosed  within  
$r$=1 kpc ranges from  6--30  
$ \times  10^{9}$ $M_{\tiny \sun}$ (Fig. 12), 
and would  increase by  6--30 \%. 
In practice, the mass buildup may be slower due to 
starburst-driven gas outflows  which can be $\sim$ 
1--a few $M_{\tiny \sun}$ yr$^{-1}$  
(e.g., Jogee et al. 1998 for the starburst NGC 2782; 
Heckman, Armus, \& Miley 1990).

As discussed in  $\S$ 8 and 10, after gas crosses the  OILR 
it settles into a more  axisymmetric distribution with predominantly 
circular motions.  Thus, in the later stages  of bar-driven inflow, a barred galaxy
is expected to become a Type II non-starburst  (Fig. 11).  The instability analysis in 
$\S$ 8 suggests that   gas concentrations which 
build up near the ILRs  will undergo SF once the gas 
densities exceed the Safronov (1960)/Toomre (1964) 
critical density  $\Sigma_{\rm crit}$ = ($ \alpha \kappa \sigma$/$\pi G $). 
In the inner kpc, the very high  epicyclic frequency  
(several 100--1000  km s$^{-1}$  kpc$^{-1}$)
will naturally force  $\Sigma_{\rm crit}$ to be 
high  (e.g., several 100-1000 M$_{\tiny \sun}$ pc$^{-2}$), 
thereby enabling SF to occur only at high gas densities. 
Under this scenario,   
\it 
some type II non-starbursts  are pre-starbursts  
\rm 
which  may eventually become starbursts  (Fig. 11) if 
the gas density  subsequently becomes  super-critical.

The intense starbursts in our samples   with circumnuclear 
SFRs of 3--11  M$_{\tiny \sun}$ yr$^{-1}$  will build 
compact  young stellar disks or annuli  inside the OILR. 
The interplay  between the SFR, the outflow rate driven by
starburst winds or jets, and the  molecular gas supply  will  
determine how massive such compact stellar components 
eventually  will  be. 
\it 
These compact stellar components , which form from  
cold gas within the inner kpc, may belong to the  class of 
so-called pseudo-bulges 
\rm 
(Kormendy 1993) whose light distribution and kinematics 
are more consistent with a disk than with a spheroidal 
bulge component.   Several studies have in fact 
reported that the central  light distribution
of many late type galaxies are better represented by
an exponential disk-like  component than an  $r^{1/4}$ 
profile  (e.g., de Jong 1996; Courteau, de Jong, \&  Broeils 1996)

As a typical  starburst in our sample converts  its 
molecular gas  reservoir into such a compact disky 
stellar component,
\it  
it is expected to  transition into a Type II  non-starburst
\rm  
as the gas density in part of the wide gas annulus  
will soon becomes  sub-critical. 
One possible outcome is  that SF shuts off  first at lower radii 
(where  $\Sigma_{\rm crit}$ is higher, 
the  instability growth 
timescale  $t_{\tiny \rm GI}$ $\sim$~$Q$/$\kappa$ is 
shorter, and  $\Sigma_{\rm SFR}$ is higher) such that 
we end up with a newly-built  compact disky stellar component 
lying inside  a thin star-forming gas ring.
In our own sample, the Type II non-starburst NGC 3351 
may provide a  striking  example of this process.
Inside its thin CO ring, NGC 3351  hosts  a central stellar 
component which has a high
ratio of rotational to random  motions (Kormendy 1983)
Isophotal fits to the $K$ and $R$ band images of NGC 3351 (Fig. 13) 
shows that this central component has an ellipticity and P.A  
similar to those of the outer disk. 
\it  \rm 
Both the light 
distribution and kinematics would suggest that a compact
stellar component with disk-like properties has built up
in the central $r$=$6\arcsec$ (300 pc) of NGC 3351.
Thus, in this framework 
\it 
 some Type II non-starbursts may also be post-starbursts 
 hosting  a compact disky  stellar component 
\rm 
(Fig. 11).
It should be noted that  
vertical  instabilities  (e.g., Sellwood  1993; Merritt  \& Sellwood 1994), 
vertical  ILRs associated with the  bars  (Combes et al. 1990) 
can scatter stars in the initially flat disky component  to large scale heights, 
thereby enhancing or building a stellar bulge.

The overall schematic  picture  (Fig. 11) which emerges is that 
an episode of bar-driven gas inflow causes a barred galaxy
to go through successive phases  from a Type I non-starburst 
to a Type II non-starburst,   which under the right conditions, 
can undergo  a starburst that will eventually evolve back into  
a Type  II non-starburst.
This picture would naturally  explain why a 
\it one-to-one   \rm correlation
is not seen between bars and  powerful central starbursts although
in a statistical sense  relatively luminous starbursts 
tend to be  preferentially barred compared to normal galaxies   
(Hunt \& Malkan 1999).  
Over its lifetime, a disk galaxy can undergo  numerous 
episodes of bar-driven gas inflow characterized by these 
phases and gradually build up its  central mass concentration 
and  bulge, provided   an adequate gas 
supply is maintained  inside the corotation radius of the bar. 
This can be achieved by accreting gas-rich satellites  in a 
barred disk galaxy  or/and  by  externally inducing  a new bar  whose  
corotation radius encloses pre-existing gas reservoirs at large radii.
Minor (1:10) to moderate  (1:4)  mass-ratio  tidal interactions 
and mergers provide a natural way of doing both (Fig. 11) and 
are believed to be frequent  over the last 8 Gyrs since z=1  
(Jogee \etal  2003; Mobasher \etal 2004).

\section{ Individual Galaxies}

The  molecular gas distribution and kinematics of 
individual  starbursts and non-starbursts  are  described below.
Figures 14a--j and 15a--f show the   total intensity (moment 0) 
and intensity-weighted velocity  (moment 1)  maps.
For selected cases, p-v plots are shown if they can better 
the represent complex kinematics.

\subsection {The Starbursts}
\bf 
NGC 4102, NGC 4536, NGC 3504, and NGC 470: 
\rm 
Within the  inner  kpc radius of  the 
starbursts NGC  470, NGC  4102, NGC  3504, 
and NGC  4536,  the molecular gas distribution (Figs. 14a--h) 
observed in 
\it the sky plane \rm  looks relatively symmetric about the
line of nodes and  is elongated along that position angle 
with an aspect ratio consistent with 
projection effects caused by the inclination of the galaxies.
This implies that  within  the galactic plane of these four  
starbursts, most of the circumnuclear molecular gas has a 
relatively axisymmetric  distribution of 300-600 pc radius. 
The moment 1 isovelocity contours (Figs. 14a--h) 
in the the inner 500 pc radius  
trace the characteristic spider diagram, 
suggesting the gas motions are predominantly circular.
Several starbursts (NGC 4102, NGC 4536) show 
faint gas streams which extend from the relatively axisymmetric 
gas concentrations and curve along the leading edges of the 
large-scale stellar bar, connecting to the large-scale 
dust lanes.
The gas  streams show  non-circular motions and are likely 
inflowing along the bar into the inner kpc region. 

\bf NGC 2782 :  \rm 
The starburst galaxy NGC 2782  
differs in several respects from  the other starbursts.
The CO  morphology in the sky plane is non-axisymmetric, 
double peaked and bar-like in the inner kpc radius (Fig. 14i--j). 
Since the galaxy has a low inclination of $\sim$ 30 $\deg$, 
this elongated morphology is not  due to projection effects, 
but  reflects the intrinsic gas distribution in the 
galactic plane.  Jogee et al. (1999) have presented evidence 
that the CO properties are consistent with the molecular  gas 
responding to a nuclear stellar bar of radius $\sim$ 1.3 kpc, 
identified  via isophotal fits to  NIR images.
The CO bar-like feature  in NGC 2782 is offset in a leading sense 
with respect  to the nuclear stellar bar and  shows some 
non-circular kinematics  (Jogee et al. 1999).  
The nuclear stellar bar appears  
to be fueling gas from the CO lobes  located  a kpc radius 
down into  the central 200 pc radius, where a  powerful central 
starburst is forming  stars at a rate $\ge$  3 
$M_{\tiny \sun}$ yr$^{-1}$  (Jogee et al. 1998). 
It is striking that NGC 2782  is the only starburst in our sample 
where $HST$  images  show the SF peaking at the center. In contrast, 
the other starbursts do not  show  evidence for  strong nuclear 
stellar bars  and their $HST$ images  show that SF does not
extend into the inner 100  pc region.

\subsection {The Non-Starbursts}

\noindent
\bf  NGC 6951:  \rm
\rm  In NGC 6951, there are two CO  peaks at a radius of $6\arcsec$ (550 pc) 
lying  nearly perpendicular to the large-scale stellar bar 
whose  P.A. is  85 $\deg$ (Fig. 3b or 10a). 
Fainter emission  around the peaks  form two  spiral-shaped  
streams  which extend out and curve along the dust lanes on the  leading 
edge of the large-scale bar.
The molecular gas at  the CO peaks and in the  spiral streams  
shows complex non-circular motions  (Figs. 15a--b).  
In  NGC 6951, NGC 3351, and  NGC 4314,  the CO peaks lie near 
the OILR of the large-scale stellar bar ($\S$ 10).

\noindent
\bf NGC 3351:  \rm  
In NGC 3351, the CO emission shows two peaks at
a radius  of $\sim$ 7'' (350 pc).
As in NGC 6951, the two CO peaks lie  nearly perpendicular to the 
large-scale stellar bar whose  P.A. is  110  $\deg$  (Fig. 3b). 
The  CO emission  around the  peaks  appears to connect to 
the dust lanes which lie on  the leading edge of the  bar.
The isovelocity contours curve near the CO peaks, 
suggesting non-circular streaming motions (Fig. 15c--d). 
Interior to the CO peaks, the p-v plot  along the 
kinematic minor axis,  where we are sensitive to the radial in-plane motion  
and vertical out-of-plane motion, shows  
complex  non-circular kinematics.

\noindent
\bf  NGC 4314:   \rm  
In  NGC 4314, most of the gas is concentrated in a 
multiple-peaked  relatively circular ring of radius 
$5\arcsec$--$7\arcsec$ (240 and 400 pc). 
Two faint CO spurs  extend from  the ring and 
intersect the dust lanes which lie on the leading edge of 
the large-scale stellar whose P.A.  is   143  $\deg$  (Fig. 3b). 
Non-circular motions of  10 to 65 km s$^{-1}$  are detected 
in the CO spurs (Fig. 15e--f). 
The CO peaks  lie near 
the OILR of the large-scale stellar bar ($\S$ 10). 
Furthermore, inside  the clumpy CO  ring, ground-based 
(Friedli et al. 1996) and $HST$   images of NGC 4314 reveal a nuclear stellar bar 
of radius  $4\arcsec$ (200 pc) at a position angle of -3 to -12 $\deg$. 
In contrast  to NGC 2782, the CO emission in  NGC 4314 
is not primarily concentrated along 
the nuclear stellar bar, but instead forms  a clumpy  ring at 
the end of the nuclear bar. 
The reason for such different gas and SF distributions in the 
two nuclear bars may be  evolutionary in nature. In the case of 
NGC 4314, we may be seeing gas which is outside the 
corotation of a  nuclear bar that formed earlier  
(Friedli \& Martinet  1993). Conversely, in NGC 2782 we may be witnessing 
the  early decoupling  phase of a nuclear stellar bar induced by 
gas inside the OILR of the large-scale bar.

\bf  NGC 4569:   \rm  
NGC 4569  (Figs. 6a--d)  differs from the above three non-starbursts 
in that it has a  highly extended molecular gas distribution with 
complex kinematics and large linewidths  of $\sim$ 150 km s$^{-1}$.  
A  large fraction of the gas lies outside the inner kpc radius, 
extends out to a  radius of 20~$''$  (1.7 kpc),  
at a similar position angle as the large-scale stellar bar 
(Figs. 6a--b). This gas  has  highly  non-circular 
kinematics, as shown in the p-v plots along the 
kinematic major and   minor axes (Figs. 6c--d). 
Along the kinematic major axis, 
velocities are generally positive (i.e., above 
the systemic value of -235  km~s$^{-1}$) on the northeastern side, 
and  negative on the southwest side. 
However,  at a radius of $5''$ (400 pc), near the feature
marked `N1' in  Fig. 6c, the velocities 
change from +100  km~s$^{-1}$ to a forbidden velocity
of -75 km~s$^{-1}$.  This indicates the presence 
of non-circular motions caused by in-plane 
azimuthal streaming motions or/and vertical motions out of
the plane. As discussed in $\S$ 3, the CO  properties 
of NGC 4569 are consistent with 
the early phases of bar-driven/tidally-driven gas inflow.

\bf  NGC 3359:   \rm  
In NGC 3359,  the interferometric observations 
captured less than 30 \%  of the single dish flux, in contrast to 
the other sample galaxies  (Table 5). The moment 0 map of NGC 3359 
in Fig. 4b shows a very patchy CO distribution made up of a few clumps 
which lie away from  the center. 
The present observations indicate that there is no bright 
centrally peaked CO distribution in the inner 500 pc of NGC 3359.
The  missing  single dish  flux may 
come from extended gas and/or diffuse low surface brightness CO in 
the central regions. 
As discussed in $\S$ 3, several authors have proposed that  
the large-scale   stellar bar  in NGC 3359 is young and 
may have  been tidally triggered and  it is  possible that there are 
significant amounts of gas further out  along the bar.  
To resolve this issue, future interferometric observations  of a 
larger field  of view, using a more compact array configuration  
needs to be carried out.

\section{ Summary and Conclusions}

There is compelling  observational and theoretical evidence that  bars 
efficiently redistribute  angular momentum in galaxies and 
drive gas inflows into the circumnuclear (inner 1--2 kpc) region. 
However, only few  high resolution studies, based on 
a large sample of galaxies, have been carried out 
on the fate of gas in this region. 
In this study, we  characterize  the molecular environment, the onset 
of starbursts, and the secular evolution  in  
the  circumnuclear  region of barred galaxies. We use  
a sample of   local ($D<$ 40 Mpc) 
barred  non-starbursts  and starbursts having   high resolution  
($\sim$~200 pc) CO  ($J$=1$\rightarrow$0), 
optical, NIR, H$\alpha$, RC,  Br$\gamma$, and archival $HST$ 
observations. Our results are summarized below.

\indent
\bf (1) \rm
The circumnuclear  regions  of barred galaxies  host
$3 \times  10^{8}$ to $ 2 \times 10^{9}$ M$_{\tiny \sun}$ of 
molecular gas and have developed  a molecular environment that 
\it  
differs markedly 
\rm  
from that  in the outer disk of galaxies. It includes 
molecular gas  surface densities of 
500-3500  M$_{\tiny \sun}$ pc$^{-2}$, gas mass 
fractions of  10 to 30 \%,  epicyclic frequencies of several 
100 to several 1000  km s$^{-1}$  kpc$^{-1}$, and 
velocity dispersions of 10 to 40  km s$^{-1}$. 
In this environment, gravitational instabilities 
set in only at very high gas densities (few 100-1000 M$_{\tiny \sun}$ pc$^{-2}$),  
but  once triggered, they grow rapidly on a timescale of a few Myrs.
This    high density,  short timescale, `burst ' mode  
may explain why the most intense starbursts tend to be in the 
central parts of galaxies.
Furthermore, the  high pressure, high turbulence ISM 
can lead to the formation of  clouds with high internal dispersion and  
mass, and hence may favor the formation of massive clusters 
as suggested by Elmegreen et al (1993).
The  molecular environment  in the  inner kpc of  the 
ULIRG galaxy  Arp 220  is a scaled-up version of the one in 
these barred galaxies, suggesting  that interactions build up even more extreme conditions.  \\
\indent
\bf (2) \rm
We suggest that the wide variety in  
CO morphologies is  due to different stages  of bar-driven inflow.
A non-starburst like NGC 4569   which is  in the \it early \rm
stages of bar-driven inflow 
has a highly extended molecular gas distribution where  
a large fraction of the circumnuclear gas is still along the large-scale stellar bar,  
outside the outer inner Lindblad resonance (OILR).
This gas   shows large non-circular kinematics and  is not forming stars efficiently. 
Several other galaxies  studied by others such as NGC 7479 
(Laine et al. 1999), NGC 7723 (Chevalier \& Furenlid 1978), 
NGC 1300,  and NGC  5383 (Tubbs 1982) may be in a similar 
phase.
In contrast, we present   dynamical and morphological that
the other non-starbursts and starbursts are in \it later \rm stages
of  bar-driven inflow.  Most of their circumnuclear  gas is  inside  or
near the OILR  of the bar and has predominantly
circular motions.  Across the sample, we estimate upper limits
in the range 43 to 115  km s$^{-1}$  kpc$^{-1}$ for the bar pattern speed
and an OILR radius of  $>$ 500 pc.\\
\indent
\bf (3) \rm
The barred  starbursts and  non-starbursts have   circumnuclear   
SFRs of  3 to 11   and  0.1-2 M$_{\tiny \sun}$ yr$^{-1}$, respectively.
For a given amount of molecular hydrogen ($M_{\tiny \rm H2}$) 
in the inner 1--2 kpc (assuming a standard  CO-to-H$_{\rm 2}$ conversion factor),  
barred galaxies can show an order of magnitude variation in the 
SFR/$M_{\tiny \rm H2}$ over this region. 
This range  seems related to the fact that the  gas surface densities in
the starbursts are larger (1000--3500 $M_{\tiny \sun}$ pc$^{-2}$)  and
close to the  Toomre  critical density over  a large region.
The Toomre $Q$ parameter  reaches its minimum value of
$\sim$  1--2 in the region of star formation, despite
an order of magnitude  variation
in the gas surface density  and epicyclic frequency.
This   suggests that the
onset of gravitational instabilities, as characterized
by $Q$,  plays an important role even in the inner kpc region.\\
\indent
\bf (4) \rm
The  dynamical mass  enclosed  within  the inner kpc radius of 
the barred galaxies in our sample  is 6--30  $ \times  
10^{9}$ $M_{\tiny \sun}$. Molecular gas makes up 10\%--30%
of the dynamical mass  and in the circumuclear starbursts, it is fueling  
a SFR of 3--11  M$_{\tiny \sun}$ yr$^{-1}$ in the inner kpc.
As these starbursts use up their gas and evolve into the post-starburst phase, 
they  will  build  young, massive, high $V/\sigma$  stellar components  
within the inner kpc, inside the OILR of the large-scale bar.
\it 
Such compact stellar components will likely belong to the  class of 
pseudo-bulges 
\rm 
(Kormendy 1993) whose light distribution and kinematics 
are more consistent with a disk than with a spheroidal 
bulge component. We present evidence of such a component 
in  NGC 3351, which seems to be in a post-burst phase.
The observations are consistent with the idea that over
a galaxy's lifetime, it can experience 
numerous episodes of bar/tidally driven gas inflows, which
lead to  a  gradual buildup of its  central mass concentration, the 
formation of pseudo-bulges, and perhaps even 
secular  evolution along the Hubble sequence.

\section{ Acknowledgments}

Support for this  work was generously provided by 
NSF grant  AST 99-81546,  and a grant 
from the K. T. and E. L. Norris Foundation.
SJ  also acknowledges support from  the National Aeronautics
and Space Administration (NASA) under  LTSA Grant  NAG5-13063
issued through the Office of Space Science, a Grant-in-Aid of 
Research from Sigma Xi (The  Scientific Research Society), 
an AAUWEF Fellowship,  and a Zonta International Amelia Earhart 
Fellowship.  


{}
                                                                                



\clearpage

\setcounter{figure}{0}
\begin{center}
\vspace{15 mm}
\includegraphics [height=6.2in]{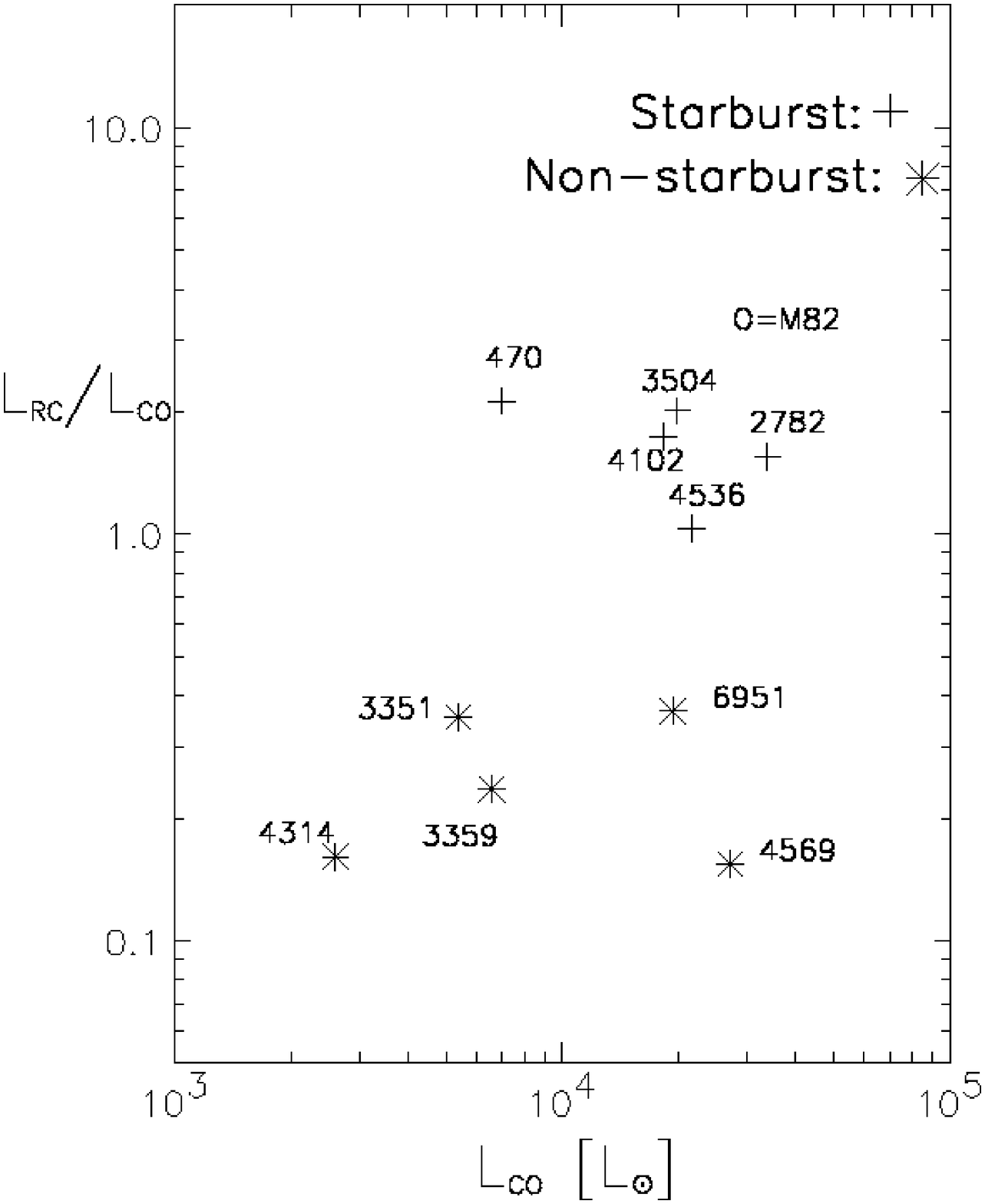}
\end{center}
\vspace{6 mm}
\begin{description}
\item[]
\noindent
\bf
Fig. 1--- 
\bf 
The sample of barred starbursts and non-starbursts:
\rm
\noindent
L$_{\tiny RC}$ is the RC luminosity at 1.49 GHz 
and $L_{\tiny \rm CO}$ is the single dish CO luminosity. Both 
are measured in the central $45''$ diameter or inner 1-2 kpc radius. 
The galaxies in our sample (designated by their NGC numbers) 
as well as the prototypical starburst M82  are plotted. 
Note that for  a given L$_{\tiny CO}$, the 
star formation  rate per unit mass  of molecular hydrogen  
as characterized by L$_{\tiny RC}$/L$_{\tiny CO}$ can 
vary by  an order of magnitude.
\end{description}

\clearpage
\setcounter{figure}{1}
\vspace{5 mm}
\begin{center}
\includegraphics [height=6.0in]{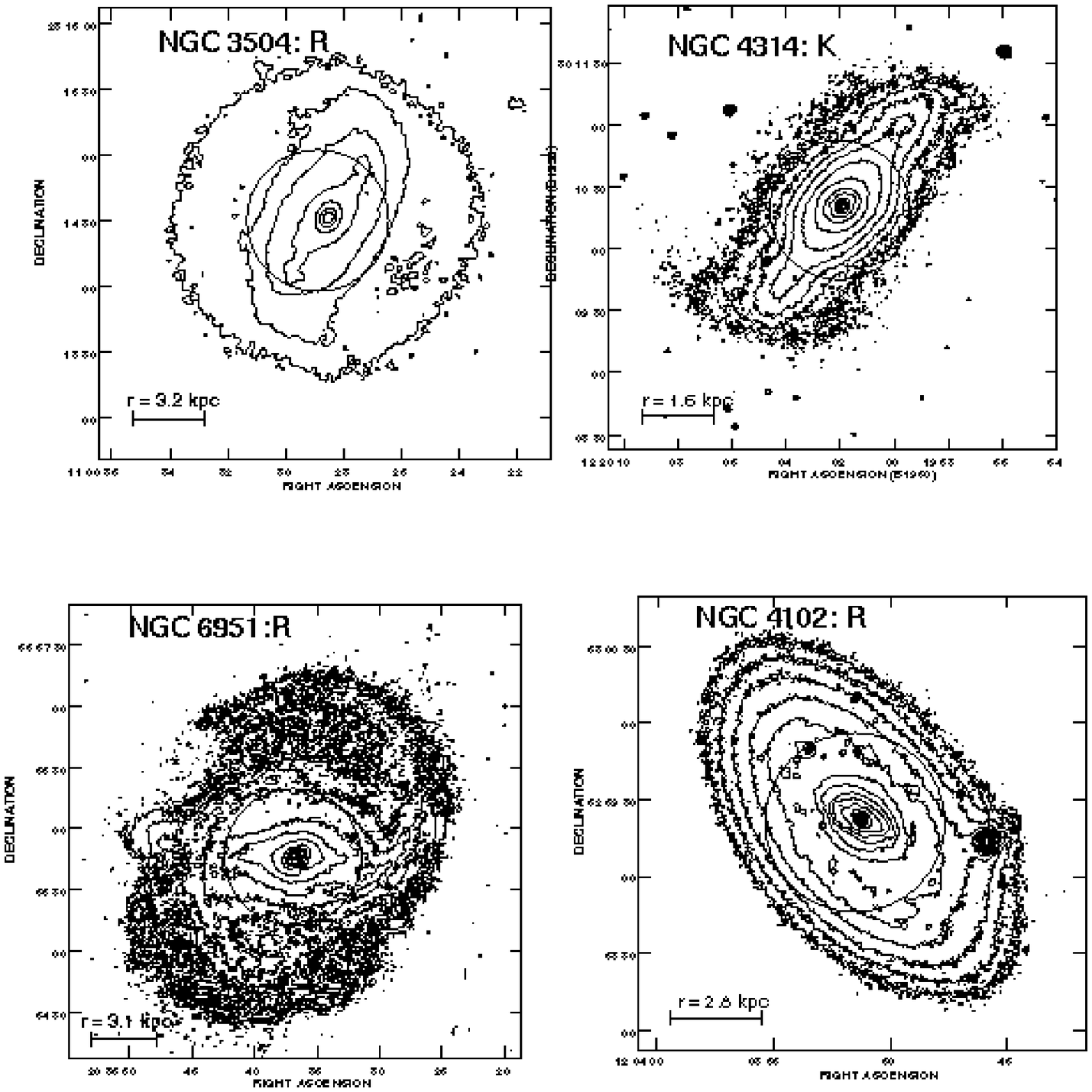}
\end{center}
\vspace{3 mm}
\begin{description}
\item[]
\noindent
\bf
Fig. 2--- 
\bf 
Evidence for large-scale stellar bars/oval distortions: 
\rm
\bf (a) \rm  $R$-band image
of NGC 3504 
\bf  (b) 
\rm $K$-band image of NGC 4314, 
\bf
(c) 
\rm
$R$-band image of NGC 6951, and 
\bf
(d) 
\rm 
the $R$-band image of NGC 4102. 
These images illustrate the  large-scale stellar bars and oval 
distortions in typical sample galaxies.
The dotted circle shows the OVRO CO(1--0) half power beam 
width ($65''$)  at 115 GHz and illustrates the region where
molecular gas is mapped.
\end{description}

\clearpage
\setcounter{figure}{2}
\vspace{5 mm}
\begin{center}
\includegraphics [height=6.5in]{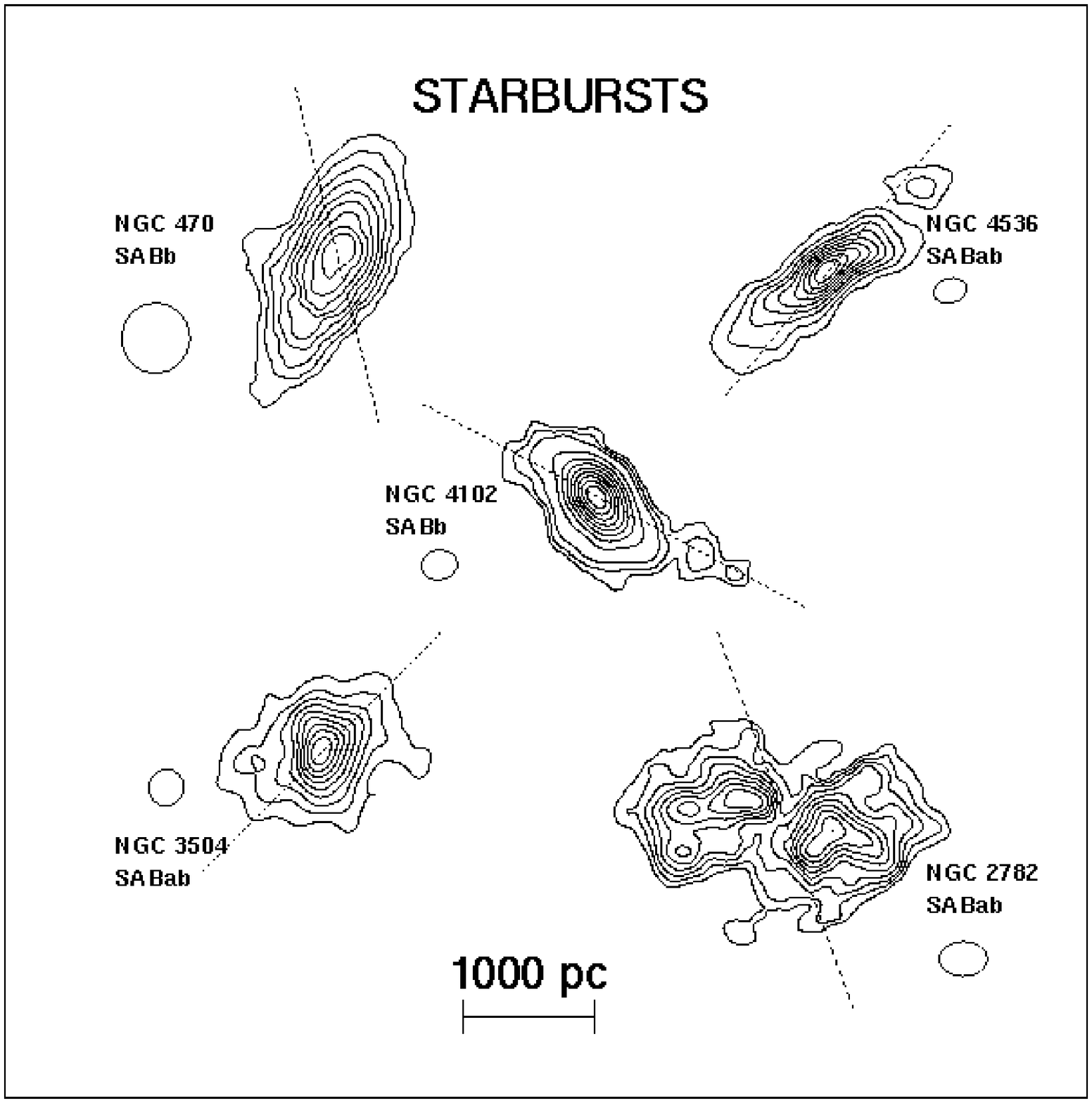}
\end{center}
\vspace{15 mm}
\begin{description}
\item[]
\noindent
\bf
Fig. 3a--- 
\bf 
The circumnuclear molecular gas morphology: 
\rm 
The CO total intensity (moment 0) maps of  barred starbursts and 
non-starbursts are shown. 
The size of the synthesized beam is shown next 
to each map. The dotted line shows the position angle of the 
large-scale stellar bar/oval. 
The contour levels plotted are specified in Table 10. 
There is a wide variety of  circumnuclear  CO morphologies 
including    relatively axisymmetric annuli or disks 
(starbursts NGC 4102, NGC 3504, NGC 4536,  and 
non-starbursts NGC 4314), elongated double- peaked  and spiral  morphologies 
(starburst NGC 2782 and non-starbursts NGC 3351 and NGC 6951), and 
extended distributions elongated along the bar (non-starburst NGC 4569). 
The relationship between the gas distribution and the resonances of the
bar is discussed  in $\S$ 10 and Fig. 9.
\end{description}

\vspace{0 mm}
\begin{center}
\includegraphics [height=6.5in]{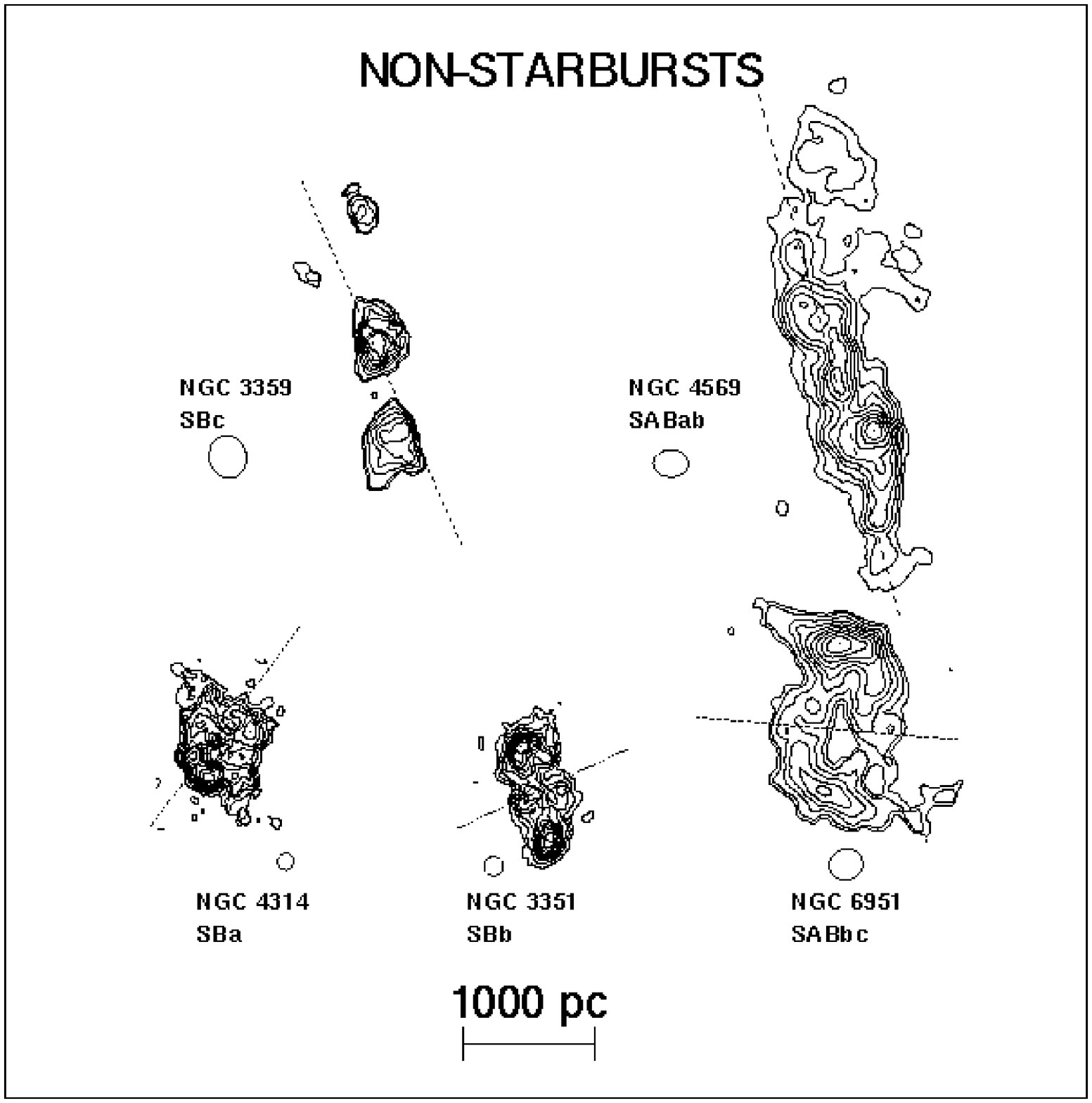}
\end{center}
\vspace{30 mm}
\begin{description}
\item[]
\noindent
\bf
Fig. 3b--- 
\bf 
The circumnuclear molecular gas morphology: 
\rm 
As in 3a, but  showing the barred non-starbursts.
\end{description}

\clearpage\setcounter{figure}{3}
\vspace{5 mm}
\begin{center}
\includegraphics [height=6.6in]{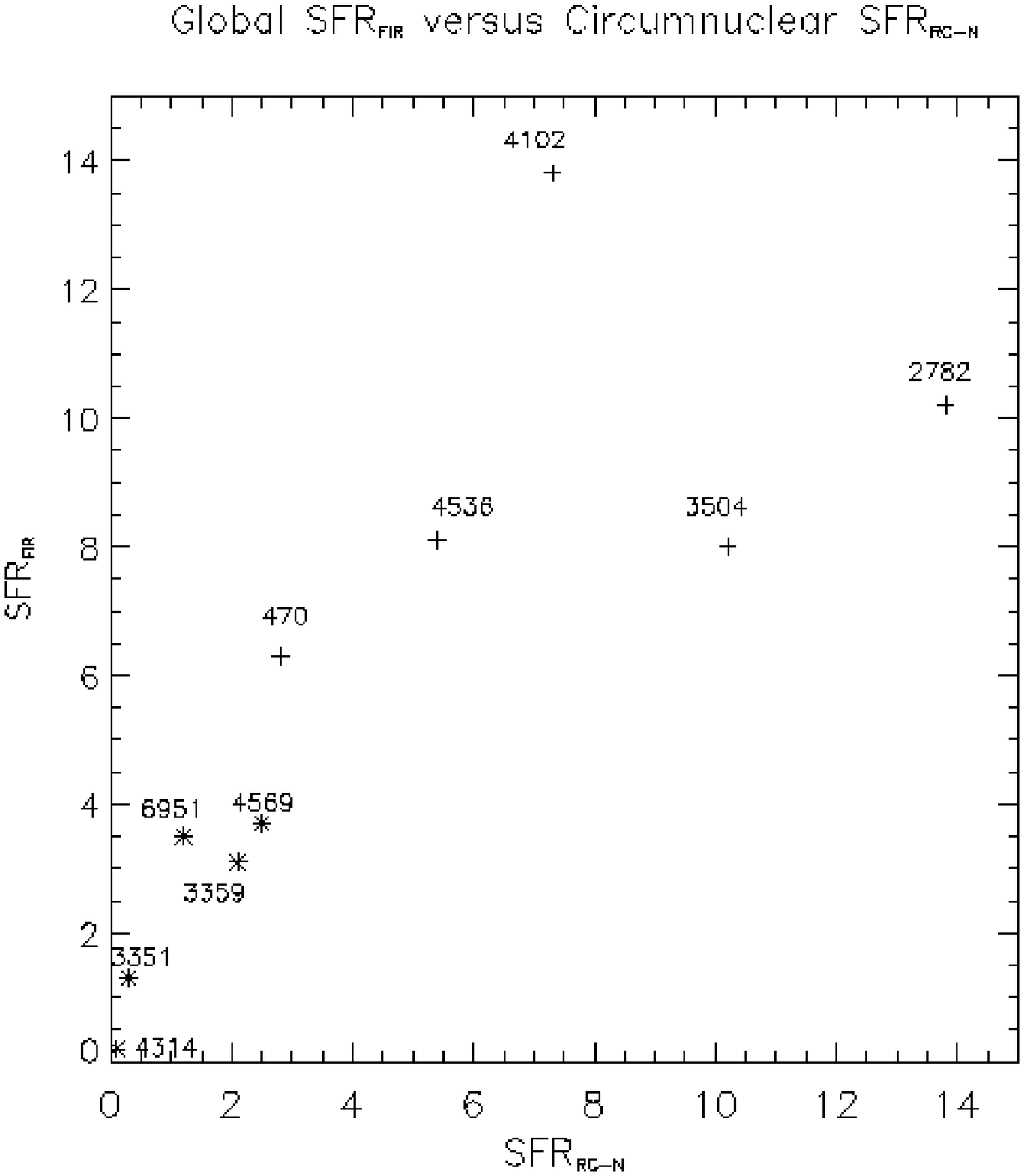}
\end{center}
\vspace{3 mm}
\begin{description}
\item[]
\noindent
\bf
Fig. 4--- 
\bf 
A comparison of the  SFRs estimated from different tracers: 
\rm
The star formation  rate (SFR$_{\tiny \rm RC-N}$) 
estimated from the non-thermal RC in $45''$ is plotted against 
the star formation  rate(SFR$_{\tiny \rm FIR}$) estimated from the 
total FIR luminosity.
SFR$_{\tiny \rm RC-N}$  is less than  SFR$_{\tiny \rm FIR}$  as 
expected,  except in NGC 2782 and NGC 3504, where SFR$_{\tiny \rm RC-N}$ 
is 10 to 30 \% higher.
\end{description}

\clearpage
\setcounter{figure}{4}
\vspace{0 mm}
\begin{center}
\includegraphics [height=6.7in]{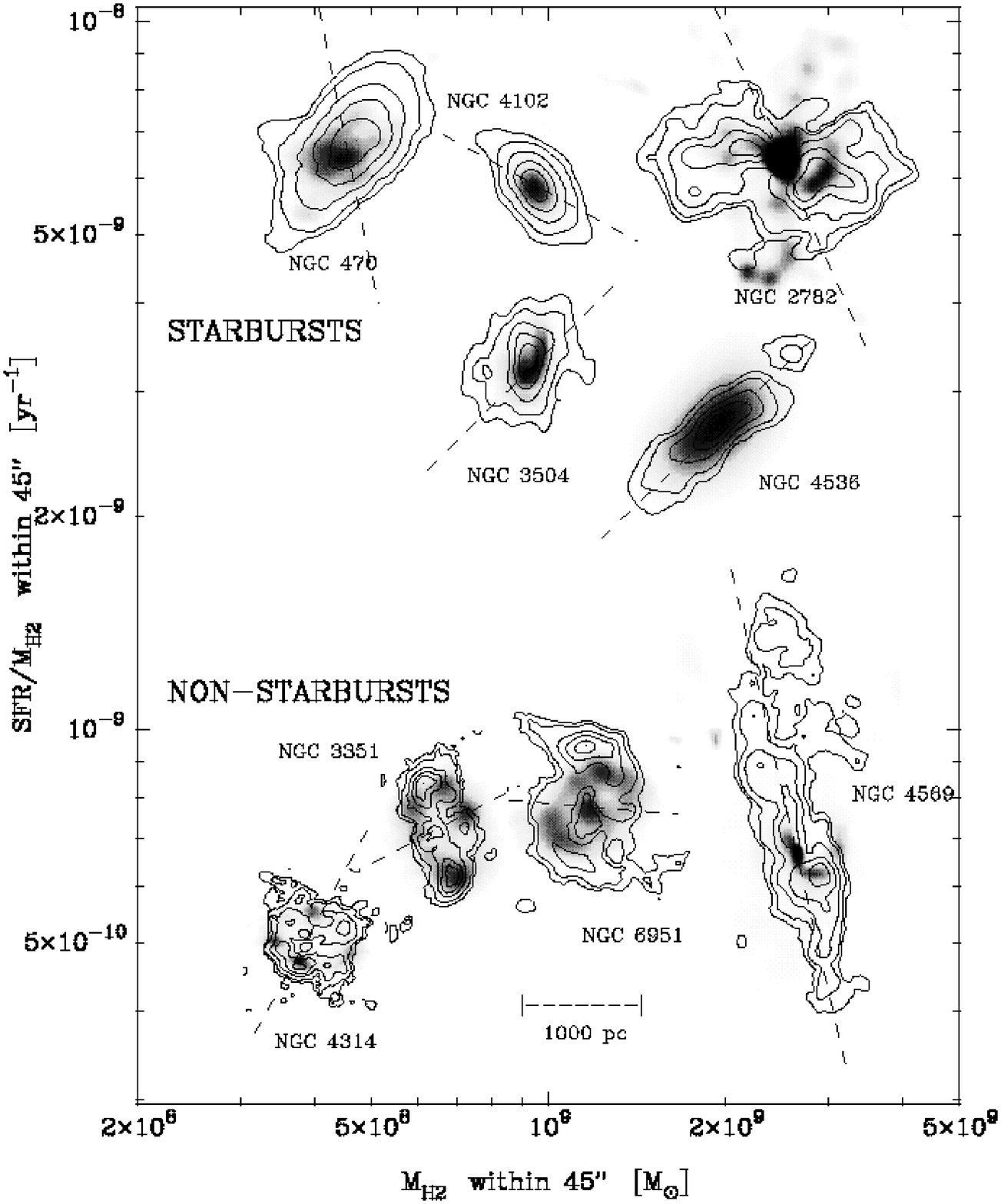}
\end{center}
\vspace{3 mm}
\begin{description}
\item[]
\noindent
\bf
Fig. 5--- 
\bf 
Distribution of molecular gas and SF in the barred starbursts 
and non-starbursts: 
\rm
In the SFR/M$_{\mathrm{H2}}$ versus M$_{\mathrm{H2}}$ plane, 
the CO  intensity  (contours) is  overlaid on the  
1.5 or 4.9 GHz radio continuum map   (greyscale)  of 
NGC 4102, NGC 2782, and NGC 6951  and on the H$\alpha$  map  
for the other  galaxies. 
The greyscale traces the star formation activity  and in the 
case of NGC 6951, it also shows the Seyfert 2 nucleus in the 
inner 100 pc.
The H$\alpha$  and RC maps have a resolution  of $1.0-1.5''$.
The synthesized CO beam is typically $2''$ or 100--200 pc.
The dotted line is  the P.A. of the  large-scale stellar bar/oval.
The contour levels plotted are specified in Table 10.
\end{description}

\clearpage
\setcounter{figure}{5}
\vspace{4 mm}
\begin{center}
\includegraphics [height=6.0in]{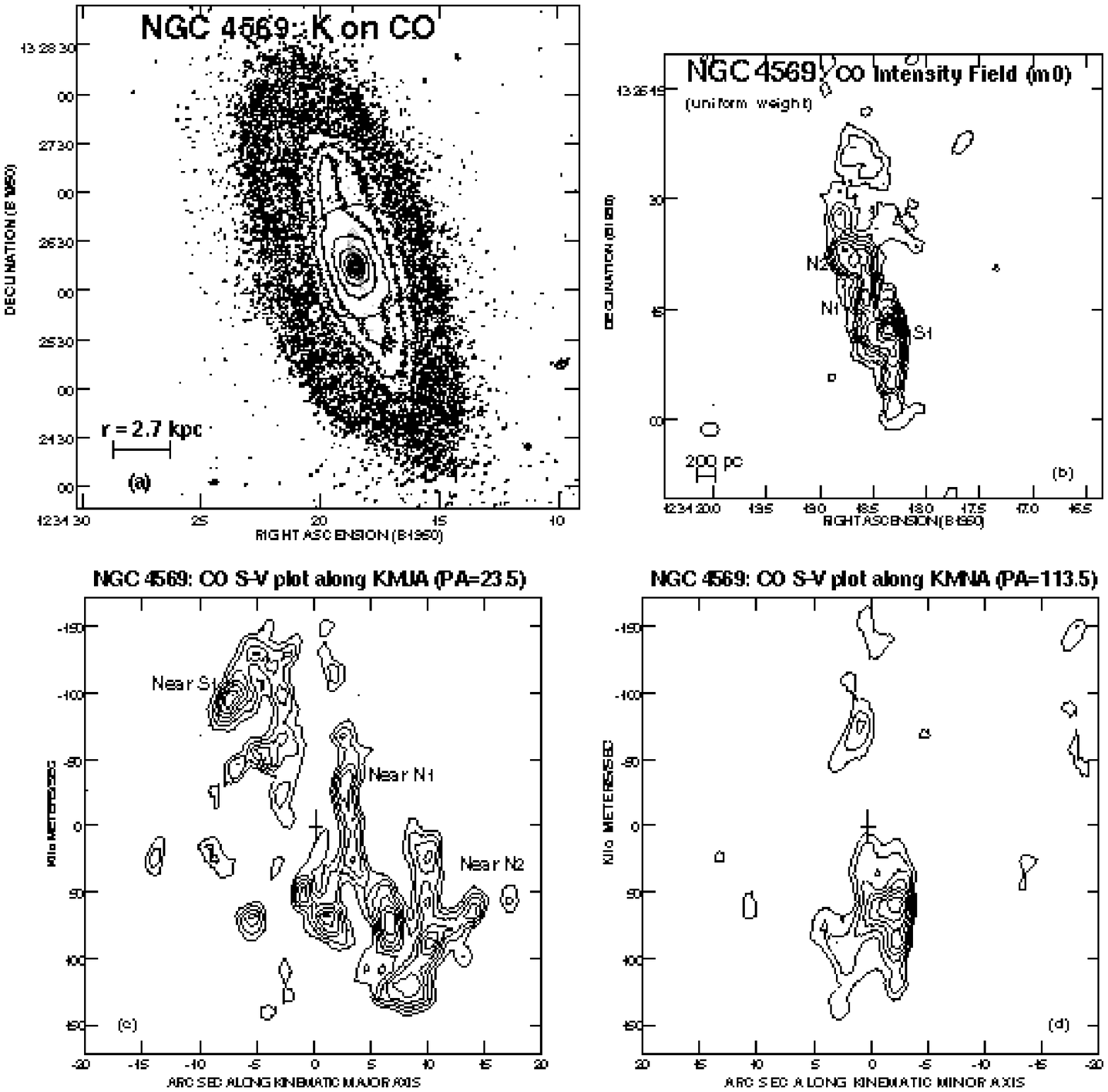}
\end{center}
\vspace{4 mm}
\begin{description}
\item[]
\noindent
\bf
Fig. 6--- 
\bf 
Gas distribution and kinematics in NGC 4569: 
\rm 
\bf
(a)
\rm
The $K$-band image (contours) with a  field of 
view  of  $5'$  (25.9 kpc) shows an asymmetric 
stellar bar.  The overlaid circle shows the $65''$  
half power beam width of the CO interferometric observations. 
The CO distribution inside the beam is shown in greyscale.
\bf
(b)
\rm
The  uniformly weighted  ($2.65'' \times 2.01''$ or $ 223 \times 169$ pc) 
CO total intensity map  in the central  $6''$ (5.0 kpc) shows 
an extended  molecular gas distribution in NGC~4569. 
Contour levels plotted are 
3.20 Jy  beam$^{-1}$~km s$^{-1}$~$\times$~(1, 2, 3, 4, 5, 6, 7, 8, 9, 10).
The gas extends out to a large (20 $''$ or 1.7 kpc) radius, and is elongated 
along the direction of the large-scale stellar bar. 
\bf 
(c), (d) 
\rm
show the p-v plots  along the kinematic major  
and minor axes.
The molecular gas extends out to  a 2  kpc radius and shows 
very disturbed kinematics.  
Along the kinematic major axis, velocities are generally positive 
(i.e above the systemic value) on the northeastern side. 
However, at a radius of $5''$ (400 pc), near the feature
marked `N1' in    (b), the velocities 
change from $\sim$ +100  km~s$^{-1}$ to a forbidden velocity
of -75 km~s$^{-1}$.
\end{description}

\clearpage
\setcounter{figure}{6}
\vspace{0 mm}
\begin{center}
\includegraphics [height=6.7in]{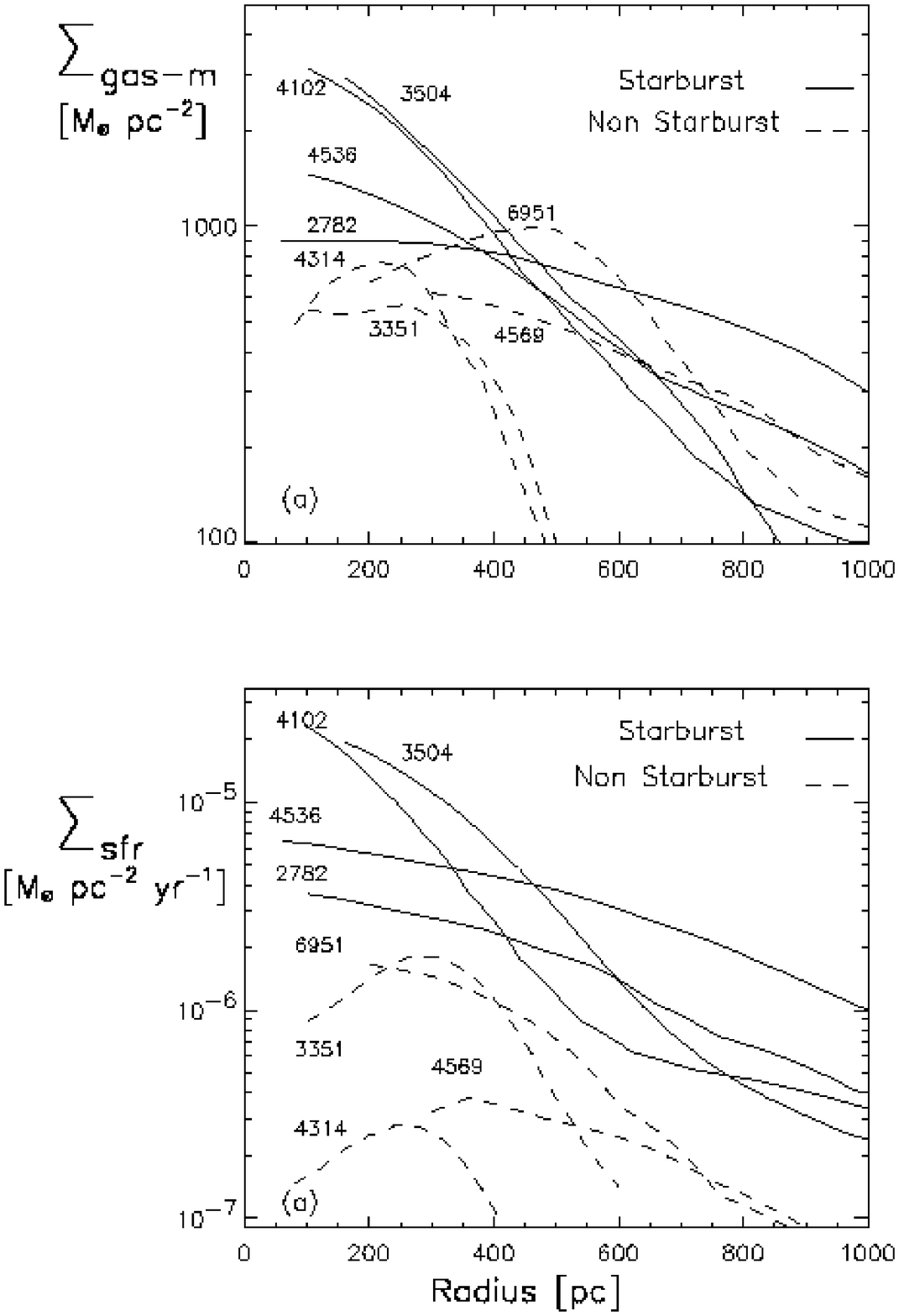}
\end{center}
\vspace{3 mm}
\begin{description}
\item[]
\noindent
\bf
Fig. 7--- 
\bf 
(a), (b) 
\rm
show the azimuthally averaged molecular gas surface density 
($\Sigma_{gas}$) and  the SFR per unit area ($\Sigma_{\tiny SFR}$).
The  extinction-corrected profiles are  convolved to a similar 
resolution of 100-200 pc  for all the galaxies. 
Quantities are plotted starting at a radius $\ge$ half 
the size of the synthesized beam (typically $2''$ or 150 pc).  
Most of the  starbursts have developed larger 
molecular gas surface densities (1000-3500 $M_{\tiny \sun}$ 
pc$^{-2}$) in the inner 500 pc radius  than the  
non-starbursts,   for a given CO-to-H$_{\rm 2}$ conversion factor. 
In the inner 500 pc radius of the  starbursts, 
both $\Sigma_{\rm gas-m}$ and $\Sigma_{\tiny SFR}$  
increase toward the center.
\end{description}

\clearpage
\setcounter{figure}{7}
\vspace{0 mm}
\begin{center}
\includegraphics [height=6.7in]{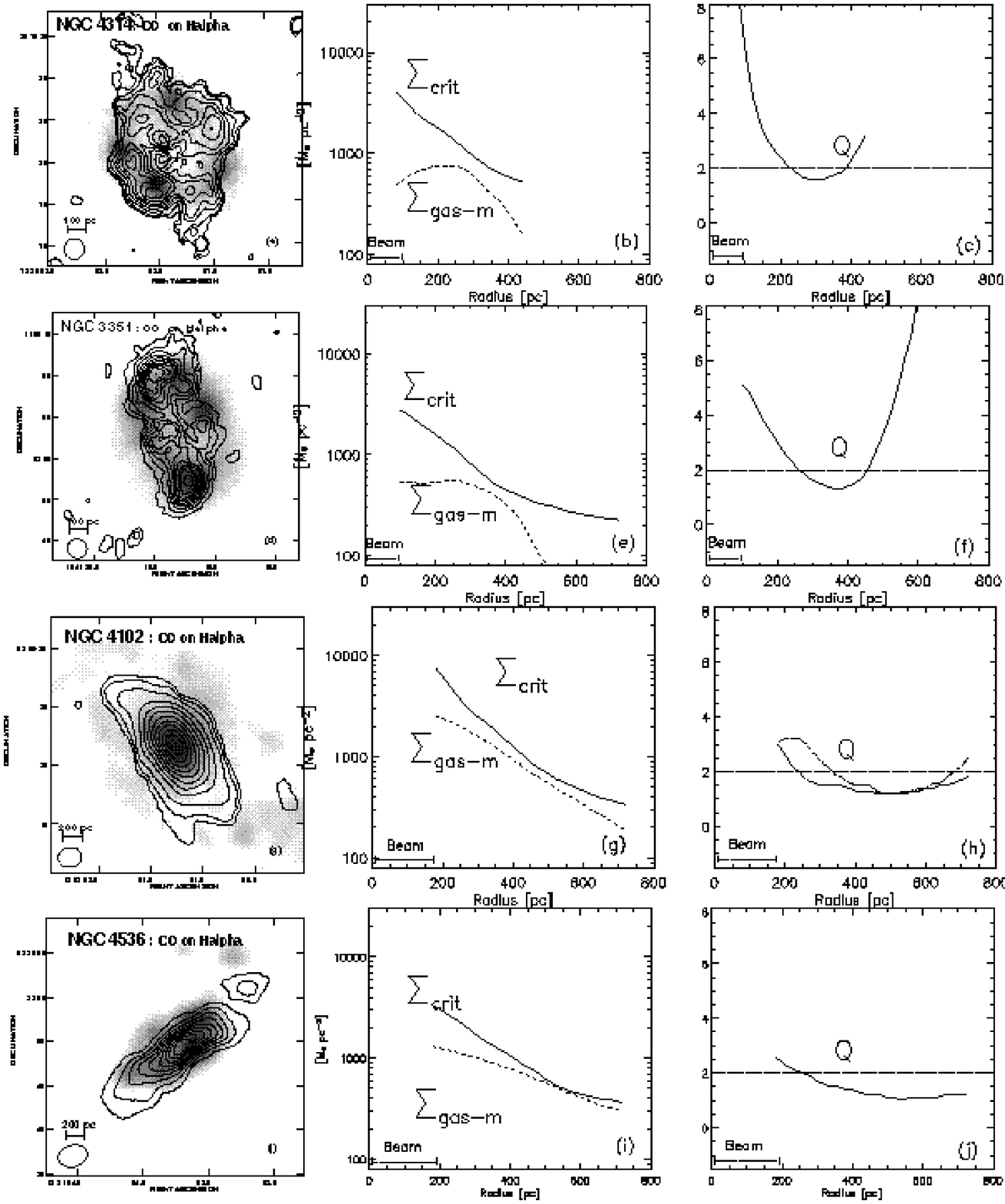}
\end{center}
\vspace{-4 mm}
\begin{description}
\item[]
\noindent
\bf
Fig. 8--- Comparison of the gas density with the Toomre critical density 
in the non-starbursts and starbursts: 
\bf Column 1:~\rm
The  CO  distribution  (contours) on 
the  H$\alpha$ (greyscale) for the non-starbursts NGC 4314 and NGC 3351,
and the starbursts NGC 4102 and NGC 4536.  
\bf Column 2:~\rm
The molecular gas surface density $\Sigma_{\rm gas-m}$ and  
the critical density  ($\Sigma_{\rm crit}$ =$ \alpha \kappa \sigma$/$\pi G $)
for the onset of gravitational instabilities assuming  $\alpha$ =1.
Quantities are plotted starting at a radius equal to the 
CO beam  size ($\sim$2$\arcsec$). 
\bf Column 3:~\rm
The Toomre $Q$ parameter ($\Sigma_{\rm crit}$/$\Sigma_{\rm gas}$). 
The horizontal line denotes the value of $Q$ in the outer disk of
spirals where  Kennicutt (1989)  finds $\alpha$ =0.7. 
In the non-starbursts  NGC 4314 and NGC 3351,  HII regions are 
concentrated in a gas-rich annulus of  radius  350 pc ($7\arcsec$) while 
further in, the star formation activity drops sharply 
despite large gas surface densities  above 400 $M_{\tiny \sun}$~pc$^{-2}$.  
In these  non-starbursts, $Q$ reaches its lowest value (1--2) in 
the ring  of HII regions while at lower radii  
$Q$ increases to  $\sim$  6, suggesting sub-critical gas densities.
In the starbursts NGC 4102 and NGC 4536, 
$Q$ remains  remarkably close to   1--2 
over a wide region,  between a radius of 250 to 700 pc,     
although the gas 
surface density and the epicyclic frequency 
both vary by nearly an order of magnitude.
\rm 
\end{description}

\clearpage
\setcounter{figure}{8}
\vspace{0 mm}
\begin{center}
\includegraphics [height=6.7in]{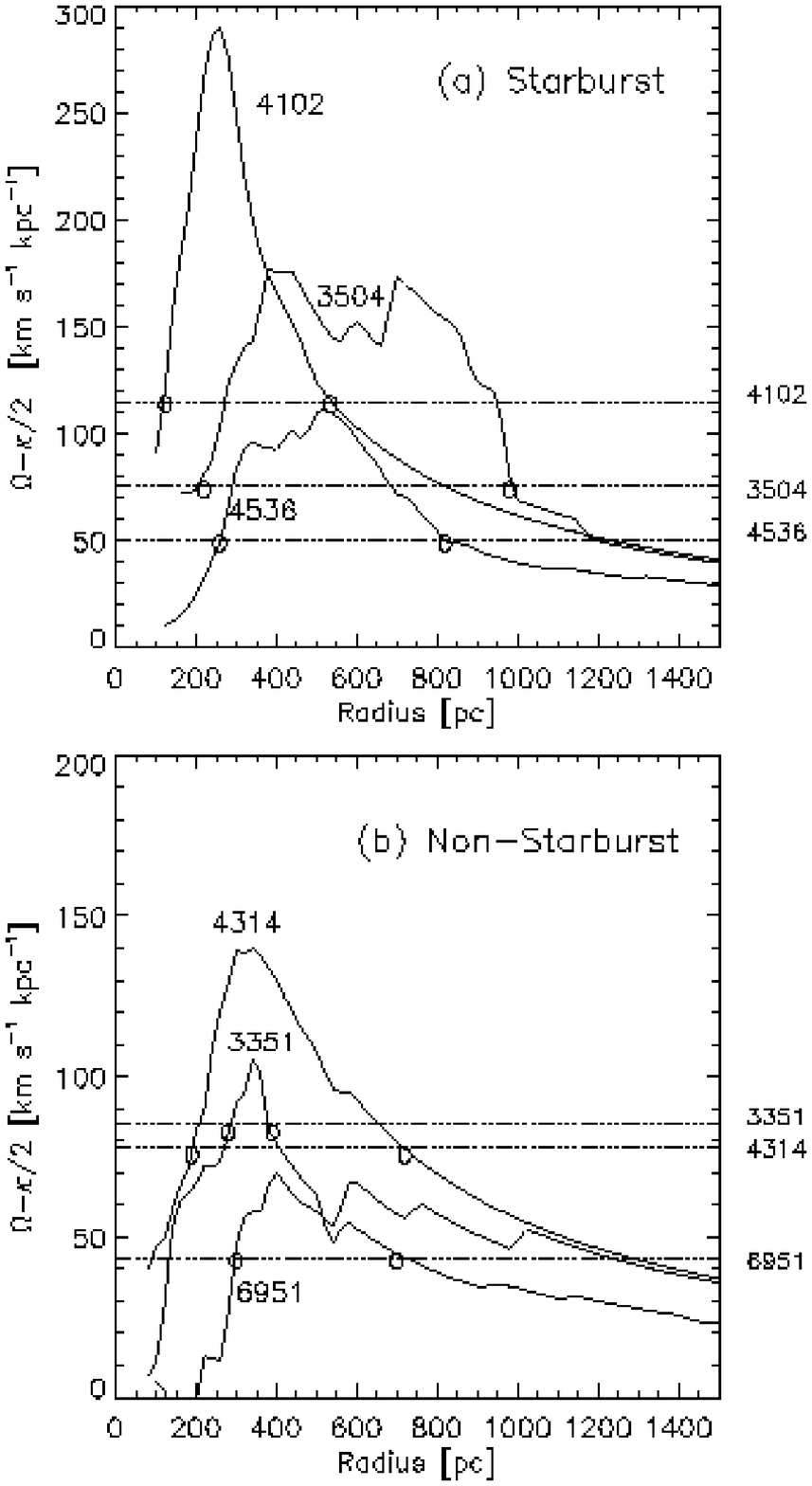}
\end{center}
\vspace{3 mm}
\begin{description}
\item[]
\noindent
\bf
Fig. 9--- 
\bf 
Inner Lindblad resonances in the barred starbursts and non-starbursts: 
\rm
[$\Omega$ - $\kappa$/2] is plotted against radius 
for select \bf (a) \rm starbursts and \bf (b) \rm non-starbursts. 
The bar pattern speed $\Omega_{\rm p}$  is drawn as horizontal  lines 
and estimated  by assuming that the corotation 
resonance is  near the end of the large-scale stellar bar (see Table 9). 
Under the epicycle theory for a weak bar, the intersection of 
[$\Omega$ - $\kappa$/2] with $\Omega_{\rm p}$ defines the 
locations of the ILRs. In these systems, the circumnuclear 
gas is concentrated inside the outer ILR of the large-scale stellar 
bar/oval.
\end{description}

\clearpage
\setcounter{figure}{9}
\vspace{0 mm}
\begin{center}
\includegraphics [height=6.0in]{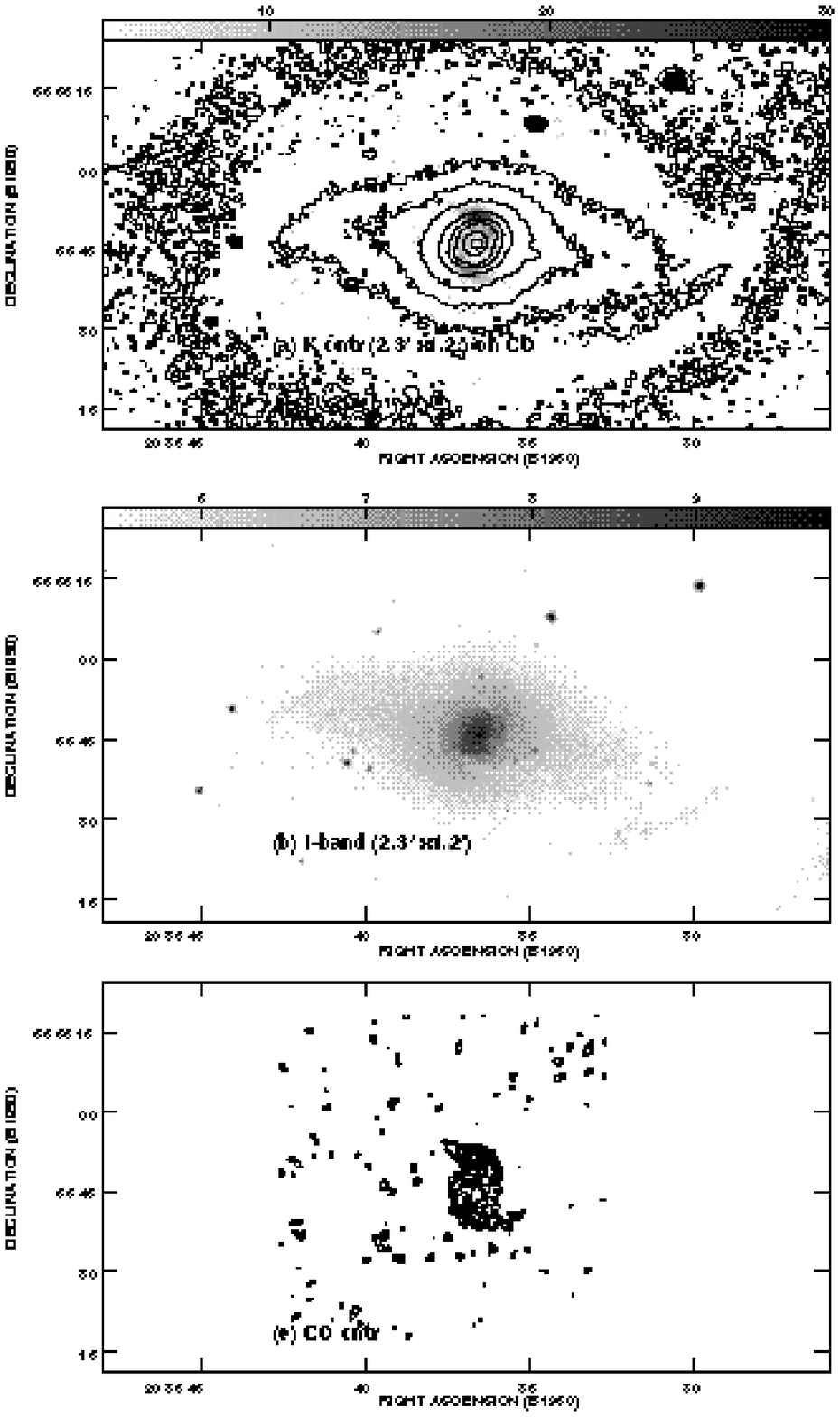}
\end{center}
\vspace{-4 mm}
\begin{description}
\item[]
\noindent
\bf
Fig. 10--- 
\bf 
Relationship between the  large-scale stellar bar, 
large-scale dust lanes, and  circumnuclear CO morphology in 
NGC 6951: 
\rm 
\bf
(a)
\rm
The $K$-band image (contours)  of NGC 6951 with a $2.3'\times 1.2'$ 
(12.7 $\times$ 6.9 kpc)  field of view shows the inner 
region of the large-scale stellar bar which has a 
semi-major axis of  $28''$ (5.2 kpc). 
The circumnuclear CO  (J=1-$>$0) total intensity map (greyscale) 
is superposed, but note that its  half power beam  is only 
$65''$ (6.1 kpc).
\bf
(b)
$I$-band image with the same field of view  as the  K-band image  in (a). 
Two relatively straight dust lanes 
extend along the major axis of the large-scale 
stellar bar and are offset towards its leading edge. 
These dust lanes do not cross the center of the galaxy, 
but rather cross the bar minor axis and connect 
to a ring of star formation of radius  $\sim 4''$  (380 pc).
\bf
(c)
\rm 
The CO  (J=1-$>$0) total intensity map (contours) is displayed 
on the same scale as (a) and (b) to facilitate comparisons. 
NGC 6951 has two spiral-shaped CO arms where the emission peaks 
at a  radius r  $\sim$ $6''$ (570 pc). The two  CO peaks lie 
almost along the minor axis of the bar and the CO arms connect to the 
two relatively straight dust lanes on the leading edge of the 
large-scale stellar bar. 
The CO and dust morphology are consistent with the piling up of gas 
near the OILR.
\end{description}

\clearpage
\setcounter{figure}{10}
\vspace{0 mm}
\begin{center}
\includegraphics [width=5.7in]{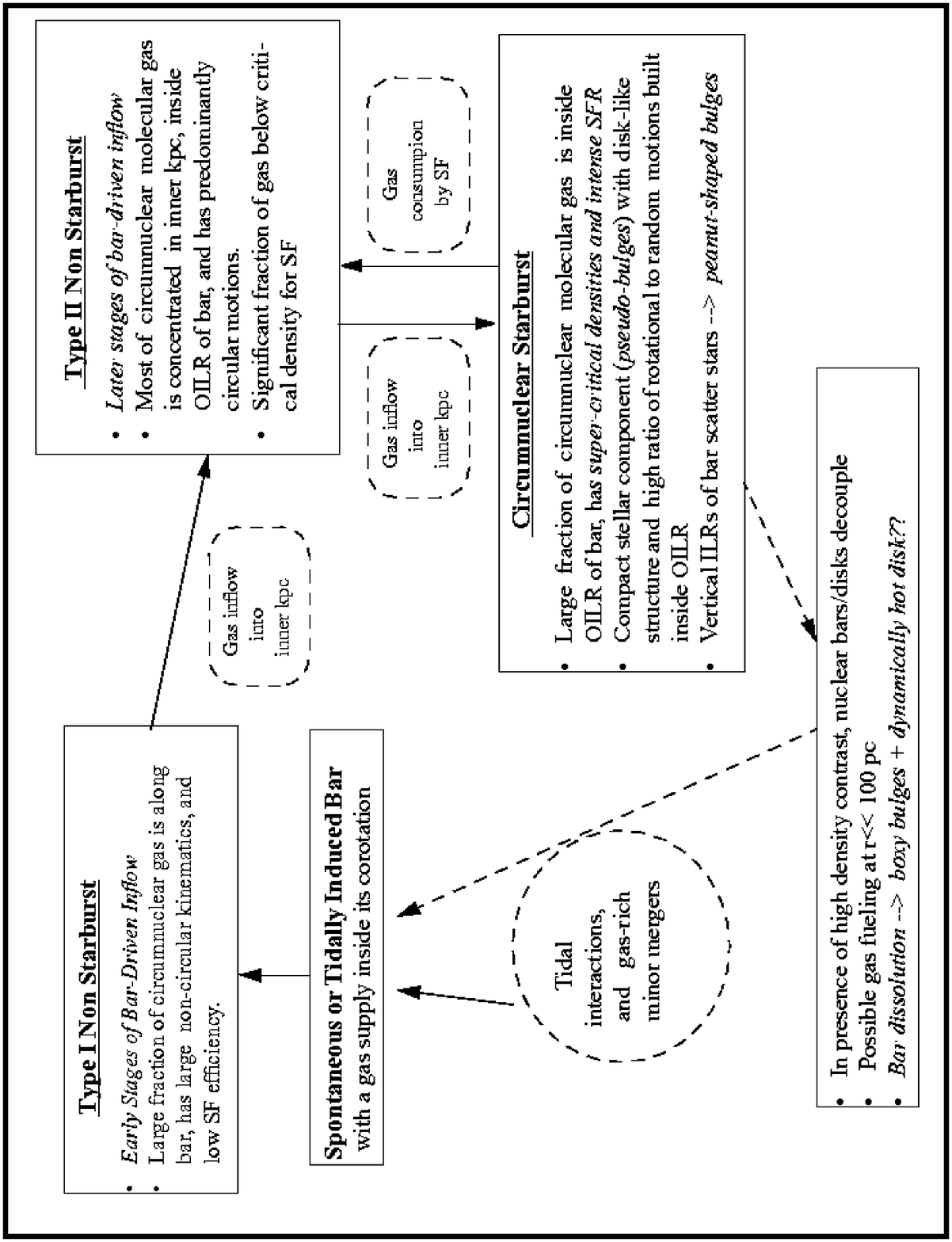}
\end{center}
\vspace{0 mm}
\begin{description}
\item[]
\noindent
\bf
Fig. 11--- 
\bf 
Bar-Driven Secular Evolution: 
\rm
This  figure sketches  possible scenarios 
for bar-driven dynamical evolution and  postulates 
potential  evolutionary  connections between the 
barred starbursts and non-starbursts  in our sample.
We suggest that a strongly barred galaxy would show up as a  
Type I non-starburst in the early stages of  bar-driven inflow
where  large amounts of gas are still inflowing along the bar, 
have large non-circular motions, and are not forming stars 
efficiently. 
In the later stages where most of the circumnuclear gas 
has piled up inside the OILR of the bar,  we may see a Type 
II non-starburst or  even a  circumnuclear starburst  if  most 
of the gas exceeds a critical density. The latter 
seems well represented by the Toomre value in our sample.  
Pseudo-bulges (Kormendy 1993)  or 
compact stellar components with  disk-like  properties  
can be built in the inner kpc, as seen in our sample galaxy 
NGC 3351. Over its lifetime, a disk galaxy can undergo  numerous 
episodes of bar-driven gas inflow and gradually build up its  
central mass concentration and  bulge, provided   an adequate gas 
supply is maintained  inside the corotation radius of the bar.  
However, if the central mass concentration becomes large enough, bar
dissolution may occur, leaving behind a dynamically hot disk.
It is as yet unclear if tidal interactions and gas-rich accretions 
can lead to viable recurrent bar formation in such a disk. 
\end{description}

\clearpage
\setcounter{figure}{11}
\vspace{7 mm}
\begin{center}
\includegraphics [height=6.7in]{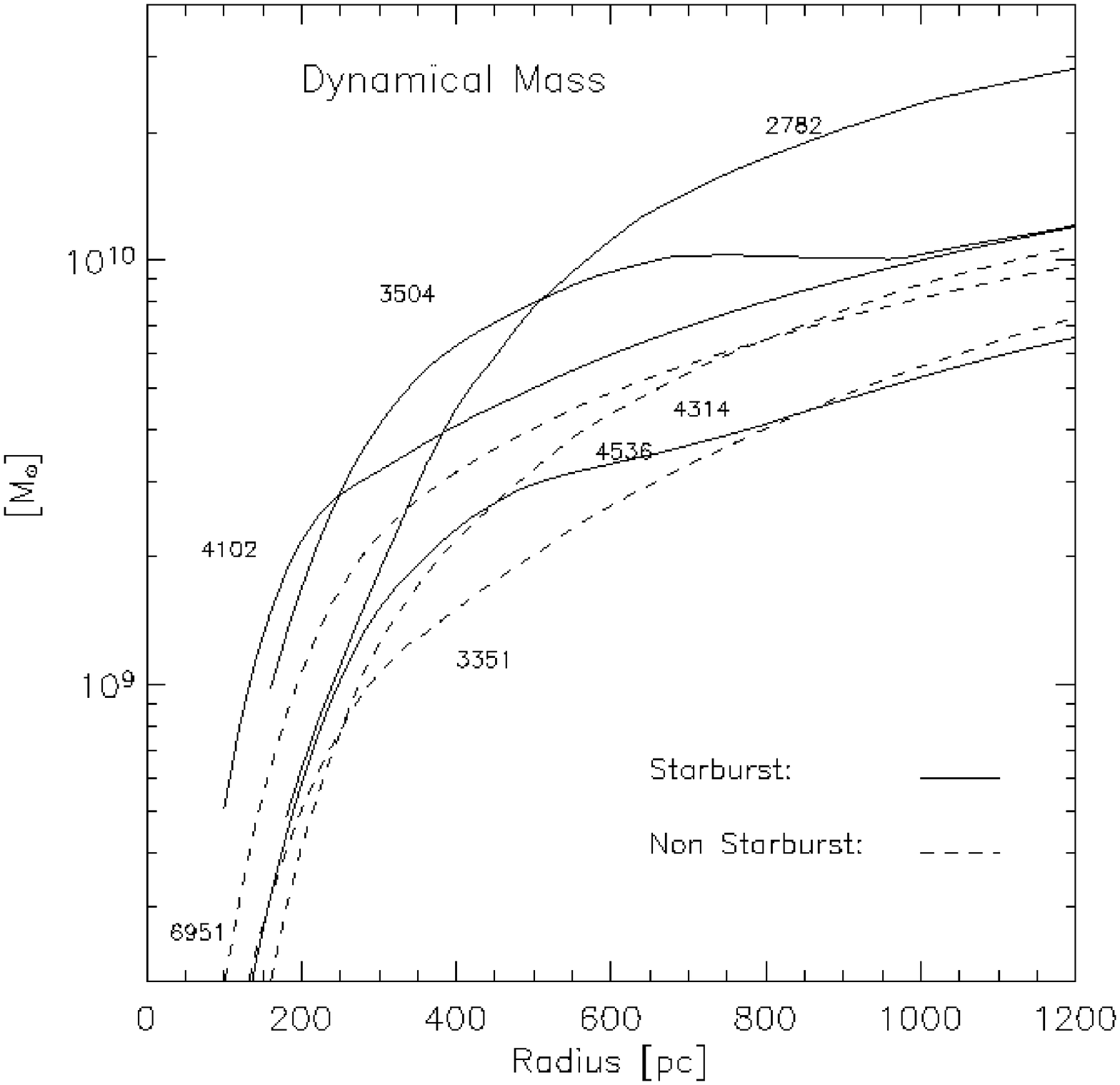}
\end{center}
\vspace{3 mm}
\begin{description}
\item[]
\noindent
\bf
Fig. 12--- 
\bf 
The  enclosed dynamical mass within the inner kpc: 
\rm
The dynamical mass (M$_{dyn}$) enclosed within a given radius $R$ is 
estimated from the rotation curve, assuming a  
spherically symmetric gas distribution. Values are shown only  
for sample  barred  galaxies  where a reliable rotation curve  can be derived.
For $R$=1 kpc, M$_{dyn}$ has a range of  6--30  $ \times  
10^{9}$ $M_{\tiny \sun}$ 
\end{description}

\clearpage
\setcounter{figure}{12}
\vspace{0 mm}
\begin{center}
\includegraphics [height=5.8in]{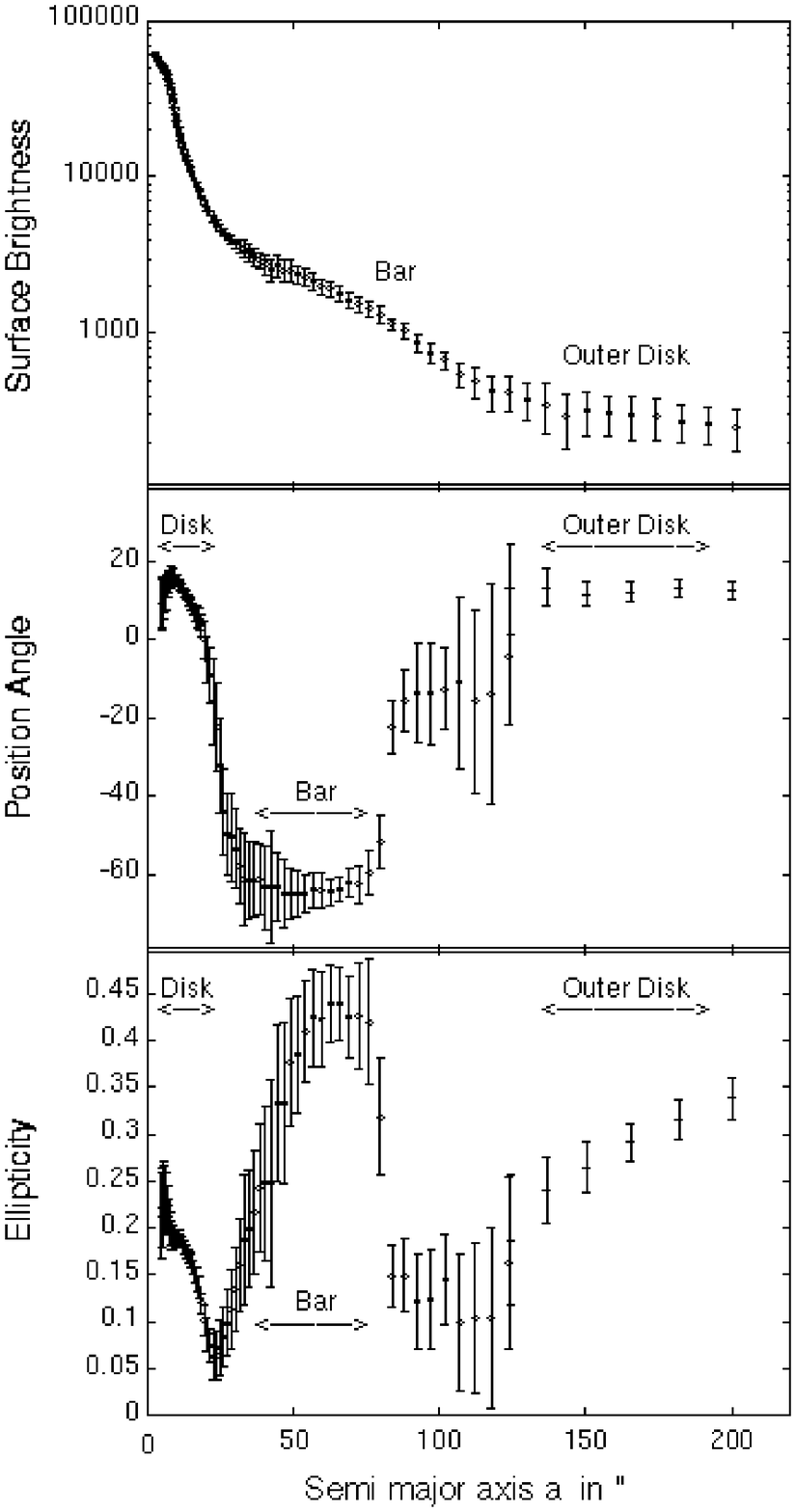}
\end{center}
\vspace{50 mm}
\begin{description}
\item[]
\noindent
\bf
Fig. 13--- 
\bf 
Evidence for a pseudo-bulge or compact disk in NGC 3351?  
\rm
The isophotal analysis of the $K$-band 
and $R$-band images of NGC~3351 is shown.
Inside the inner kpc radius, the ellipticity 
rises from a minimum value of 0.05  at r $\sim$ 1 kpc 
to a maximum of $\sim$  0.2 at a radius of  $5''$ (250 pc).
The position angle (P.A.) of the isophotes  change 
from -60$\deg$ in the region of the 
large-scale stellar bar to a value between 5 and 20$\deg$ 
in the central  $5''$ (250 pc) radius. 
The central component has an ellipticity and P.A.
which is similar to that of the outer disk, and 
a published high ratio of rotational to random  
motions.
\end{description}

\clearpage
\setcounter{figure}{13}
\vspace{5 mm}
\begin{center}
\includegraphics [height=6.0in]{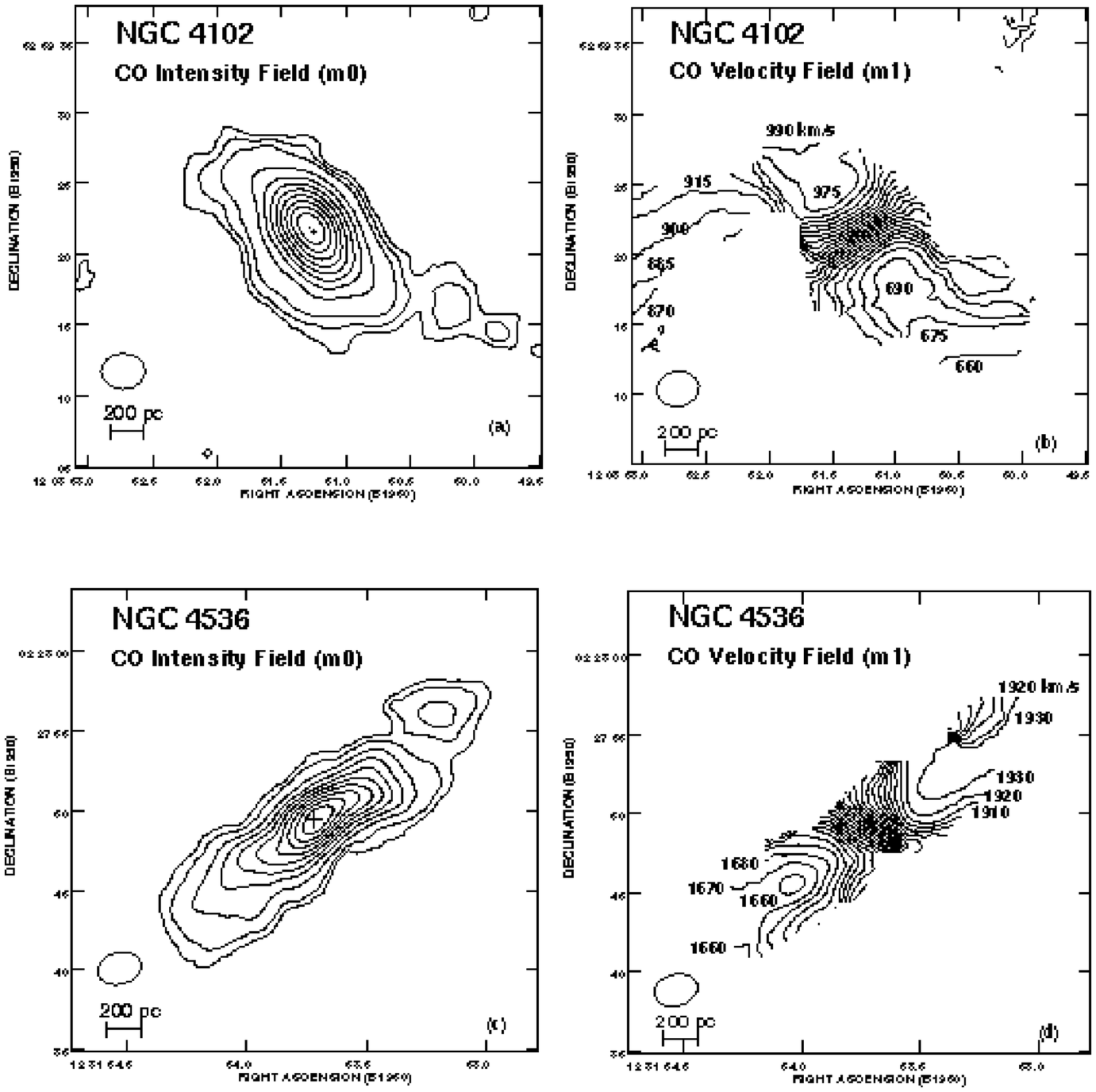}
\end{center}
\vspace{3 mm}
\begin{description}
\item[]
\noindent
\bf
Fig.14--- 
Molecular gas kinematics and distribution in individual starbursts: 
\rm
The CO  intensity-weighted velocity (moment 1) fields 
and the CO total intensity (moment 0) maps 
of the starbursts NGC 4102  (a-b), NGC 4536 (c-d), NGC 3504 (e-f), 
NGC 470 (g-h), and NGC 2782 (i-j) are shown.
The size of the synthesized beam is shown next to each map. 
The contour levels plotted are specified in Table 10. 
Relatively axisymmetric annuli or disks (NGC 4102, NGC 3504, NGC 4536) 
and  elongated double-peaked  morphologies (NGC 2782) are seen. 
The velocity field in the inner 500 pc radius is generally 
dominated by circular motions, with  weaker non-circular components. 
In NGC 4102 and NGC 4536,  there are  faint gas streams which 
extend out, intersect the  dust lanes on  the leading edges of the 
large-scale stellar bar, and show  non-circular motions. 
In NGC 2782, there are weak bar-like streaming motions in the inner 400 
pc radius, on the leading side of a nuclear 
stellar bar.
\end{description}

\clearpage
\setcounter{figure}{13}
\vspace{5 mm}
\begin{center}
\includegraphics [height=6.0in]{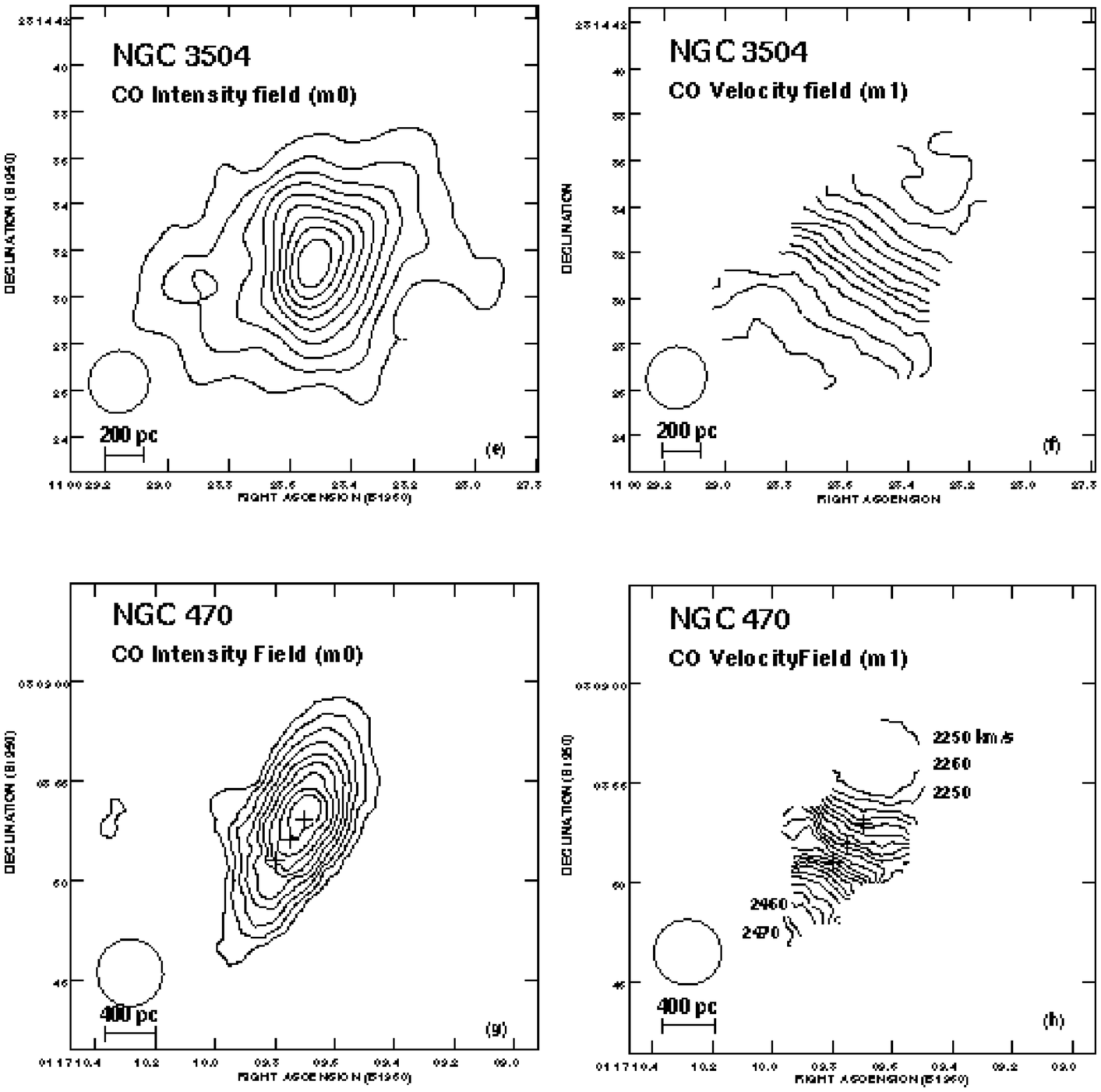}
\end{center}
\vspace{3 mm}
\begin{description}
\item[]
\noindent
\bf
Fig.14--- Continued
\bf
\end{description}

\clearpage
\setcounter{figure}{13}
\begin{figure}
\includegraphics[width=2.5in]{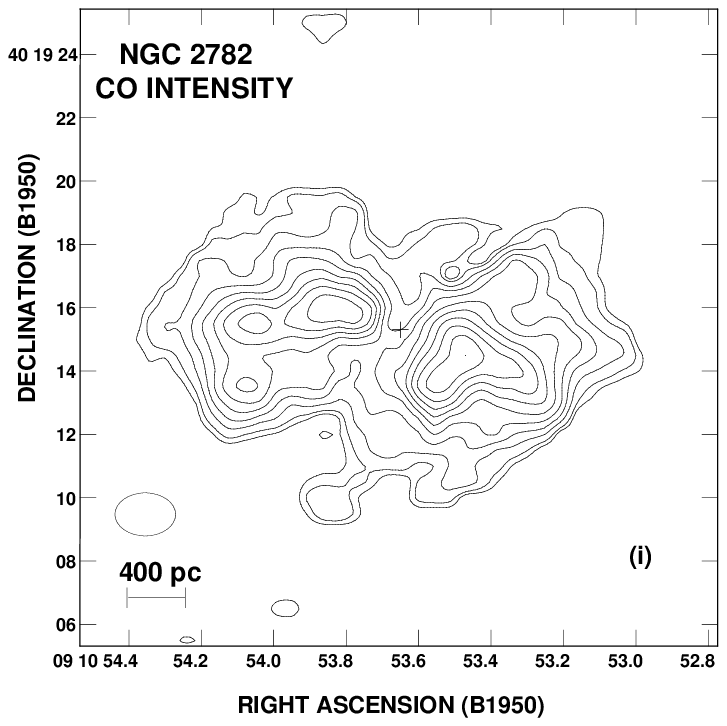}
\includegraphics[width=2.7in]{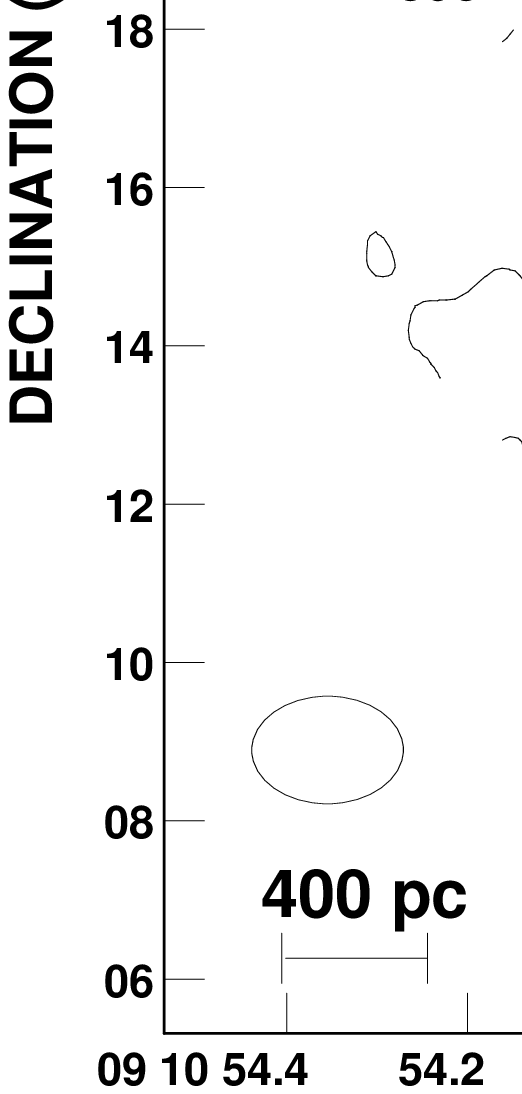}
\caption{Continued}
\end{figure}

\clearpage
\setcounter{figure}{14}
\vspace{5 mm}
\begin{center}
\includegraphics [height=6.0in]{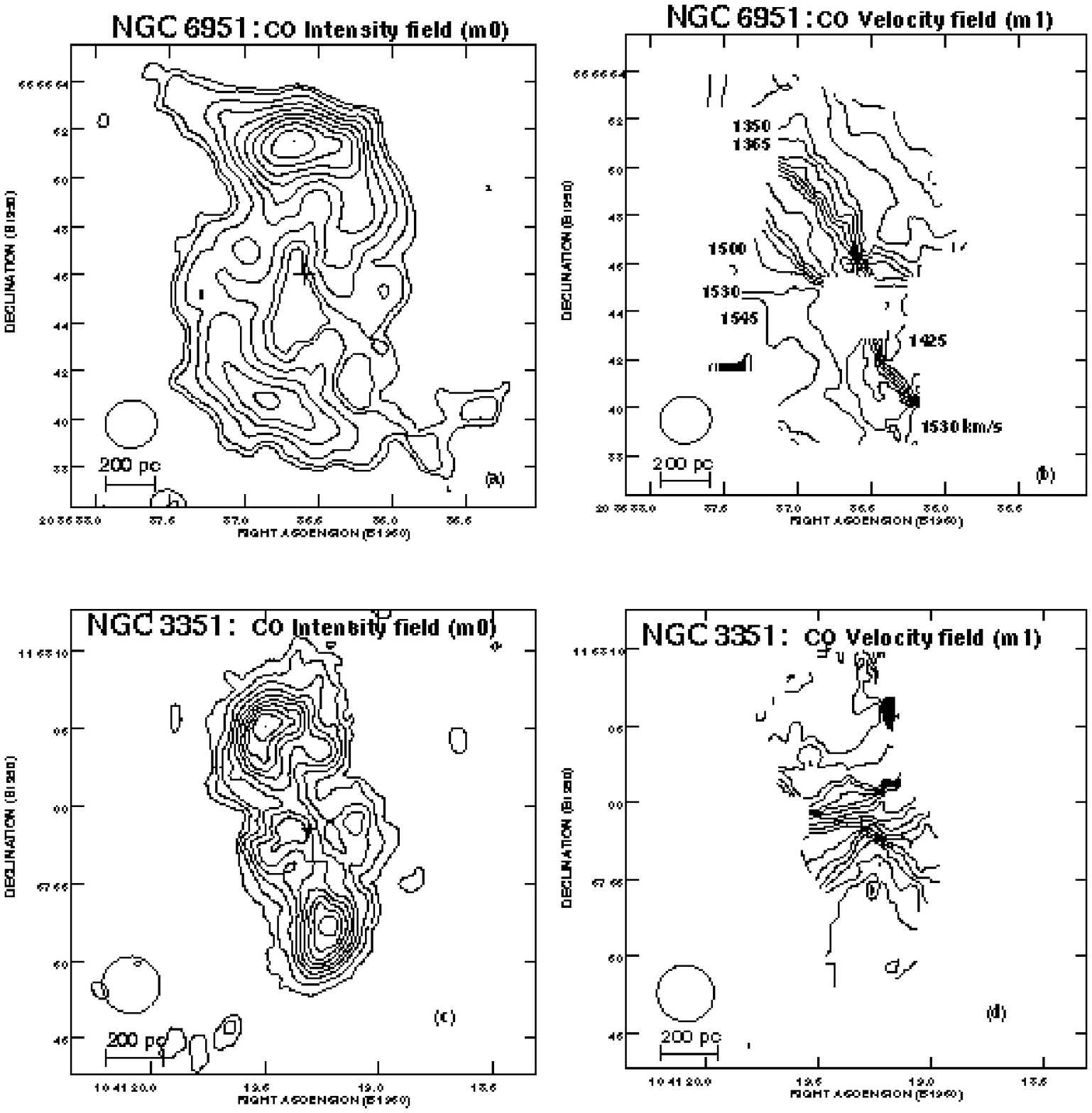}
\end{center}
\vspace{3 mm}
\begin{description}
\item[]
\noindent
\bf
Fig. 15--- 
\bf 
Molecular gas  kinematics  and distribution in the individual non-starbursts: 
\rm
As in Fig. 14, but for non-starbursts 
NGC 6951 (a-b), NGC 3351 (c-d), and  NGC 4314 (e-f). 
The contour levels plotted are specified in Table 10. 
NGC 6951 has two spiral-shaped CO arms where the CO emission peaks 
at a  radius r  $\sim$ $6''$ (570 pc). 
NGC 3351 hosts  two CO peaks at  a  radius r  $\sim 7''$ 
(350 pc). In  NGC 4314, there is 
a relatively circular CO ring of $8''$ (400 pc) 
radius from which extend two CO spurs/streams. 
The CO peaks in both NGC 6951 and NGC 3351 lie almost 
along the minor axis of the large-scale  stellar bar. 
The CO arms in NGC 6951, the fainter emission around 
the CO peaks in NGC 3351, and the two CO spurs in NGC 4314 
intersect the dust lanes  on the 
leading edge of the large-scale stellar bar and show 
non-circular motions.
\end{description}

\clearpage
\setcounter{figure}{13}
\vspace{5 mm}
\begin{center}
\includegraphics [width=6.0in]{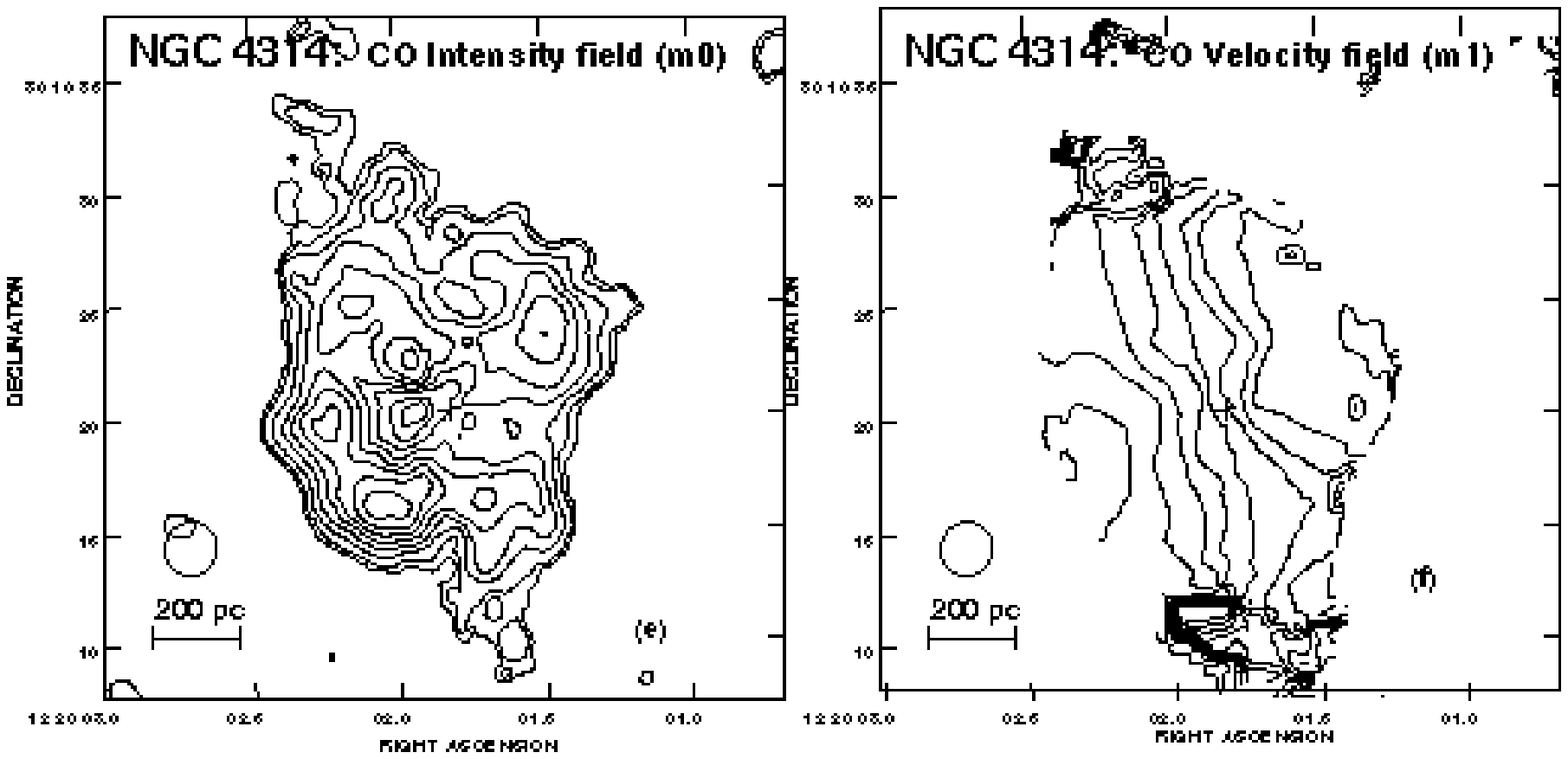}
\end{center}
\vspace{3 mm}
\begin{description}
\item[]
\noindent
\bf
Fig.15--- Continued
\bf
\end{description}



%
%
%
%
%
%

\clearpage
\setcounter{table}{0}
\begin{deluxetable}{lcccccccc}
\tabletypesize{\scriptsize}
\tablewidth{0pt}
\tablecaption{The Sample}
\tablehead{
\colhead {NGC} & 
\colhead {R.A. } &
\colhead {Dec} &
\colhead {Type } &
\colhead {$i$ } &
\colhead { $D$ } &
\colhead { $D_{\rm\tiny 25}$  } &
\colhead { $m_{\rm \tiny B}$  } &
\colhead {  Notes } \\
\colhead{}   & 
\colhead{ (1950.0) }   & 
\colhead{ (1950.0) }   & 
\colhead{ (RC3) }   & 
\colhead{ ($\deg$) }   & 
\colhead{ (Mpc)  }   & 
\colhead{ (')  }   & 
\colhead{  }   &
\colhead{  }   \\
\colhead {(1)} & \colhead {(2)} & \colhead {(3)} & \colhead {(4)} &
\colhead {(5)} & \colhead {(6)} & \colhead {(7)} & 
\colhead {(8)} & \colhead {(9)} \\
}
\startdata
\multicolumn {9}{c} {\small CIRCUMNUCLEAR STARBURSTS} \\
NGC 470& 01:17:10.5 & 03:08:53 &SABb(rs) & 57&30.5&3.2 &12.37&D,HII   \\
NGC 2782 & 09:10:54.1 & 40:19:18 &SABa(rs)pec & 30 &34 &3.9 &12.0&D,HII  \\
(Arp 215)&  &  &  &   &  &  & &  \\
NGC 3504&11:00:28.1&28:14:35&SABab(s)&22&20.0&2.6&11.74& D,HII \\
NGC 4102&12:03:51.6&52:59:23&SABb(s) & 52& 17.0 &3.1 &12.13& D,L \\
NGC 4536&12:31:53.6&02:27:50 &SABbc(rs)& 66&18.0&6.3&10.60&HII,L  \\
\hline\\
\multicolumn {9}{c} {\small CIRCUMNUCLEAR NON-STARBURSTS} \\
NGC 3351 &10:41:19.7 &11:58:00 &SBb(r) & 46 &10.1& 7.5&10.28& HII  \\
NGC 3359 & 10:43:21.1&63:29:11  &SBc(rs) & 55 &19.2& 7.2& 10.83& HII \\
NGC 4314&12:20:2.1&30:10:25&SBa(rs)& 21&9.7&4.2&11.28 & L  \\
NGC 4569& 12:34:18.7 &13:26:20  &SABab(rs) & 64&16.8&8.8 &9.86&HII,L \\
NGC 6951 &20:36:37.7&65:55:48 &SABbc(rs) & 44 &18.8& 3.7&11.18 &SEY2 \\
%
\enddata
\tablecomments{ Columns are : 
(1)  NGC number; 
(2) (3) Right Ascension and Declination (Equinox 1950.0) 
of the optical center, from  the Third Reference Catalog of Bright 
Galaxies,  hereafter RC3 (de Vaucouleurs et al. 1991);
(4) Hubble types from  RC3; 
(5) \it i \rm  in degrees, the inclination of the galaxy. Sources are :  
NGC 2782 (Jogee et al.  1998, 1999), 
NGC 3351 (Grosbol 1985), NGC 3359 (Martin \& Roy 1995), 
NGC 6951 (Grosbol 1985),  NGC 4314 (Benedict et al. 1996), 
and the Nearby Galaxies (NBG) Catalogue (Tully 1988) for 
the remaining galaxies.
(6)  $D$, the adopted distance in Mpc. Sources are : 
NGC 4536 (Saha et al. 1996), 
NGC 3504 (Kenney et al. 1993),  NGC 2782 (Jogee et al. 1998, 1999), and  
for the remaining galaxies, the NBG Catalogue (Tully 1988) which assumes a 
Hubble constant of 75 km~s$^{-1}$~Mpc$^{-1}$; 
(7) $D_{\tiny \rm 25}$ in arcminutes, the diameter  of the isophote where the 
surface brightness is 25 magnitude arcsecond$^{-2}$ 
in blue light. Values are from the NBG Catalogue (Tully 1988).
(8)  $m_{\tiny \rm B}$, the blue apparent magnitude  corrected for reddening, 
from the NBG Catalogue (Tully 1988); 
(9)  Description of the nuclear activity and optical 
emission line spectrum of the galaxy.
D denotes a galaxy belonging to 
Devereux's (1989)  sample of circumnuclear starbursts.
H denotes an HII region-like spectrum (Giuricin et al. 1994), 
L denotes a LINER  (Guiricin et al. 1994), and SEY2
denotes a Seyfert 2 (Guiricin et al. 1994; Wozniak et al. 1995).
}
\end{deluxetable}

\setcounter{table}{1}
\begin{deluxetable}{lcccccc}
\tabletypesize{\scriptsize}
\tablewidth{0pt}
\tablecaption{Single Dish CO Luminosities and Tracers of Star Formation}
\tablehead{
\colhead {NGC} & 
\colhead { $S_{\tiny \rm CO,S}$ } &
\colhead { $L_{\tiny \rm CO}$ } & 
\colhead { $L_{\tiny \rm RC}$ } & 
\colhead { $L_{10\micron}$  } & 
\colhead { $L_{\tiny \rm FIR}$} & 
\colhead { $L_{\tiny \rm RC}$/L$_{\tiny \rm CO}$ } \\
\colhead {  } &
\colhead {$45\arcsec$  } &
\colhead {$45\arcsec$  } &
\colhead {$45\arcsec$  } &
\colhead {$5''-10''$  } &
\colhead {Global  } &
\colhead {$45\arcsec$   } \\
\colhead {  } &
\colhead {(Jy kms$^{-1}$)  } &
\colhead { ($L_{\tiny \sun}$) } &
\colhead { ($L_{\tiny \sun}$) } &
\colhead {($L_{\tiny \sun}$)  } &
\colhead { ($L_{\tiny \sun}$) } &
\colhead {  } \\
\colhead {(1)} & \colhead {(2)} & \colhead {(3)} & \colhead {(4)} &
\colhead {(5)} & \colhead {(6)} & \colhead {(7)} \\
}
\startdata
\multicolumn {7}{c} {\small CIRCUMNUCLEAR STARBURSTS} \\
470 &68 & 3.83 & 4.17 & 9.12 &10.21 & 2.10 \hfill  \\
2782 & 266 & 4.50 &   4.72 &9.02 & 10.42& 1.60 \hfill  \\
3504 & 450  & 4.30 & 4.60 & 9.11 & 10.31& 1.99  \\ 
4102 & 572 & 4.26 & 4.50& 9.23  & 10.55& 1.74  \\
4536 & 607  & 4.33 & 4.35& 8.77 & 10.32& 1.03  \\  
\hline\\
\multicolumn {7}{c} {\small CIRCUMNUCLEAR NON-STARBURSTS} \\
3351 & 479  & 3.73  & 3.28& 7.51  &9.52& 0.35  \\
3359  & 162  & 3.82 & 3.19& --  & 9.90 & 0.23  \\
4314 & 251  & 3.41 & 2.62& $<$7.38 & 8.78& 0.16  \\
4569 & 870  & 4.43 & 3.62& 8.48  & 9.98& 0.15  \\
6951 & 496  & 4.28 & 3.85& --  & 9.95 & 0.37  \\
%
\enddata
\tablecomments{ Columns are : 
(1)  NGC number;
(2) $S_{\tiny \rm CO,S}$  in Jy  km s$^{-1}$, the single dish CO flux 
measured in the  central $45\arcsec$  for all galaxies, except 
NGC 4314 where the aperture is  $55''$.
S$_{\tiny \rm CO,S}$  was estimated from   observations
with the  Five College Radio Observatory (FCRAO) 14-m dish  
(Young et al. 1995) and  NRAO 12-m single dish observations 
 (Sage 1993). 
To derive flux densities S$_{\tiny \rm CO,S}$ in units of  
Jy  km s$^{-1}$ from the integrated intensity 
I$_{\tiny \rm CO}$ in  units of   K(T$_A$$^{*}$) km s$^{-1}$, 
we adopted a main beam efficiency of  42 Jy  K$^{-1}$   and 
35 Jy  K$^{-1}$    respectively for the 
FCRAO (Young et al. 1995), and  NRAO (e.g., Maiolino et al. 1997) 
observations,   and a  source-beam coupling efficiency of 0.7; 
(3) $L_{\tiny \rm CO}$ in logarithmic   units of  $L_{\tiny \sun}$, 
  the single dish CO luminosity determined from   $S_{\tiny \rm CO,S}$;
(4)  $L_{\tiny \rm RC}$ in  logarithmic units of  $L_{\tiny \sun}$, 
the  RC  luminosity  at 1.5 GHz in the central $45\arcsec$  
(Condon et al. 1990);
(5) L$_{\tiny 10 \micron}$ in logarithmic units of $L_{\tiny \sun}$, 
the nuclear 10 
$\micron$ luminosity in a central $5''$-$10''$ aperture 
(Giuricin et al. 1994); 
(6) $L_{\tiny \rm FIR}$ in logarithmic  units of  $L_{\tiny \sun}$, 
 the far-infrared  
luminosity   from the whole galaxy , measured in a bandpass 80 $\micron$ 
wide centered on a wavelength of  82.5 $\micron$, from 
the IRAS Bright Galaxy Sample (Soifer et al. 1989); 
(7) $L_{\tiny \rm RC}$/$L_{\tiny \rm CO}$, the ratio  of the RC  
luminosity at 1.49 GHz to the  CO luminosity. Both luminosities are  
measured within the same  $45\arcsec$ aperture, except for NGC 4314. 
}
\end{deluxetable}

\setcounter{table}{2}
\begin{deluxetable}{lcl}
\tabletypesize{\scriptsize}
\tablewidth{0pt}
\tablecaption{External Disturbances  and Large-Scale Stellar Bars} 
\tablehead{
\colhead {NGC} & 
\colhead {Bar Type$^{f}$   } & 
\colhead {Evidence  for External Disturbances } \\
\colhead {(1)} & \colhead {(2)} & \colhead {(3)} \\
}
\startdata
\multicolumn {3}{c} {\small CIRCUMNUCLEAR STARBURSTS} \\
470 & SAB & Outer disk has asymmetric  H$\alpha$ emission$^{h}$  and shows evidence\\
   &   &for warping. A recent interaction with 474$^{b}$ is likely.  \\
2782 & SAB& Outer galaxy shows HI tails, optical tails, and ripples.$^{c,d,e}
$. A close \\
 &  &  and intermediate mass-ratio interaction with a disk galaxy$^{c}$ is likely. \\
3504  & SAB & Belongs to a group. $^{k}$ Possibly interacting with the small 
galaxy  3512.\\
4102  & SAB  & Belongs to a group  whose  brightest  member is NGC~3992.$^{j}$ \\
4536 & SAB & Is in the Southern Extension of the Virgo cluster.$^{l}$ \\
     &  & Possibly interacting with 4527. \\
\hline\\
\multicolumn {3}{c} {\small CIRCUMNUCLEAR NON-STARBURSTS} \\
3351 &SB  & Belongs to a group. $^{k}$  \\
3359 & SB & Shows evidence$^{m,n}$  for a bar which is young and possibly \\
   &   & tidally triggered. Has a possible small bound companion.$^{g}$ \\
4314 & SB& Belongs to a   group.$^{11}$ Outer disk is unusually gas poor.  \\
     &   & The total HI to H$_{2}$ mass ratio is unusually low ($\sim$ 0.02) 
\\
4569 & SAB & Is located in projection within 2 $\deg$ of the center of the Virgo Cluster.\\
     &   & Has anemic spiral structures$^{o}$ and a truncated  HI disk.$^{i}$ 
 Optical images $^{a}$  \\
      &   & suggest an outer warped disk. In the inner 3.5 kpc radius, the K-band  \\ 
     &   & image$^{h}$ shows an asymmetric bar, and  the molecular gas has very  \\
     &    & disturbed kinematics$^{h}$ .\\
6951 & SAB & Isolated galaxy.  \\
\enddata
\tablecomments{ Columns are : 
(1)  NGC number; 
(2) Bar Type according to RC3 (de Vaucouleurs et al. 1991); 
(3) Evidence  for External Disturbances
}
\tablerefs{
a. Sandage \& Bedke 1994; b. Schweizer \& Seitzer 1988; c. Smith 1994; 
d. Jogee et al. 1998 ; e. Jogee et al. 1999; 
f. RC3 (de Vaucouleurs et al. 1991); 
g. Ball 1986; 
h. Jogee 1999 ;
i. Giovanelli  \& Haynes 1983;
j. Geller \& Huchra 1983 ; k.  Tully 1988;
l. Bingelli, Sandage, \& Tammann 1985; m. Friedli, Benz, \& Kennicutt 1994;  
n. Martin \& Roy 1995; 
o. van den Bergh 1976; 
}
\end{deluxetable}

\setcounter{table}{3}
\begin{deluxetable}{lcccc}
\tabletypesize{\scriptsize}
\tablewidth{0pt}
\tablecaption{ CO (J=1-$>$ 0) observations \& channel maps}
\tablehead{
\colhead {NGC} & 
\colhead { Synthesised } & 
\colhead {Weight  } & 
\colhead {R.m.s.  } & 
\colhead {T$_{\tiny \rm B}$ for  } \\
\colhead {  } & 
\colhead {Beam  } & 
\colhead {  } & 
\colhead {noise  } & 
\colhead {1 Jy/Beam } \\
\colhead {  } & 
\colhead {  } & 
\colhead {  } & 
\colhead { (mJy/Beam) } & 
\colhead { (K) } \\
\colhead {(1)} &
\colhead {(2)} &
\colhead {(3)} & \colhead {(4)} &
\colhead {(5)} \\
}
\startdata
470    & 
3.61''x3.46''= 550x530 pc & U &
22.0 & 7.4 \\ 
2782 &   
2.14''x1.52''= 364x258 pc  & U & 
13.0 &28.2  \\
                         & 2.62''x1.98''= 445x337 pc  & N& 
10.5 & 17.7  \\
3504 & 
2.50''x2.40''= 250x240 pc  & U & 
35.0 &15.3 \\
4102 & 
2.40''x1.96''= 204x167 pc  & U & 
15.0 &19.5  \\
                         &  3.12''x2.56''= 265x218 pc  & N & 
13.0 &11.5  \\
4536 &  
2.63''x1.88''= 235x170 pc  & U & 
15.0 & 18.6 \\
                         & 2.82''x2.37''= 255x215 pc  &  N & 
12.0 & 13.7  \\
3351  & 
2.38''x2.34''= 120x118 pc  & U & 
30.0 & 16.5 \\
3359  & 
3.38''x2.91''= 324x279 pc  & N & 
16.0 & 9.3  \\
4314 & 
2.29''x2.17''= 110x105 pc &  U & 
28.0 & 18.5 \\
                     & 3.12''x2.70''= 150x130 pc  &  N & 
22.0 & 10.9  \\
4569 & 
2.65''x2.01''= 225x170 pc  &  U & 
 16.0 & 9.7  \\
                      & 3.22''x2.95''= 275x251 pc  & N &
14.0 & 17.4  \\
6951 & 
2.28''x2.11''= 110x102 pc & U & 
19.2  & 19.1  \\
%
\enddata
\tablecomments{ Columns are : 
(1)  NGC number; 
(2)  The size of the synthesized beam ;
(3) The weighting used for the channel maps: `U' refers to 
uniform and `N' to natural weighting; 
(4)  The r.m.s. noise in mJy beam$^{-1}$ for the channel maps;
(5) The equivalent  brightness temperature  (T$_{\rm \tiny B}$) in K  for 
1 Jy beam $^{-1}$.
}
\end{deluxetable}

\setcounter{table}{4}
\begin{deluxetable}{lcccc}
\tabletypesize{\scriptsize}
\tablewidth{0pt}
\tablecaption{Cicumnuclear Molecular Properties}
\tablehead{
\colhead {NGC} & 
\colhead { $S_{\tiny \rm CO,I}$  } &
\colhead { $M_{\rm \tiny H2}$  } &
\colhead {$V_{\rm sys}$   } &
\colhead { $f_{\rm \tiny SD}$  } \\
\colhead {  } & 
\colhead { (Jy kms$^{-1}$)   } & 
\colhead { ($M_{\tiny \sun}$)  } & 
\colhead {  (km s$^{-1})$) } & 
\colhead { (\%)   } \\
\colhead {(1)} & \colhead {(2)} & \colhead {(3)} & \colhead {(4)} &
\colhead {(5)}\\
}
\startdata
\multicolumn {5}{c} {\small CIRCUMNUCLEAR STARBURST} \\
470 & 53 &  5.0E8 & 2350  & 80 \hfill  \\
2782 & 184 &  2.3E9 & 2555  &65 \hfill \\
3504& 280 & 1.2E9 & 1535 & 71 \hfill \\
4102 & 450 & 1.4E9 & 835  &80 \hfill \\ 
4536 & 355 &1.2E9 &  1800 & 60 \hfill \\
\hline\\
\multicolumn {5}{c} {\small CIRCUMNUCLEAR NON-STARBURSTS} \\
3351  & 241 & 2.7E8& 780 & 60 \hfill \\
3359 & 20 & 8.0E7 & - &  12$^{1}$  \hfill \\
4314 & 248 & 2.6E8 &  983 & 92  \hfill \\
4569 & 550 & 1.7E9 & -235 & 65  \hfill \\
6951 & 299 & 1.2E9 & 1431 & 60  \hfill \\
%
\enddata
\tablecomments{ Columns are : 
(1)  NGC number; 
(2) $S_{\rm \tiny CO,I}$ in Jy  km s$^{-1}$, the flux detected by the 
OVRO interferometric observations; 
(3) M$_{\rm \tiny H2}$  in solar units,  the mass of molecular hydrogen  
corresponding to $S_{\rm \tiny CO, I}$,  if one  assumes a standard Galactic
CO-to-H$_{\rm 2}$  conversion factor; 
(4) $V_{\rm sys}$  in  km s$^{-1}$, the  systemic velocity 
estimated  from the  moment 1 map.
(5)  $f_{\tiny \rm SD}$, the fraction  of the single dish flux detected by 
the interferometric observations.  See text for a discussion  of 
the low $f_{\tiny \rm SD}$ for NGC 3359.
}
\end{deluxetable}

\setcounter{table}{5}
\begin{deluxetable}{lccc}
\tabletypesize{\scriptsize}
\tablewidth{0pt}
\tablecaption{Molecular Environments in the  Circumnuclear Region 
vs. the Outer Disk}
\tablehead{
\colhead{Quantities $^{a}$  }   & \colhead{Outer Disk}      &
\colhead{Inner 500 pc radius }          & \colhead{Inner 500 pc radius } \\
\colhead{}   & \colhead{ of Normal}      &
\colhead{of Sample Starbursts }          & \colhead{of ULIRG }\\ 
\colhead{}   & \colhead{Sa-Sc Spirals}      &
\colhead{and Non Starbursts}  & \colhead{Arp 220}\\
}
\startdata
(1) $M_{\rm gas,m}$ [$M_{\tiny \sun}$] & 
$\le$ few $\times 10^{9}$ $^{b}$ & 
Few $ \times$ (10$^{8}$-10$^{9}$) $^{\tiny c}$  & 
$ 3 \times 10^{9}$   $^{\tiny d}$   \nl
(2) M$_{\rm gas}$/M$_{\rm dyn}$  [\%] & 
$<$ 5  $^{e}$         & 
10 to 30  $^{c}$       &  
40 to 80  $^{d}$       \\
(3) SFR [$M_{\tiny \sun}$ yr$^{-1}$] & -  & 
0.1-11 $^{\tiny c}$  
& $>$ 100$^{g}$  \nl
(4) $\Sigma_{\rm gas,m}$ [$M_{\tiny \sun}$ pc$^{-2}$] & 
1-100 $^{f}$ & 
500 to  3500  $^{c}$  &  
$ 4 \times 10^{4}$  $^{d}$ \\
(5)  $\sigma$  [km s$^{-1}$] & 
6-10$^{\tiny h}$ & 
10 to 40$^{i} $ 
& 90$^{g}$ \nl
(6) $\kappa$ [km s$^{-1}$  kpc$^{-1}$] & 
$<$ 100$^{j}$ & 
800 to 3000  $^{c}$ &  
$>$ 1000$^{g}$ \\
(7) $\Sigma_{\rm crit}$ [$M_{\tiny \sun}$ pc$^{-2}$]& 
$<$ 10   $^{ k}$  & 500-1500  $^{l}$ & 2200 $^{m}$ \nl
(8) t$_{\tiny \rm GI}$ [Myr]& 
$>$10 $^{k}$   & 0.5-1.5  $^{l}$  & 0.5  $^{m}$ \nl
(9) $\lambda_{\tiny \rm J}$ [pc]&
Few $\times$ (100-1000)  $^{ k}$  & 100-300  $^{l}$ & 90 $^{m}$ \nl
(10) $M_{\tiny \rm J}$ [$M_{\tiny \sun}$] &
 Few  $\times$ (10$^{7}$-10$^{6}$) $^{k}$  &
 Few $\times$ (10$^{7}$-10$^{8}$)  $^{l}$ 
& $8 \times 10^{8}$ $^{m}$ \nl
%
%
\enddata
\tablecomments{Rows are : 
(1) $M_{\rm gas,m}$ in $M_{\tiny \sun}$, the molecular gas 
content including hydrogen and helium. 
A  standard CO-to-H$_{\rm 2}$ conversion factor and a solar metallicity 
are assumed;
(2) M$_{\rm gas,m}$/M$_{\rm dyn}$ in \%, the ratio of  molecular gas mass 
to dynamical mass; 
(3) $\Sigma_{\rm gas,m}$ in $M_{\tiny \sun}$ pc$^{-2}$, the molecular 
gas surface density; 
(4) $\Sigma_{\tiny SFR}$  in $M_{\tiny \sun}$ yr$^{-1}$ pc$^{-2}$, 
the SFR per unit area; 
(5) $\sigma$  in km s$^{-1}$, the gas velocity dispersion; 
(6) $\kappa$  in km s$^{-1}$  kpc$^{-1}$,  the epicyclic frequency; 
(7) $\Sigma_{\rm crit}$  in $M_{\tiny \sun}$ pc$^{-2}$, the critical 
density for the onset of gravitational instabilities as defined 
in $\S$ 5.3.1, with $ \alpha$ = 0.7 ; 
(8) t$_{\tiny \rm GI}$ = Q/$\kappa$  in Myr,  the growth  
timescale of the most unstable wavelength associated with 
gravitational instabilities, assuming  Q $\sim$ 1;
(9) $\lambda_{\tiny \rm J}$ in pc, the Jeans length 
(10) $M_{\tiny \rm J}$ in $M_{\tiny \sun}$, the  Jeans mass.
}
\tablecomments{ 
$^{a}$  Typical values are quoted for the quantities. Individual 
cases may vary;
$^{b}$ Young \& Scoville 1991;
$^{c}$ This work; 
$^d$  Sakamoto et al. (1999), for a  
self-gravitating gas disk. Quoted mass is the sum for  Arp 220 W and E; 
$^e$  Binney \& Tremaine 1987; 
$^f$ Devarheng et al. 1994; 
$^g$ Scoville, Yun, and Bryant 1997
$^h$  Dickey et al. 1990; 
$^i$ The upper limit for the  velocity dispersion  is quoted 
after correcting for beam smearing; 
$^j$  Larson 1988; 
$^k$ For the outer disk, (7) to (10) are computed 
for 
$\kappa$ $<$ 100 km s$^{-1}$ kpc$^{-1}$, 
$\sigma$ $\sim$ 6  km s$^{-1}$, 
and $\Sigma_{\rm gas,m}$ = 10-100 $M_{\tiny \sun}$ pc$^{-2}$; 
$^l$ 
Quantities (7) to (10)  are computed at r=250 pc in 
the circumnuclear starbursts and non-starbursts 
NGC 4102, NGC 4536, NGC 3504, NGC 3351 and NGC 4314; 
$^m$ Quantities (7) to (10) are computed  for Arp 220 W
assuming $\kappa$ $\sim$ 2000  km s$^{-1}$ kpc$^{-1}$, 
$\sigma$ $\sim$ 90  km s$^{-1}$, 
and $\Sigma_{\rm gas,m}$ $\sim 4 \times 10^{4}$  
$M_{\tiny \sun}$ pc$^{-2}$.  
}
%
%
\end{deluxetable}

\setcounter{table}{6}
\begin{deluxetable}{lcccc}
\tabletypesize{\scriptsize}
\tablewidth{0pt}
\tablecaption{  SF Rates from FIR, RC, and Br$\gamma$ data}
\tablehead{
\colhead {NGC} & 
\colhead {Based on global $L_{\tiny \rm FIR}$    } & 
\colhead {Based on RC    } & 
\colhead {Based on Br$\gamma$  } \\
\colhead {   } & 
\colhead { $L_{\tiny \rm FIR}$, SFR$_{\tiny \rm  FIR}$  } & 
\colhead { $\nu$, $S_{\nu}$, $R_{\tiny \rm RC}$, SFR$_{\tiny \rm RC-N}$     } & 
\colhead {  $R_{\tiny Br\gamma}$, $N_{\tiny \rm Ly}$, SFR$_{\tiny Br\gamma}$  } \\
\colhead {   } & 
\colhead {  ($L_{\tiny \sun}$, $M_{\tiny \sun}$ yr$^{-1}$)  } &
\colhead {(GHz, mJy, pc, $M_{\tiny \sun}$ yr$^{-1}$)   } & 
\colhead { (pc, s$^{-1}$, $M_{\tiny \sun}$ yr$^{-1}$)  } \\
\colhead {(1)} & \colhead {(2)} & \colhead {(3)} & \colhead {(4)} \\
}
\startdata
\multicolumn {4}{c} {\small CIRCUMNUCLEAR STARBURSTS} \\
470 & 10.21, 6.3  & 1.5, 26.6$^{a}$, 1680, 2.8 &  - \hfill \\
2782 & 10.42, 10.2 & 1.5, 107$^{a}$, 2210, 13.8 & 1650, 5.0E53$^{g}$, 5.6  \hfill\\
     &               &  4.9, 40$^{e}$, 1700, 11.6 &  \hfill \\ 
3504 & 10.31, 8.0 & 1.5, 230$^{a}$, 800, 10.2 & 945, 1.5E53$^{g}$, 2.8 \hfill\\
4102 & 10.55, 13.8 & 1.5, 227$^{a}$, 1105, 7.3 & 945, 9.2E52$^{g}$, 1.1 \hfill\\
4536 & 10.32, 8.1 & 1.5, 149$^{a}$, 1800, 5.4 & 840, 4.0E53$^{g}$, 4.8  \hfill\\
     &              & 4.9, 61$^{f}$, 900, 4.9 &   \hfill \\
\hline\\
\multicolumn {4}{c} {\small CIRCUMNUCLEAR NON-STARBURSTS} \\
3351  & 9.52, 1.3 & 1.5, 28$^{b}$, 1000, 0.3 & 490, 6.0E52$^{g}$, 0.5 \hfill \\
3359 & 9.90, 3.1 & 1.5, 50$^{a}$, 20160, 2.1 & -  \hfill \\
4314 & 8.78, 0.2 & 1.5, 12.5$^{c}$, 3640, 0.1 & - \hfill \\
4569 & 9.98, 3.7 & 1.5, 83$^{c}$, 8420, 2.5 & - \hfill \\
6951 & 9.95, 3.5 & 4.9, 14$^{e}$, 750, 1.2  & - \hfill \\
\enddata
\tablecomments{ Columns are : 
(1) NGC number; 
(2) $L_{\tiny \rm FIR}$   in logarithmic units of $ L_{\tiny \sun}$, 
the  global FIR luminosity; 
SFR$_{\tiny \rm FIR}$ in $M_{\tiny \sun}$  yr$^{-1}$,  the  global 
SFR for the entire galaxy; 
(3) $ \nu$   in GHz, the frequency of the RC  observations; 
$S_{\nu}$ in mJy, the RC  flux density; 
$R_{\tiny RC-N}$  in  pc, the radius within which more than 
90 \% of the RC emission is concentrated; 
SFR$_{\tiny RC-N}$  in $M_{\tiny \sun}$  yr$^{-1}$,  the  SFR within
$R_{\tiny RC}$, estimated from  the non-thermal  component of the radio 
continuum flux density; 
(4) $R_{\tiny Br\gamma}$  in pc, the aperture  radius  for the  Br$\gamma$  
observations; $N_{\tiny Ly}$    in s$^{-1}$,  the number 
of Lyman continuum photons per seconds from Puxley et al. (1990); 
SFR$_{\tiny Br\gamma}$ in $M_{\tiny \sun}$  yr$^{-1}$, the  SFR  within 
$R_{\tiny Br\gamma}$ 
}
\tablerefs{a. Condon et al. 1990;  b. \rm Condon, Anderson, 
\& Broderick 1995; c. \rm Condon 1987 ; d. \rm  Hummel et al. 1985; 
e.  \rm Saikia et al. 1994; f. \rm Villa et al. 1990;   g. Puxley et al. 1990}
\end{deluxetable}

\setcounter{table}{7}
\begin{deluxetable}{lccccc}
\tabletypesize{\scriptsize}
\tablewidth{0pt}
\tablecaption{Adopted Circumnuclear SFRs over Radius of CO Emission}
\tablehead{
\colhead {NGC} & 
\colhead { $R_{\tiny \rm CO}$   } & 
\colhead {  $M_{\tiny \rm H2}$  } & 
\colhead { SFR$_{\rm CO}$  } &
\colhead {  SFR$_{\rm CO}$/$M_{\tiny \rm H2}$ } & 
\colhead { $t_{\tiny \rm SFR}$  } \\
\colhead {  } & 
\colhead { (pc) } & 
\colhead { ($M_{\tiny \sun}$) } & 
\colhead { ($M_{\tiny \sun}$ yr$^{-1}$) } & 
\colhead {  (yr$^{-1}$) } & 
\colhead { (10$^{8}$ yrs) } \\
\colhead {(1)} & \colhead {(2)} & \colhead {(3)} & \colhead {(4)} &
\colhead {(5)} & \colhead {(6)}\\
}
\startdata
\multicolumn {5}{c} {\small CIRCUMNUCLEAR STARBURST} \\
470 & 1100 & 5.0E8 &  3  & 6E-9  & 2 \hfill \\
2782 &1530 &  1.8E9 & 11 &  6E-9 & 2 \hfil \\
3504 & 1300 & 1.2E9 &  10 & 8E-9 & 1 \hfill \\
4102 & 1275 & 1.4E9 &  7 & 5E-9  & 2 \hfill \\
4536 & 1300 & 1.2E9 & 5  & 4E-9 &  2 \hfill \\
\hline\\
\multicolumn {5}{c} {\small CIRCUMNUCLEAR NON-STARBURSTS} \\
3351  & 600 &  5.3E8 & 0.5 & 9E-10  & 11 \hfill \\
4314 &  590 &  2.3E8 & $<$0.1 & $<$ 4E-10 & $>$ 25  \hfill \\
4569 &  1680 & 1.6E9 & $<$1.2 &   $>$ 7E-10  & $>$ 14 \hfill \\
\enddata
\tablecomments{ Columns are : 
(1)  NGC number; 
(2) $R_{\tiny \rm CO}$  in pc, the radius of CO emission;
(3) $M_{\tiny \rm H2}$  in $M_{\tiny \sun}$, the mass of molecular hydrogen 
within a radius  $R_{\tiny \rm CO}$; 
(4) SFR$_{\rm CO}$  in $M_{\tiny \sun}$  yr$^{-1}$,  the  SF rate 
within $R_{\tiny \rm CO}$;
(5) (SFR$_{\rm CO}$/$M_{\tiny \rm H2}$) in  yr$^{-1}$, the SFR per unit mass of 
molecular hydrogen;
(6)  $t_{\tiny \rm SFR}$) in  10$^{8}$ years, the average gas consumption 
timescale by  SF given by ($M_{\tiny \rm H2}$/SFR$_{1}$)
}
\end{deluxetable}

\setcounter{table}{8}
\begin{deluxetable}{lccccccccc}
\tabletypesize{\scriptsize}
\tablewidth{0pt}
\tablecaption{ Primary Bar Properties and Dynamical Resonances}
\tablehead{
\colhead {NGC} & 
\colhead {Bar } &
\colhead { $\epsilon_{1}$ } &
\colhead {PA$_{1}$ } &
\colhead {a$_{1}$ } &
\colhead { $M_{\rm gas,m}$ } &
\colhead { $\frac { M_{\rm gas,m}}{M_{\rm dyn}}$ } &
\colhead {$\Omega_{\rm p}$} &
\colhead {$R_{\tiny OILR}$ } &
\colhead {$R_{\tiny IILR}$ } \\
\colhead {(1)} & \colhead {(2)} & \colhead {(3)} & \colhead {(4)} &
\colhead {(5)} & \colhead {(6)} & \colhead {(7)} & 
\colhead {(8)} & \colhead {(9)} &
\colhead {(10)} \\
} 
\startdata
470  & AB & 0.55$^{a}$  & 14$^{a}$ & 4.8$^{a}$ & 2E8 & 30 & 
 \nodata  &  \nodata  &  \nodata \\
3504 &  AB & 0.58 $^{b}$ &  135$^{b}$ &  3.0$^{e}$ & 8E8 & 20 & 
  $<$ 76 & $>$ 900 & $<$ 200 \\
4102 & AB  & 0.45$^{e}$ &  62$^{e}$  &  1.8$^{e}$  &  8E8 & 25 & 
  $<$ 115 & $>$500 & $<$ 250 \\
4536 &  AB  & 0.40$^{b}$  & 140$^{b,c}$  & 3.1$^{c}$ & 1E9 & 30 & 
 $<$ 50 & $>$ 800 & $<$ 200 \\
3351 & B  & 0.57 $^{d}$ & 115$^{c}$ &  2.2$^{c,d}$ & 2E8 & 17 & 
$<$ 85 & $>$ 300 & $<$ 300 \\
3359 & B   &  0.68 $^{d}$ &  24 $^{c}$ &  2.9$^{d}$ & \nodata   & \nodata  & 
 \nodata  & \nodata   & \nodata  \\
 4314 & B  & 0.69$^{a}$ &  143$^{a}$ &  3.6$^{a}$ & 2E8 & 9 & 
 $<$ 78 & $>$ 700 & $<$ 200 \\
 4569 & AB  &  \nodata  & 15$^{e}$ & \nodata & 2E8 &  \nodata & 
 \nodata  & \nodata  & \nodata \\
6951 &  AB  & 0.59$^{a}$  & 85$^{a}$  &  5.2$^{a}$ &  3E8  & 21 & 
  $<$ 43 &  $>$ 650 & $<$ 300 \\
\enddata
\tablecomments{ Columns are : 
(1)  NGC number; 
(2) Bar type from  RC3 (de Vaucouleurs et al. 1991):  B denotes a strong 
bar, and AB a bar of intermediate strength;
(3) (4) (5)  $\epsilon_{1}$, PA$_{1}$  in $\deg$, a$_{1}$ in kpc, 
the ellipticity,   position angle, and  semi-major axis of the 
large-scale stellar bar;
(6) M$_{\rm gas,m}$ [r=300]           in $M_{\tiny \sun}$, the molecular gas 
mass enclosed within a 300 pc radius.
A  standard CO-to-H$_{\rm 2}$ conversion factor and a solar metallicity 
are assumed;
(7) $\frac { M_{\rm gas,m}}{M_{\rm dyn}}$ [r=300], the ratio of  molecular 
gas mass to dynamical mass  within a 300 pc radius; 
(8) Upper limit on $\Omega_{\rm p}$ in km s$^{-1}$ kpc$^{-1}$, 
the primary bar pattern speed; 
(9) Upper limit on $R_{\tiny OILR}$ in pc, the  radius of the Outer Inner Lindblad Resonance;
(10) Lower limit on $R_{\tiny IILR}$ in pc, the  radius of the Inner Inner Lindblad Resonance.
}
%
\tablerefs{
a. Friedli et al. 1996; b. Pompea \& Rieke 1990; 
c. Shaw et al. 1995; d. Martin 1995; 
e. This work.
}
\end{deluxetable}

\setcounter{table}{9}
\begin{deluxetable}{lll}
\tabletypesize{\scriptsize}
\tablewidth{0pt}
\tablecaption{ Contour Levels Plotted in Figures}
\tablehead{
\colhead {NGC} & 
\colhead {Figure} & 
\colhead {Contour levels }\\
} 
\startdata
\multicolumn {3}{c} {\small CIRCUMNUCLEAR STARBURSTS} \\
470  &  Figs. 3, 5, 14  & 
Levels = 2.48 Jy  beam$^{-1}$~km s$^{-1}$~$\times$~(1, 2, 3, 4, 5, 6, 7, 8, 9, 10)  \\
470  &  Fig. 14 & Levels = (2200, 2210, 2220, ....., 2470, 2480, 2490)~km s$^{-1}$  \\
2782  &  Figs. 3, 5, 14  & 
Levels = 7.15 Jy  beam$^{-1}$~km s$^{-1}$~$\times$~(1, 2, 3, 4, 5, 6, 7, 8, 9, 10)  \\
2782  &  Fig. 14 & Levels = (2275, 2295, 2315, ....., 2815, 2835, 2855)~km s$^{-1}$  \\
3504x  &  Figs. 3, 5, 14  & 
Levels = 8.33 Jy  beam$^{-1}$~km s$^{-1}$ ~$\times$~(0.5, 1, 2, 3, 4, 5, 6, 7, 8, 9, 10)  \\
3504 x &  Fig. 14 & Levels = (585, 600, 615, ....., 990, 1005, 1020)~km s$^{-1}$  \\
4102  &  Figs. 3, 5, 14  & 
Levels = 8.33 Jy  beam$^{-1}$~km s$^{-1}$~$\times$~(0.5, 1, 2, 3, 4, 5, 6, 7, 8, 9, 10)  \\
4102  &  Fig. 14 & Levels = (585, 600, 615, ....., 990, 1005, 1020)~km s$^{-1}$  \\
4536  &  Figs. 3, 5, 14  & 
Levels = 3.05 Jy  beam$^{-1}$~km s$^{-1}$~$\times$~(1, 2, 3, 4, 5, 6, 7, 8, 9, 10)  \\
4536  &  Fig. 14 & Levels = (1640, 1650, 1660, ....., 1910, 1920, 1930)~km s$^{-1}$  \\
\hline\\
\multicolumn {3}{c} {\small CIRCUMNUCLEAR NON-STARBURSTS} \\
3351  &  Figs. 3, 5, 15  & 
Levels = 1.84 Jy  beam$^{-1}$~km s$^{-1}$ ~$\times$~(1, 2, 3, 4, 5, 6, 7, 8, 9, 10)  \\
3351  &  Fig. 15 & Levels = (660, 670, 680, ....., 930, 940, 950)~km s$^{-1}$  \\
4314  &  Figs. 3, 5, 15  & 
Levels = 1.34 Jy  beam$^{-1}$~km s$^{-1}$~$\times$~ (1, 2, 3, 4, 5, 6, 7, 8, 9, 10)  \\
4314  &  Fig. 15 & Levels = (915, 930, 945, ....., 1050, 1065, 1080)~km s$^{-1}$  \\
4569  &  Figs. 3, 5, 15  & 
Levels = 3.20 Jy  beam$^{-1}$~km s$^{-1}$~$\times$~(1, 2, 3, 4, 5, 6, 7, 8, 9, 10)  \\
6951  &  Figs. 3, 5, 15  & 
Levels = 2.92 Jy  beam$^{-1}$~km s$^{-1}$~$\times$~(1, 2, 3, 4, 5, 6, 7, 8, 9, 10)  \\
6951  &  Fig. 15 & Levels = (1275, 1290, 1305, ....., 1680, 1695, 1710)~km s$^{-1}$  
\enddata
\tablecomments{ Columns are : 
(1)  NGC number; (2) Figures ; (3) Countour levels  plotted.
}
%
\end{deluxetable}

\end{document}